\documentclass[12pt]{article}

\global\arraycolsep=1pt
\usepackage{graphicx}
\usepackage{wrapfig}
\usepackage{amssymb}
\usepackage{amsmath}
\usepackage{amscd}
\usepackage {pstricks}
\newcommand{\beq}{\begin{equation}}
\newcommand{\eeq}{\end{equation}}
\newcommand{\beqa}{\begin{eqnarray}}
\newcommand{\eeqa}{\end{eqnarray}}
\newcommand{\CR}{\nonumber \\}

\newcommand{\lam}{\lambda}

\newcommand{\ep}{\epsilon_1}
\newcommand{\es}{\epsilon_2}

\newcommand{\Langle}{ \langle\!\langle}
\newcommand{\Rangle}{ \rangle\!\rangle}
\newcommand{\nbf}[5]{n_f^S[{#1}, {#2}] ({#3}, {#4}~; {#5})}

\newcommand{\floor}[1]{\lfloor#1\rfloor}
\def\IZ{\mathbb {Z}}

\def\IC{\mathbb {C}}
\newcommand{\IP}{{\relax{\rm I\kern-.18em P}}}
\newcommand{\IF}{{\relax{\rm I\kern-.18em F}}}

\newdimen\tableauside\tableauside=1.0ex
\newdimen\tableaurule\tableaurule=0.4pt
\newdimen\tableaustep
\def\phantomhrule#1{\hbox{\vbox to0pt{\hrule height\tableaurule width#1\vss}}}
\def\phantomvrule#1{\vbox{\hbox to0pt{\vrule width\tableaurule height#1\hss}}}
\def\sqr{\vbox{%
  \phantomhrule\tableaustep
  \hbox{\phantomvrule\tableaustep\kern\tableaustep\phantomvrule\tableaustep}%
  \hbox{\vbox{\phantomhrule\tableauside}\kern-\tableaurule}}}
\def\squares#1{\hbox{\count0=#1\noindent\loop\sqr
  \advance\count0 by-1 \ifnum\count0>0\repeat}}
\def\tableau#1{\vcenter{\offinterlineskip
  \tableaustep=\tableauside\advance\tableaustep by-\tableaurule
  \kern\normallineskip\hbox
    {\kern\normallineskip\vbox
      {\gettableau#1 0 }%
     \kern\normallineskip\kern\tableaurule}%
  \kern\normallineskip\kern\tableaurule}}
\def\gettableau#1 {\ifnum#1=0\let\next=\null\else
  \squares{#1}\let\next=\gettableau\fi\next}

\makeatletter
 \renewcommand{\theequation}{%
       \thesection.\arabic{equation}}
 \@addtoreset{equation}{section}
\makeatother

\makeatletter
\def\eqnarray{%
 \stepcounter{equation}%
 \let\@currentlabel=\theequation
 \global\@eqnswtrue
 \global\@eqcnt\z@
 \tabskip\@centering
 \let\\=\@eqncr
 $$\halign to \displaywidth\bgroup\@eqnsel\hskip\@centering
 $\displaystyle\tabskip\z@{##}$&\global\@eqcnt\@ne
 \hfil$\displaystyle{{}##{}}$\hfil
 &\global\@eqcnt\tw@$\displaystyle\tabskip\z@{##}$\hfil
 \tabskip\@centering&\llap{##}\tabskip\z@\cr}
\makeatother

\renewcommand{\theequation}{\thesection.\arabic{equation}}
\newcommand{\Section}{\setcounter{equation}{0} \section}
\renewcommand{\thefootnote}{\fnsymbol{footnote}}
\begin{document}
%
\begin{titlepage}
\begin{flushright}
{August 2010 \\
Revised version, October 2010}
\end{flushright}
\vspace{0.5cm}
\begin{center}
{\Large \bf
Localization with a Surface Operator, 
 Irregular Conformal Blocks
 and Open Topological String}
\vskip1.0cm
{\large Hidetoshi Awata${}^\ast$, Hiroyuki Fuji${}^\diamond$, Hiroaki Kanno${}^{\ast, \star}$ \\
Masahide Manabe${}^\ast$ and Yasuhiko Yamada${}^\dagger$}
\vskip 1.0em
{\it 
${}^\ast$%
Graduate School of Mathematics \\
${}^\diamond$%
Department of Physics \\
${}^\star$%
Kobayashi-Maskawa Institute
for the Origin of Particles and the Universe (KMI) \\
Nagoya University, Nagoya, 464-8602, Japan \\
and \\
${}^\dagger$%
Department of Mathematics, Faculty of Science \\
Kobe University, Hyogo, 657-8501, Japan}
\end{center}
\vskip1.5cm

\begin{abstract}
Following a recent paper by Alday and Tachikawa, 
we compute the instanton partition function in the presence of the surface
operator by the localization formula on the moduli space.
For $SU(2)$ theories we find an exact agreement 
with CFT correlation functions with a degenerate operator
insertion, which enables us to work out the decoupling limit of the superconformal theory
with four flavors to asymptotically free theories at the level of differential equations
for CFT correlation functions (irregular conformal blocks).  
We also argue that the K theory (or five dimensional) lift of these
computations gives open topological string amplitudes on local Hirzebruch
surface and its blow ups, which is regarded as a geometric engineering of the surface operator.
By computing the amplitudes in both A and B models we collect convincing
evidences of the agreement of the instanton partition function with surface operator and
the partition function of open topological string.

\end{abstract}
\end{titlepage}


\renewcommand{\thefootnote}{\arabic{footnote}} \setcounter{footnote}{0}


\Section{Introduction}

In the problem of the non-perturbative physics of four dimensional gauge
theory the connection to two dimensional theory has been an useful idea. 
For instanton effects in the low energy effective action ($F$-term)
of $\mathcal{N}=2$ supersymmetric gauge theories the seminal work of Nekrasov \cite{Nekrasov:2002qd} 
gives a combinatorial formula of the instanton partition function, which reminds us
of the theory of free fermions and bosons in two dimensional
conformal field theory (CFT). Last year this expectation was made quite
explicit by AGT relation \cite{Alday:2009aq}. 
The holomorphic version of their proposal tells a relation of the homological 
(four dimensional) instanton partition function of $\mathcal{N}=2$ (quiver) 
gauge theories and appropriate conformal blocks. 
Subsequently this correspondence was extended to incorporate loop and surface operators 
in four dimensional gauge theory \cite{Alday:2009fs} (see also \cite{Drukker:2009tz, Drukker:2009id}).

In this paper we consider the instanton partition function in the presence of 
a surface operator and its relation to CFT correlation function 
with a degenerate field insertion. In a last few months there appeared several works 
where related ideas have been developed \cite{Kozcaz:2010af, Alday:2010vg, Dimofte:2010tz, Maruyoshi:2010iu, Taki:2010bj}.
We note that most of them (except \cite{Alday:2010vg}) assume the extension of
AGT relation proposed in \cite{Alday:2009fs} and discuss the partition function with
surface operators by computing the corresponding CFT correlation functions and/or 
topological string amplitudes. However, as is clearly explained in \cite{Alday:2010vg}
the computation of the instanton partition function can be made more directly
by localization formula on the gauge theory side,
if we consider the moduli space of instantons  which involves a certain type of surface operator. 
In a sense this is a natural extension of the method which was used by Nekrasov
to derive his formula of the instanton partition function. 
Based on the equivariant character formula derived by Feigin et.~al. \cite{Feigin}, 
we first present a few examples of direct computations of the instanton partition function with 
a surface operator. Precisely speaking the formula in \cite{Feigin} is expected to hold
when the residual gauge symmetry on the surface is the maximal abelian
subgroup $U(1)^N \subset U(N)$, which was called the full surface operator in \cite{Alday:2010vg}. 
But the surface operator which was argued to correspond to the degenerate operator insertion
is the simple surface operator on which the gauge symmetry is 
reduced to $U(1) \times U(N-1) \subset U(N)$. Fortunately for the gauge group $U(2)$
these two types of the surface operator coincide. Since we rely on this coincidence, we only consider $U(2)$
gauge theories in this paper. After decoupling the diagonal $U(1)$ part, they describe $SU(2)$ theories.

The original AGT relation was proposed for the superconformal gauge theories,
which are obtained by compactifying the world volume theory of $M5$ branes 
on an appropriate Riemann surface with punctures \cite{Gaiotto:2009we}. Recent papers on the extension of
AGT relation with surface operator  \cite{Kozcaz:2010af, Alday:2010vg, Dimofte:2010tz, Maruyoshi:2010iu, Taki:2010bj}
mainly considered the superconformal case. 
However, the AGT relation can be generalized to asymptotically  free theories
\cite{Gaiotto:2009ma, Marshakov:2009gn}. In this paper we will focus on $SU(2)$ theories where the number
of flavors is in the region $0 \leq N_f \leq 3$. According to \cite{Alday:2009fs}
for superconformal theories we should look at the conformal blocks with
a degenerate primary operator $\Phi_{1,2}$ insertion. On the other hand in the nonconformal case
we have to replace the Virasoro highest weight states with the so-called Gaiotto states \cite{Gaiotto:2009ma},
or an analogue of the Whittaker vector for the Virasoro algebra \cite{Braverman:2004vv,Braverman:2004cr}. 
We derive the differential equations  for the one point function of  $\Phi_{1,2}$ operator 
with respect to the Gaiotto states in a systematic manner following 
the appendix of \cite{Awata:2009ur}.  In contrast to the differential equations 
for the usual conformal blocks, our differential equations have irregular singularities. 
We then obtain solutions to the differential equations which can be compared with 
the instanton partition function, namely those in the form of a power series in the scale parameter 
$\Lambda$ which appears in the definition of the Gaiotto state on the CFT side. 
We show that they agree to the results from the localization formula on the moduli space.
We emphasize that the agreement is established beyond the semi-classical limit which was argued in 
\cite{Alday:2009fs}. That is we do not have to take the limit $\ep, \es \to 0$ for the equality.
This becomes possible, since we are able to compute an exact instanton partition function by the localization formula. 
On the gauge theory side the asymptotically free theories are obtained rather easily 
by taking the decoupling limit of $\mathcal{N}=2$ $SU(2)$ theory with four flavors, 
where we take some of the masses of matter hypermultiplets into infinity and redefine the parameter $\Lambda$ of instanton expansion. 
However, it is not straightforward to achieve the corresponding limit at the level of differential equations on the CFT side. 
Hence, we carefully work out the degeneration of the differential equations with irregular singularities, which describes 
the reduction of the number of flavors. Note that the irregular singularities appear as a consequence of 
the congruence of regular singularities. As a byproduct we can also see how the Gaiotto state arises from
a degeneration of two Virasoro primaries.

As is expected from the idea of geometric engineering \cite{Katz:1996fh}
the instanton partition function without surface operator is
related to the topological string amplitudes \cite{Iqbal:2003ix, Iqbal:2003zz, 
Eguchi:2003sj, Eguchi:2003it, Hollowood:2003cv}. 
Namely when the equivariant parameters (or the $\Omega$ background) $(\epsilon_1, \epsilon_2)$
satisfy the self-dual condition $\epsilon_1 + \epsilon_2 =0$,
the five dimensional lift ($K$ theory version) of the instanton partition function
agrees exactly with the closed topological string partition function
on the local toric Calabi-Yau manifold whose toric diagram is dictated by the geometric engineering. 
Since the closed topological string amplitudes compute the index of 
BPS states (the Gopakumar-Vafa invariants),
the Nekrasov partition function in general $\Omega$ background is expected to
give a refinement of the BPS state counting in topological string theory \cite{Hollowood:2003cv, Awata:2008ed}.
As the presence of surface operators breaks half of the supersymmetry and
the semi-classical part of the partition function with surface operator
is identified with the twisted superpotential \cite{Alday:2009fs}, a natural generalization of the above
geometric engineering is to look at open topological strings, which has
been advocated by Gukov \cite{Gukovtalk}. In the second half of the paper 
we explore the idea of geometric
engineering of the surface operator in $\mathcal{N}=2$ gauge theories. 
As was proposed by Ooguri and Vafa \cite{Ooguri:1999bv} the open topological string amplitudes
(open BPS invariants) give the knot and link invariants via the relation to the Chern-Simons
theory with the Wilson loop operator. As the dimensional reduction to
three dimensions reduces the surface operator to the loop operator,
the relation to the open topological string is natural also
from the view point of three dimensional Chern-Simons theory. 


\begin{figure}[tbp]
\begin{center}
%
\setlength{\unitlength}{1mm}
\begin{picture}(140,70)(-10,-1.5)
\put(-10,40){\framebox(60,20){\shortstack{
Instanton Partition Function \\ \\
with surface operator \\ \\
{
\red
Localization formula 
}
}}}
\put(-10,5){\framebox(60,20){\shortstack{
CFT Correlation Function \\ \\
with degenerate operator \\ \\
{
\red
Differential Equation
}
}}}
\put(75,0){\framebox(60,65)}
\put(80,40){\framebox(50,20){\shortstack{
Topological A model \\ \\
with Lagrangian brane \\ \\
{
\red
Topological Vertex
}
}}}
\put(80,5){\framebox(50,20){\shortstack{
Topological B model \\ \\
with mirror curve \\ \\
{
\red
Topological Recursion
}
}}}
\thicklines
\put(51,49.5){$\Longleftarrow\!=\!=\!=\!=\!\Longrightarrow$}
\put(16,31){$\displaystyle{\Uparrow\atop\Downarrow}\hskip-7.819pt\|$}
\put(105.5,32){\vector(0,1){5}}
\put(105.5,32){\vector(0,-1){5}}
\put(32.5,51){\makebox(60,7.5){Geometric}}
\put(32.5,42.5){\makebox(60,7.5){Engineering}}
\put(23,22.5){\makebox(20,20){AGT-relation}}
\put(108.5,22.5){\makebox(20,20){Mirror Sym.}}
\put(143,5){.}
\end{picture}
  \end{center}
 \caption{Correspondence among instanton counting, CFT and topological string}
 \label{GE}
\end{figure}
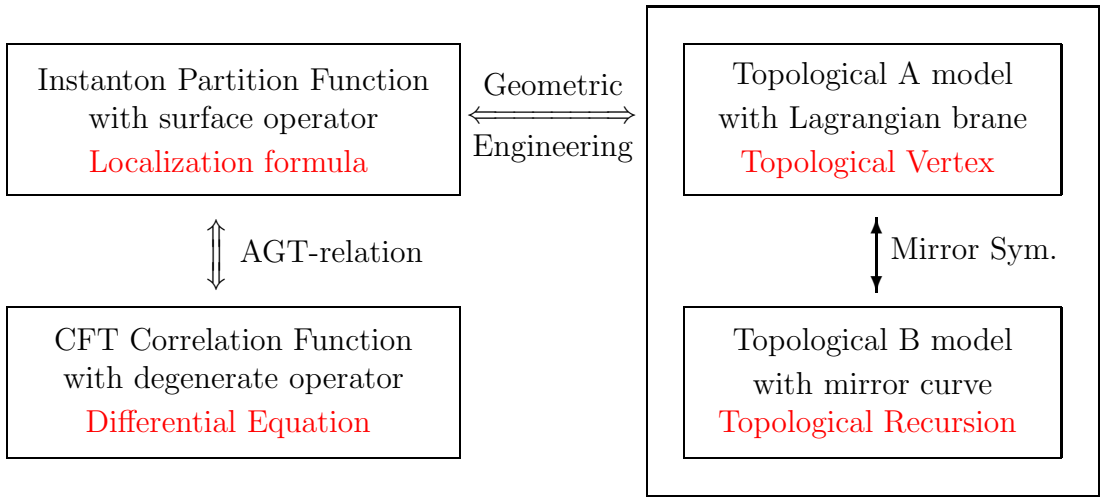


 Pure $SU(2)$ Seiberg-Witten theory is geometrically engineered by the local Hirzebruch surface $K_{{\bf F}_0}$ (the total space
of the canonical bundle of ${\bf F}_0 = {\bf P}^1_{b} \times {\bf P}^1_{f}$). 
The local Calabi-Yau manifold $K_{{\bf F}_0}$ has two moduli parameters $t_b$ and $t_f$, 
which represent the K\"ahler parameters of the base  ${\bf P}^1_{b}$ 
and the fiber  ${\bf P}^1_{f}$, respectively. The parameter of the instanton expansion (the dynamical mass scale) 
$\Lambda$ and the vacuum expectation value $a$ of the scalar field in the prepotential of $\mathcal{N} =2$ theory 
are related to these moduli parameters by $(\beta\Lambda)^4 \sim e^{-t_b}$ and $2\beta a \sim t_f$ with $\beta$ being 
a scale parameter of length.  By blowing up at $N_f$ points in toric geometry 
we can add $N_f$ matter hypermultiplets in the fundamental representation. 
The corresponding geometry is described by the local toric del Pezzo surfaces. 
It has been argued that the (simple) surface operator is geometrically engineered by a toric Lagrangian brane 
inserted on the inner edge of the toric diagram which corresponds 
to the base ${\bf P}^1_{b}$ of the surface \cite{Dimofte:2010tz}.
We compute topological open string amplitudes on this local toric Calabi-Yau geometry with a brane 
in both A and B model perspectives. 
The disk amplitude, which corresponds to the superpotential, is most easily computed by the B  model approach, since
it is naturally related to the period integral. We first use the Seiberg-Witten curve which can be associated to  the semi-classical
limit  of the expectation value of the energy-momentum tensor on the CFT side. We also make computations based on the 
mirror curve of the local Calabi-Yau geometry. In both cases we can show an agreement with CFT correlation functions
with a degenerate field insertion. We employ the method of remodeling \cite{Marino:2006hs, Bouchard:2007ys} in our B model computations. 
One of the advantages of this method is that we can easily increase the number of holes (boundaries) of the world sheet 
by the topological recursion relation coming from the matrix model \cite{Eynard:2007kz}. 
Motivated by a recent suggestion in \cite{Kozcaz:2010af}, we also compare annulus amplitude and three hole amplitude with
CFT correlation functions with multiple insertion of $\Phi_{1,2}$ operator. 
We again find a matching of both computations as far as the comparison is possible.

For the A model computation we use the powerful method of the topological vertex \cite{Aganagic:2003db}. 
We first look at the decoupling limit of four dimensional gauge theory from the two dimensional theory on the surface.
As argued in \cite{Dimofte:2010tz} in this limit the partition function is reduced to the generating function of the vortex counting. 
We show that the vortex counting in \cite{Dimofte:2010tz} can be successfully recovered from the localization formula on the affine Laumon space.
From the viewpoint of four dimensional theory only the sector of vanishing instanton number survives in this decoupling limit. 
Thus the next task is to examine the sector of instanton number one. The corresponding part of topological string amplitudes is
the first order term in the K\"ahler moduli parameter $t_b$ of the base ${\bf P}^1_{b}$. 
We check that in this order the open topological string amplitude 
on the local Hirzebruch surface exactly agrees with the instanton partition function with a surface operator 
modulo a partial shift of the K\"ahler moduli $t_f$ of the fiber ${\bf P}^1_{f}$
by the parameter of the $\Omega$ background. 
We conjecture this shift becomes trivial in the limit $\es \to 0$, while keeping $\ep$ finite. Note that such a limit
appears in the recent proposal of a quantization of the integrable system associated with the Seiberg-Witten geometry \cite{Nekrasov:2009rc}
(see also \cite{Mironov:2009uv, Mironov:2009dv, Nekrasov:2010ka} and a more recent discussion \cite{Maruyoshi:2010iu}). 
It is desirable to understand the origin of the shift as an effect of 
the presence of the surface operator, or the insertion of $\Phi_{1,2}$ operator to CFT correlation functions. 
The computations of topological string amplitudes in this paper are subjected to the condition $\ep + \es =0$.
In the A model the amplitudes in general $\Omega$ background $(\ep, \es)$ can be computed
by the refined topological vertex \cite{Awata:2005fa, Iqbal:2007ii},  but the computation gets rather involved. 
The validity of the above conjecture should be checked by computing
the refined topological string amplitudes. We leave these issues to future works.

The paper is organized as follows: 
In the next section we introduce the instanton partition function with surface operator and review some of
mathematical background for the relevant moduli space. 
In section 3, following the prescription described in \cite{Alday:2010vg}, 
we compute the instanton partition function for pure $SU(2)$ theory as a basic example 
of the application of the localization formula. We also consider $N_f=4$ theory from which asymptotically
free theories with $N_f <3$ are obtained by the decoupling limit. The instanton partition functions 
computed by the localization formula are compared with the corresponding CFT correlation functions
in section 4. We have to multiply appropriate overall factors for the matching. The origin of the factor is clarified in section 5, 
where the degeneration of the differential equations for irregular conformal
blocks is derived from the consistency with the decoupling of the hypermultiplets on the gauge theory side.
The latter half of the paper is devoted to the geometric engineering of the half-BPS 
surface operator in ${\mathcal N} =2$ theories. In sections 6 and 7 we take the B model approach
based on the topological recursion relation. In section 8 we compute the A model amplitude 
by the method of the topological vertex. Basic formulas and some of technical details are collected in Appendices.


\Section{Instanton partition function with surface operator}

In \cite{Alday:2009fs} the semi-classical matching of the instanton partition 
function in the presence of a surface operator and the conformal block 
with the insertion of a degenerate field was pointed out. To establish 
a full agreement beyond the semi-classical limit we have to set up an 
appropriate framework of the instanton counting that incorporates 
the surface operator. In this section we review a few mathematical 
backgrounds following  \cite{Alday:2009fs, Alday:2010vg} and try 
to make the definition of the partition function as clear as possible, 
since a proper definition of the moduli space is required to justify 
the computation of the partition function by the equivariant localization.

Recall that one of the ways to define the surface operator is to prescribe a singular behavior 
of the gauge field \cite{Gukov:2006jk} (see also \cite{Gukov:2007ck, Gaiotto:2009fs} 
for the surface operators in $\mathcal{N}=2$ theories and \cite{KM, Tan:2010dk} for more mathematical formulation).
Let us consider a gauge field $A_\mu$ on $\mathbb{R}^4 \simeq \mathbb{C}^2$ with complex coordinates $(z_1, z_2)$
and assume that there is a surface operator at $z_2 =0$ which fills the $z_1$-plane. 
If $\theta$ is the angular coordinate of the transverse plane (the $z_2$-plane) to the surface operator, the gauge field diverges as
\beq
A_\mu dx^\mu \sim \mathrm{diag}~(\alpha_1, \alpha_2, \cdots, \alpha_N)~i d\theta, \label{diverge}
\eeq
near the support $\mathcal{S} := \{ (z_1, z_2) \vert z_2 =0 \}$ of the surface operator. 
Note that the data $(\alpha_1, \alpha_2, \cdots, \alpha_N)$ which characterize the surface operator
give an element of the Lie algebra of the maximal Abelian subgroup $U(1)^N$ of the gauge group $G=U(N)$. 
Then we can associate a Young diagram with $N$ boxes (a partition of $N$):
$N = N_1 + N_2 + \cdots + N_s$, if $\mathrm{diag}~(\alpha_1, \alpha_2, \cdots, \alpha_N)$ commutes with 
$\mathbb{L}:= U(N_1) \times U(N_2) \times \cdots \times U(N_s)$. From the viewpoint of the principal $G$-bundle
this means the structure group is reduced to a Levi subgroup $\mathbb{L} \subset G$ on the surface. 
The subgroup $\mathbb{L}$ is identified with the Levi part of a parabolic subgroup $P$ 
of the complexified Lie group $G_{\mathbb{C}} = GL(N)$. 
By a gauge transformation we may assume $\alpha_i \geq \alpha_{i+1}$. 
When $\alpha_i$ are the most generic, the commutant is $U(1)^{N}$ and 
the corresponding parabolic subgroup becomes minimal one, namely  the Borel subgroup $B$ of $GL(N)$. 
The corresponding surface operator is called full surface operator in  \cite{Alday:2010vg}. 
Note that since we have fixed the ordering $\alpha_1 > \alpha_2 > \cdots > \alpha_N$,
the Weyl invariance is lost. We will see its effect on the instanton partition function in the next section. 
Following the terminology used in the context of $\mathcal{N}=4$ gauge theory \cite{Gukov:2006jk},
we call the instantons with the singular behavior \eqref{diverge} \lq\lq ramified\rq\rq\ instantons\footnote{
The name  \lq\lq ramified\rq\rq\ comes from the fact that the ramification in the (geometric) Langlands problem is 
related to the presence of a surface operator, or a codimension two singularity in gauge theory.}. 
The \lq\lq ramified\rq\rq\ instantons are anti-self-dual connections on $\mathbb{R}^4 \setminus \mathcal{S}$ and 
their topological indices are the instanton number $k$ and the monopole number
\beq
\mathfrak{m} := \frac{1}{2\pi} \int_{\mathcal{S}} F 
\in \Lambda_{\mathbb{L} } \simeq  H_2(G/\mathbb{L} , \mathbb{Z}).
\eeq 
For the full surface operator we can see the origin of the monopole number as follows:
Since the gauge group on the surface is reduced to $\mathbb{L}=U(1)^N$ in this case, 
we have $N$ abelian gauge fields or line bundles $L_1, L_2, \cdots L_N$ on the surface. 
Hence the \lq\lq ramified\rq\rq\ instanton has $N$ monopole numbers
\beq
\mathfrak{m}_i := \frac{1}{2\pi} \int_{\mathcal{S}} F_i = \int_{\mathcal{S} } c_1(L_i), \qquad i = 1,2, \cdots N.
\eeq

The generating function of the instanton counting with surface operator is
defined by
\beq
Z_{\mathrm{inst}}^{(S)} = \sum_{k=0}^\infty \sum_{\mathfrak{m}\in \Lambda_\mathbb{L} }
q^k z^{\mathfrak{m}} \int_{\mathcal{M}_{N,(k, \mathfrak{m})}} \bf{1}, \label{counting-def}
\eeq
where $q$ is a parameter of instanton expansion and $\mathcal{M}_{N,(k, \mathfrak{m})}$ is
the moduli space of \lq\lq ramified\rq\rq\  $U(N)$ instantons with instanton number $k$ and the monopole number $\mathfrak{m}$. 
If the theory is superconformal, we can relate the expansion parameter to the gauge coupling $\tau$ by $q = e^{2 \pi i \tau}$.
For asymptotically free theories it is replaced with the parameter of dynamical scale $\Lambda$ with appropriate mass dimension. 
If we put the expansion parameter $z$ associated with the monopole number to $z=e^{2\pi i t}$,
then the parameter $t$ has the following meaning. As was argued by Gukov and Witten \cite{Gukov:2006jk} in 
$\mathcal{N}=4$ gauge theories, the surface operator may be described 
by a coupling of four dimensional gauge theory to a two dimensional sigma model
on the surface $\mathcal{S}$ with the target $G/\mathbb{L} \simeq G_{\mathbb{C}}/ P$. 
Then the parameter $t$ is identified with the complexified 
K\"ahler moduli of the flag manifold $G_{\mathbb{C}}/ P$. From the view point of the sigma model
the monopole number $\mathfrak{m}$ measures the degrees of the map $\Phi : \mathcal{S} \to G/ \mathbb{L}$.
For example, when $\mathbb{L}=U(1) \times SU(N-1) \subset U(N)$,
the target space is the projective space $\mathbf{CP}^{N-1}$ and $t$ is the complexified K\"ahler moduli
of the projective space, which is one dimensional. In this case the monopole number is a single integer
and the corresponding surface operator is called simple \cite{Alday:2010vg}.

As was discussed in  \cite{Alday:2010vg} it is convenient to combine the instanton number and monopole numbers
to define a vector $\vec{k} = (k_1,k_2, \cdots , k_N)$ as follows\footnote{In \cite{Braverman:2004vv, Braverman:2004cr}
it was pointed out that it is natural to combine $k$ with $\mathfrak{m}_i$ from the viewpoint of the affine Lie algebra.}:
\beq
k_1 = k, \quad k_{i+1} - k_i = \mathfrak{m}_i.
\eeq
The moduli space $\mathcal{M}_{N,\vec{k}}$ of the  \lq\lq ramified\rq\rq\  instantons with the topological number $\vec{k}$
has real dimension $4(k_1 + k_2 + \cdots + k_N)$. 
Since we integrate $\bf 1$ over the moduli space in \eqref{counting-def}, we may expect it computes
the volume of $\mathcal{M}_{N,\vec{k}}$. However, the moduli space is highly singular and 
\lq\lq non-compact \rq\rq. Hence we have to regularize the integral. To overcome the problem we can employ
the strategy that was used to derive the Nekrasov partition function. 
We consider a natural toric action of $\mathbb T$ on the moduli space
and the integral is regularized as the equivariant integral, or the push-forward to the equivariant 
cohomology of a point $H_{\mathbb{T}}(\mathrm{pt})$. In the next section we will compute
the equivariant integral by using the localization formula. But the use of the localization theorem
is mathematically justified only when the moduli space is smooth. 
However, $\mathcal{M}_{N,\vec{k}}$ suffers from various types of singularities,
which keeps us from applying the localization formula. A standard method to handle such a problem is
to consider torsion free sheaves with an appropriate stability condition;
see  \cite{Alday:2010vg} and literatures in mathematics cited therein. 
The use of torsion free sheaves for the instanton counting without surface operators is clearly 
explained in \cite{Nakajima, NY1}.  It is shown that torsion free sheaves are also useful 
for constructing a Uhlenbeck space for the instantons with a parabolic structure \cite{FGK}.
For general gauge group $G$ the existence of a smooth moduli space is still open problem, even if we shift 
the construction of a smooth moduli space to the problem of torsion free sheaves. 
Fortunately for $G=U(N)$ a resolution of singularities $\widetilde{\mathcal{M}}_{N,\vec{k}} \to \mathcal{M}_{N,\vec{k}}$
(called small resolution in mathematics) is successfully constructed\footnote{We would like to thank
K. Nagao for explaining this fact.}.
The smooth moduli space $\widetilde{\mathcal{M}}_{N,\vec{k}}$ can be regarded as an affine version of the Laumon
space and called affine Laumon space in mathematics \cite{FFFR, Negut}. According to the description in  \cite{Alday:2010vg}
it consists of the equivalence classes of the following data up to gauge transformations:
\begin{itemize}
\item stable rank $N$ torsion free sheaves  on ${\bf P}^1 \times{\bf P}^1$ with a given topological number $\vec{k}$,
\item a fixed framing at infinity $\{ z_1 = \infty \} \cup \{ z_2 = \infty \}$,
\item a reduction of the gauge group $GL(N)$ to a parabolic subgroup $P$ 
on the surface $\{ z_2 =0 \}$, which is called a parabolic structure.
\end{itemize}
It is remarkable that in the definition of the affine Laumon space $\widetilde{\mathcal{M}}_{N,\vec{k}}$, 
a \lq\lq compactification\rq\rq\ of $\mathbb{C}^2$ is given not
by ${\bf P}^2$ but by ${\bf P}^1 \times {\bf P}^1$. The standard toric action 
$(z_1,z_2) \to (e^{i \ep} \cdot z_1, e^{i \es} \cdot z_2)$ on ${\mathbb C}^2$ survives 
after this \lq\lq compactification\rq\rq.
Thus we can consider the fixed point of the toric action of ${\mathbb T} := U(1)^2 \times U(1)^N \subset SO(4) \times U(N)$ 
on the moduli space of  \lq\lq ramified\rq\rq\  instantons, which is familiar in the computation of 
the Nekrasov partition function. In \cite{Feigin} it was shown that the fixed point is isolated and labeled 
by an $N$-tuple of Young diagrams $\vec{\lambda} = (\lambda_1, \lambda_2, \cdots \lambda_N)$. 
However, we should warn that the manner how these Young diagrams appear is rather different from the case of
the standard instanton where the moduli space is constructed by ADHM data. In fact the constraints imposed on the
$N$-tuple of Young diagrams $\vec{\lambda}$ are 
\beq
k_i = k_i(\vec{\lambda}) := \sum_{j \geq 0} \lambda_{i+j, j+1},  \label{constraints}
\eeq
where $\lambda_{i,j}$ is the length of the $j$-th row of the Young diagram $\lambda_i$ and 
we define $\lambda_{i,j}$ for $i > N$ by requiring $\lambda_{i+N} \equiv \lambda_{i}$. 
Thus the fixed points on the moduli space $\widetilde{\mathcal{M}}_{N,\vec{k}}$ are in one to one 
correspondence with $\vec{\lambda}$ that satisfies the condition \eqref{constraints}.
Because the affine Laumon space is smooth, we can consider the tangent space at 
each fixed point with complex dimension $2(k_1 + k_2 + \cdots + k_N)$. At fixed points 
the toric action induces a structure of $U(1)^2 \times U(1)^N $ module on the tangent
space. In \cite{Feigin} the structure of this module was determined and a formula of
equivariant character was provided, which allows us to compute the instanton partition 
function \eqref{counting-def} by the localization theorem.

It is amusing that a closely related moduli space was already appeared in a proof 
of the Nekrasov conjecture from the viewpoint of integrable system and the representation theory of the affine Lie algebra
\cite{Braverman:2004vv, Braverman:2004cr}, where the moduli space of the instantons with parabolic structure
was introduced. In \cite{Braverman:2004vv, Braverman:2004cr} the Uhlenbeck compactification of the moduli space \cite{FGK} 
and a sophisticated theory of the intersection cohomology were used to compute the equivariant integral. 
On the other hand the affine Laumon space provides a semi-flat resolution of singularities and we can apply 
the standard theory of the equivariant cohomology and the localization theorem to compute our partition function \eqref{counting-def}.

\Section{Equivariant localization on affine Laumon space}

In this section, following the method of computation in \cite{Alday:2010vg},
 we work out a few examples of the instanton partition function
in the presence of the surface operator by localization formula.
As was discussed in the last section if the gauge group is $U(2)$,  
the fixed points are isolated and labeled by a pair of Young diagrams
The measure of the localization formula at each fixed point is
obtained from explicit computations of the equivariant character 
of toric action on the affine Laumon space. Fortunately we have
a formula of the equivariant character derived in \cite{Feigin},
which is given in Appendix A (see also eq. (3.10) in \cite{Alday:2010vg}).

\subsection{Pure Yang-Mills theory}

Let us assume that all the fields in the theory are in the adjoint representation.
The so called $\mathcal{N}=2^*$ theory with a massive adjoint matter, which is
a deformation of $\mathcal{N}=4$ conformal theory is a typical example. 
Pure Yang-Mills theory, which can be obtained by decoupling
the adjoint matter of $\mathcal{N}=2^*$ theory, is an example of asymptotically free theories. 
In this case we only need the diagonal component of the equivariant character 
provided in Appendix A, where we set $a=b$ and $\lam=\mu$.
The fixed points of the toric action are labeled by a pair of partitions 
$\vec{\lam} := (\lam_1, \lam_2)$ and $\lambda_{m, n} = \lambda_{m+2, n}$ 
denotes the $n$-th component of the partition $\lam_m$. The vacuum expectation values of the scalar fields
$a_k$ are also defined with $a_{k+2} \equiv a_k$. 
At each fixed point $\vec\lam$ the formula in Appendix A gives many terms in general. 
But after several cancellations the final result should be a sum of $2|\vec\lam|$ monomials with 
$|\vec\lam| = \sum_{n=1}^\infty (\lam_{1,n} + \lam_{2,n})$:
\beq
\mathrm{ch} (a_k, \ep, \es) := {\mathrm{Tr}}_{{\mathrm{Ext}}(\vec\lam, \vec\lam)} g 
= \sum_{i=1}^{2|\vec\lam|} e^{s_i}, \label{char}
\eeq
where each power $s_i$ is a linear combination of $\ep, \es$ and $a_k$.

By the localization theorem the instanton partition function with a (full) surface operator is 
computed as follows \cite{Alday:2010vg}:
\beq
Z^{(S)}(x,y; \ep, \es, a,m) = \sum_{\vec\lam} x^{k_1(\vec\lam)} y^{k_2(\vec\lam)}
\frac{n_{\mathrm{matter}}(\vec\lam; a,m)}{n_{\mathrm{gauge}}(\vec\lam,a)}. \label{Zdef}
\eeq
The equivariant character \eqref{char} gives 
$n_{\mathrm{matter}}(\vec\lam; a,m) := n_{\mathrm{adj}}(\vec\lam; a,m)$ 
for the adjoint hypermultiplet with mass $m$ 
and $n_{\mathrm{gauge}}(\vec\lam,a):= n_{\mathrm{adj}}(\vec\lam; a,0)$ 
for the vector multiplet. For cohomology version we have
\beq
n_{\mathrm{adj}}(\vec\lam; a,m) = \prod_{i=1}^{2|\vec\lam|} (s_i -m),
\eeq
while for $K$-theory version it is
\beq
n_{\mathrm{adj}}(\vec\lam; a,m) = \prod_{i=1}^{2|\vec\lam|} 2 \sinh ((s_i -m)/2).
\eeq
In \eqref{Zdef} $x,y$ are (formal) expansion parameters and topological numbers are defined by
\beq
k_1(\vec\lam) := \sum_{n \geq 1} \lam_{1, 2n-1} +  \sum_{n \geq 1} \lam_{2, 2n}, \qquad
k_2(\vec\lam) := \sum_{n \geq 1} \lam_{1, 2n} +  \sum_{n \geq 1} \lam_{2, 2n-1}. \label{topnumber}
\eeq
We see that $k_1 + k_2 = |\vec\lam|$. The relation to the instanton number
$k$ and the monopole charge $\mathfrak{m}$ on the surface is given by
\beq
k= k_1, \qquad \mathfrak{m} = k_2 - k_1.
\eeq
Note that we have both positive and negative monopole charges.

In view of the comparison with CFT correlation functions let us look at the cohomology version:
\beq
Z^{(S)}_{{\cal N} = 2^*} (x,y; \ep, \es, a,m) = \sum_{\vec\lam} x^{k_1(\vec\lam)} y^{k_2(\vec\lam)}
\prod_{i=1}^{2|\vec\lam|} \frac{s_i -m}{s_i}.
\eeq
This is the instanton partition function for the mass deformed ${\cal N}=4$ theory.
In the massless limit it just counts the number of fixed points 
with weight $x^{k_1(\vec\lam)} y^{k_2(\vec\lam)}$. 
In the decoupling limit $(m \to \infty)$, by renormalizing the parameters $x, y$ by $m^2$,
we have
\beq
Z^{(S)}_{{\mathrm{N_f}} =0} (\Lambda_i ; \ep, \es, a) = \sum_{\vec\lam} \Lambda_1^{k_1(\vec\lam)} 
\Lambda_2^{k_2(\vec\lam)}\prod_{i=1}^{2|\vec\lam|} \frac{1}{s_i}, \label{Zpure}
\eeq
where $\Lambda_1 = m^2 x$ and $\Lambda_2 = m^2 y$. 
The condition of vanishing monopole charge is $k:= k_1(\vec\lam) = k_2(\vec\lam)$
and in this case $|\vec\lam| = 2k$. Restricting to this sector the partition function becomes 
\beq
Z^{(\mathfrak{m}=0)}_{{\mathrm{N_f}} =0} (\Lambda_i ; \ep, \es, a) 
= \sum_{k= k_1(\vec\lam) = k_2(\vec\lam)} (\Lambda_1 \Lambda_2)^k 
\prod_{i=1}^{4k} \frac{1}{s_i}.
\eeq

From the formula in Appendix A we have computed the characters $\mathrm{ch} (a_k, \ep, \es)$ for lower instanton numbers. 
We can see that in general the character at $(\lam_2, \lam_1)$ is obtained from that at $(\lam_1, \lam_2)$
by the transformation $(a_1 - a_2) \to (a_2 - a_1) - \es$ and $(a_2 - a_1) \to (a_1 - a_2) + \es$.
Our computation gives the following partition function for pure gauge theory:
\beqa
&& Z^{(S)}_{{\mathrm{N_f}} =0} (\Lambda_i ; \ep, \es, a) \CR
& & = 1 + \frac{1}{\ep (-2a + \ep)} \Lambda_1 + \frac{1}{\ep(2a +\ep + \es)} \Lambda_2  \CR
& & ~~~ + \frac{1}{2 {\ep}^2 (-2a + \ep)(-2a + 2\ep)} \Lambda_1^2 
+ \frac{1}{2 {\ep}^2 (2a + \ep + \es)(2a + 2\ep + \es)} \Lambda_2^2 \CR
& & ~~~ + \left( \frac{1}{\ep\es(2a)(-2a+\ep)} + \frac{1}{{\ep}^2(-2a)(2a+\es)}
+ \frac{1}{\ep\es(2a+ \ep +\es)(-2a - \es)}\right) \Lambda_1 \Lambda_2 \CR
& &~~~+ \frac{1}{6 \ep^3 (-2a + \ep)( -2a +2\ep) (-2a +3 \ep)} \Lambda_1^3 \CR
& &~~~+ \left(\frac{1}{{\ep}^2(-\ep +\es)(2a)(-2a+\ep)(-2a+2\ep)} 
+ \frac{1}{{\ep}{\es}(\ep-\es)(2a)(-2a+\ep)(-2a+\ep+\es)} \right.  \CR
& &~~~+ \left. \frac{1}{2{\ep}^3(2a -\ep +\es)(-2a)(-2a+\ep)} 
+ \frac{1}{{\ep}^2 \es(-2a)(-2a+\ep-\es)(2a+\ep + \es)}\right) \Lambda_1^2 \Lambda_2  \CR
& &~~~ + \left(\frac{1}{{\ep}^2 \es (-2a +\ep)(2a +\ep)(2a+\es)} + \frac{1}{2 {\ep}^3(-2a -\ep)(2a +\es)(2a+ \ep +\es)} \right. \CR
& &~~~ + \frac{1}{{\ep}^2(-\ep + \es)(-2a -\es)(2a+ \ep + \es)(2a + 2\ep + \es)} \CR
& &~~~ +  \left.\frac{1}{{\ep}{\es}(\ep-\es)(-2a-\es)(2a+\ep+  \es)(2a +\ep + 2\es)} \right) \Lambda_1 \Lambda_2^2  \CR
& &~~~+ \frac{1}{6 \ep^3 (2a + \ep + \es)( 2a +2\ep +\es) (2a +3 \ep +\es)} \Lambda_2^3  +  O( \Lambda_i^4),   \label{pureSU2-partition}
\eeqa
where we have set $a := a_1 = - a_2$. As we will see in the next section, 
up to this order the partition function $Z^{(S)}_{{\mathrm{N_f}} =0}
 (\Lambda_i ; \ep, \es, a)$ completely agrees to the result which is 
 obtained from the differential equation for CFT one point function with $\Phi_{1,2}$ insertion. 
 This means that $Z^{(S)}_{{\mathrm{N_f}} =0}
 (\Lambda_i ; \ep, \es, a)$ satisfies the differential equation 
in the Appendix of \cite{Awata:2009ur}, after the substitution $\Lambda_1 = - z^{-1} \Lambda^2,
\Lambda_2 = - z \Lambda^2$, where $z$ is the position of the degenerate field insertion.

The free energy is defined by 
\beq
 F^{(S)}_{{\mathrm{N_f}} =0} (\Lambda_i ; \ep, \es, a) 
 =  \log Z^{(S)}_{{\mathrm{N_f}} =0} (\Lambda_i ; \ep, \es, a).
\eeq
Using $\log (1 + x) = x - \frac{x^2}{2} + \frac{x^3}{3} + \cdots$, we find 
\beqa
&& F^{(S)}_{{\mathrm{N_f}} =0} (\Lambda_i ; \ep, \es, a) \CR
& & = \frac{1}{\ep (-2a + \ep)} \Lambda_1 + \frac{1}{\ep(2a +\ep + \es)} \Lambda_2 
- \frac{1}{2 {\ep} (-2a + \ep)^2(-2a + 2\ep)} \Lambda_1^2 \CR
&& ~~- \frac{1}{2 {\ep} (2a + \ep + \es)^2 (2a + 2\ep + \es)} \Lambda_2^2 
 - \frac{2}{\ep\es(2a+\ep +\es)(2a - \ep)} \Lambda_1 \Lambda_2 \CR
& & ~~ -\frac{2}{3 \ep (2a-\ep)^3 ( 2a- 2\ep)(2a - 3\ep)} \Lambda_1^3
+ \frac{2}{ 3\ep (2a+\ep +\es)^3 (2a + 2\ep +\es)(2a + 3\ep + \es)} \Lambda_2^3 \CR
& & ~~ - \frac{2}{\ep (2a -\ep)^2 (2a - 2\ep)(2a - \ep -\es)(2a + \ep +\es)} \Lambda_1^2 \Lambda_2 \CR
& & ~~ + \frac{2}{\ep(2a + \ep +\es)^2 (2a - \ep)(2a + 2 \ep +\es)(2a + \ep + 2\es)} \Lambda_1 \Lambda_2^2 
+ O( \Lambda_i^4).   \label{surface-free}
\eeqa
Note that the higher pole of ${\ep}^{-2}$ disappears in the free energy. 
The above expressions are not invariant under the Weyl group action $ a \to -a$ of $SU(2)$. However,
by the shift $ 2\tilde a := 2a + \frac{\es}{2} $, we may recover the invariance under $ \tilde a \to - \tilde a$.

In the free energy \eqref{surface-free} the pole structure of the terms with non-vanishing monopole number
($\Lambda_1^n \Lambda_2^m,~(n \neq m)$) is $\ep^{-1}$, while that of zero monopole part is $(\ep\es)^{-1}$.
Thus it is natural to compare the zero monopole number terms of the free energy with the Nekrasov partition function.
Up to three instantons we obtain
\beqa
F^{(\mathfrak{m}=0)}_{{\mathrm{N_f}} =0} (\Lambda_i ; \ep, \es, a) 
&=& 
\frac{2\Lambda_1 \Lambda_2}{\ep\es D_1(a)D_1(-a-\es/2)}
- \frac{N_2\Lambda_1^2  \Lambda_2^2}
{\ep\es D_2(a)D_2(-a-\es/2)}
\cr
&+& 
\frac{16}{3} \frac{N_3 \Lambda_1^3  \Lambda_2^3}{\ep \es D_3(a)D_3(-a-\es/2)}
+ O( \Lambda_1^4\Lambda_2^4) \label{nomonopole3inst}
\eeqa
with
\beqa
&&\hskip101pt
D_1(a) 
:= 
(2a + \ep + \es),
\cr
&&\hskip26pt
D_2(a) 
:= 
(2a + 2\ep + \es)
(2a + \ep + \es)^2
(2a + \ep + 2\es),
\cr
D_3(a) 
&:=&
(2a + 3\ep + \es)
(2a + 2\ep + \es)
(2a + \ep + \es)^{3}
(2a + \ep + 2\es)
(2a + \ep + 3\es),
\cr
N_2
&:=&
20 a^2 + 10 \es a + (7 \ep^2 + 11 \ep\es + 2 \es^2),
\cr
N_3 
&:=& 
144 a^4 + 144 \es a^3 +( 232 \ep^2 + 416 \ep\es + 88 \es^2) a^2 
+ (116 \ep^2 + 208 \ep \es + 26 \es^2) \es a
\CR
&+& (29 \ep^4 + 116 \ep^3 \es + 118 \ep^2 \es^2 + 13 \ep \es^3 + 3 \es^4).
\eeqa
On the other hand the free energy of the Nekrasov partition function is
\beq
F^{(\mathrm {Nek})}_{{\mathrm{N_f}} =0} (\Lambda ; \ep, \es, a) 
= \frac{2\Lambda^4}{\ep\es D_1(a) D_1(-a)} 
 - \frac{N^{(\mathrm {Nek})}_2\Lambda^8}
{\ep\es D_2(a) D_2(-a) }
+ {\displaystyle \frac {16}{3}} 
\frac{N^{(\mathrm {Nek})}_3\Lambda^{12} }
{\ep\es D_3(a) D_3(-a) }
+ O( \Lambda^{16}) \label{Nek3inst}
\eeq
with
\beqa
N^{(\mathrm {Nek})}_2
&:=&
{20 a^2 + (7 \ep^2 + 16 \ep \es + 7 \es^2)},
\cr
N^{(\mathrm {Nek})}_3
&:=&
  144a^{4} 
+ (
  232\ep^{2} 
+ 568\ep\es 
+ 232\es^{2} 
)a^{2}
\cr
&+&
  (29\ep^{4}
+ 154\ep^{3}\es 
+ 258\ep^{2}\es^{2} 
+ 154\es^{3}\ep 
+ 29\es^{4}).
\eeqa
The free energy of the Nekrasov partition function is symmetric under
both $a \to -a$ (the $SU(2)$ Weyl invariance) and $\ep \leftrightarrow \es$.
However, the existence of the surface breaks these symmetries, even 
in the vanishing monopole sector. One may argue the origin of this discrepancy from the view point of CFT correlation function
with $\Phi_{1,2}$ operator insertion. The comparison of \eqref{nomonopole3inst} and \eqref{Nek3inst} suggests
a simple rule of translation between the denominators. We will encounter a similar rule in the computation of 
open topological string amplitudes by the topological vertex. It is very curious that up to three instantons
both the free energies give the same result in the limit $\es \to 0$.
Thus we conjecture that
\beq
\lim_{\es \to 0} \ep\es~F^{(\mathfrak{m}=0)}_{{\mathrm{N_f}} =0} (\Lambda_i ; \ep, \es, a) 
= \lim_{\es \to 0} \ep\es~F^{(\mathrm{Nek})}_{{\mathrm{N_f}} =0} (\Lambda ; \ep, \es, a). \label{BEcor}
\eeq
with $\Lambda^4 = \Lambda_1 \Lambda_2$. 
If we assume a complete agreement of the instanton partition function with a surface operator and the CFT correlation function 
with a degenerate field insertion, which we confirm in lower orders in the instanton expansion, the conjecture follows from
theorem 1.6 in \cite{Braverman:2004cr}. This is because the differential equation for the CFT correlation function 
with a degenerate field insertion coincides with the one derived by  Braverman and Etingof  \cite{Awata:2009ur}.

\subsection{$N_f =4$ theory (superconformal  case)}

The equivariant character $\mathrm{Tr}_{\mathrm{Ext}(\vec\lam, \vec\mu)}
[\mathrm{diag.} (\ep,\es ; \vec{a}, \vec{b})]$ at a fixed point of the toric action
on the affine $SU(2)$ Laumon space is given in Appendix A.
After the summation over all the contributions,
the equivariant character $\mathrm{Tr}_{\mathrm{Ext}(\vec\lam, \vec\mu)}
[\mathrm{diag.} (\ep,\es ; \vec{a}, \vec{b})]$ is expressed as
a sum of $|\vec\lam| + |\vec\mu|$ monomials:
\beq
 {\mathrm{Tr}}_{{\mathrm{Ext}}(\vec\lam, \vec\mu)} 
[\mathrm{diag.} (\ep,\es ; \vec{a}, \vec{b})]
= \sum_{i=1}^{|\vec\lam| + |\vec\mu|} e^{s_i}, 
\eeq
where each power $s_i$ is a linear combination of $\ep, \es, a_k$ and $b_k$. 
Then the basic ingredient in the following computation is
\beq
\nbf{\vec\lam}{\vec\mu}{\vec{a}} {\vec{b}} {m} := \prod_{i=1}^{|\vec\lam| + |\vec\mu|} (s_i -m),
\eeq
which was originally denoted by $z_{\mathrm{bif}}^S$ in \cite{Alday:2010vg}. This is the contribution of
the bifundamental matter hypermultiplet in the localization formula of the instanton 
partition function in the presence of the (full) surface operator. 

To reformulate the instanton partition function Alday and Tachikawa \cite{Alday:2010vg}
introduced a Hilbert space $\mathcal{H}_{\vec a}^S$ with basis
$\vert \vec\lam \Rangle$. The inner product is defined by
\beq
\Langle \vec\lam \vert \vec\mu \Rangle 
= \frac{\delta_{\vec\lam, \vec\mu}}{n_{\mathrm{vec}}[\vec\lam](\vec{a})}, \label{inner}
\eeq
where $n_{\mathrm{vec}}[\vec\lam](\vec{a}) := \nbf{\vec\lam}{\vec\lam}{\vec{a}}{\vec{a}}{0}$.
We will need the operator that counts the topological number:
\beq
\hat{K}_i \vert \vec\lam \Rangle = k_i(\vec\lam) \vert \vec\lam \Rangle.
\eeq
Alday-Tachikawa also introduced  the intertwining operator $\Phi^S_{\vec{a}, m, \vec{b}}
: \mathcal{H}^S_{\vec b} \longrightarrow \mathcal{H}^S_{\vec a}$, which is defined by 
\beq
\Phi^S_{\vec{a}, m, \vec{b}} \vert \vec\lam \Rangle_{\vec b}
= \frac{1}{n_{\mathrm{vec}}[\vec\lam](\vec{b})}
\sum_{\vec\mu} \nbf{\vec\mu}{\vec\lam}{\vec{a}}{\vec{b}}{m} \vert \vec\mu \Rangle_{\vec a}. \label{intw}
\eeq
Then the instanton partition function with four flavors in the presence of the surface
operator is given by the following \lq vacuum\rq\ expectation value (eq. (3.21) of \cite{Alday:2010vg}):
\beq
 Z^{(S)}_{{\mathrm{N_f}} =4}(a, M_i, \ep, \es ; x,y)
=~ _{a_1}\!\Langle \vec\emptyset \vert \Phi^S_{a_1, m_1, a} x^{\hat K_1} y^{\hat K_2}
\Phi^S_{a, m_2, a_2} \vert \vec\emptyset \Rangle_{a_2},
\eeq
where $x$ and $y$ are formal parameters of topological (instanton-monopole) expansion. 
The mass of the hypermultiplets $M_i$ gives the parameters $a_i$ and $m_i$ in
AGT like fashion:
\beqa
&& a_1 = M_1 - M_2, \quad m_1 = M_1 + M_2, \CR
&& a_2 = M_3 - M_4, \quad m_2 = M_3 + M_4.
\eeqa
The above reformulation is convenient for identifying the partition function
with the conformal block 
on the sphere with four punctures\footnote{If we have an adjoint matter
the partition function is given by the trace over $\mathcal{H}^S_{\vec a}$,
since it should be identified with the conformal block on the torus with a 
single puncture.}.
Inserting a complete system of $\mathcal{H}^S_{a}$ in the intermediate channel
we obtain
\beqa
 Z^{(S)}_{{\mathrm{N_f}} =4}(a, M_i, \ep, \es ; x,y)
&=& \sum_{\vec\lam} x^{k_1(\vec\lam)} y^{k_2(\vec\lam)} (n_{\mathrm{vec}}[\vec\lam](\vec{a}))
~ _{a_1}\!\Langle \vec\emptyset \vert \Phi^S_{a_1, m_1, a} 
\vert \vec\lam \Rangle_{a} \cdot
~_{a}\!\Langle \vec\lam\vert 
\Phi^S_{a, m_2, a_2} \vert \vec\emptyset \Rangle_{a_2} \CR
&=& \sum_{\vec\lam} x^{k_1(\vec\lam)} y^{k_2(\vec\lam)}
\frac{ \nbf{\vec\emptyset}{\vec\lam}{a_1}{a}{m_1} \cdot \nbf{\vec\lam}{\vec\emptyset}{a}{a_2}{m_2}}
{n_{\mathrm{vec}}[\vec\lam](\vec{a})}. 
\eeqa
Hence to compute $Z^{(S)}_{{\mathrm{N_f}} =4}(a, M_i, \ep, \es ; x,y)$
we only need the equivariant character where one of the pairs of partitions is trivial.
In this case among eight types of contributions given 
in Appendix A only two terms survive, which give
\beq
 {\mathrm{Tr}}_{{\mathrm{Ext}}(\vec\lam, \vec\emptyset)} [g] 
=
- \sum_{k \geq 1} e^{a_{k} - b_{1}} e^{\ep - \es \floor{\frac{k}{2}}}
\frac{e^{-\ep\lam_{k,k}} -1}{e^{\ep} -1} 
- \sum_{k \geq 1} e^{a_{k+1} - b_{2}} e^{\ep - \es \floor{\frac{k}{2}- \frac{1}{2}}}
\frac{e^{-\ep\lam_{k+1,k}} -1}{e^{\ep} -1},  \label{ec1}
\eeq
and 
\beq
 {\mathrm{Tr}}_{{\mathrm{Ext}}(\vec\emptyset, \vec\lam)} [g] 
=
\sum_{k \geq 1} e^{a_1 - b_{k+1}} e^{\ep + \es \floor{\frac{k}{2} + \frac{1}{2}}}
\frac{e^{\ep\lam_{k+1,k}} -1}{e^{\ep} -1} 
+\sum_{k \geq 1} e^{a_2 - b_{k}} e^{\ep + \es \floor{\frac{k}{2} }}
\frac{e^{\ep\lam_{k,k}} -1}{e^{\ep} -1}.  \label{ec2}
\eeq
From \eqref{ec1} and \eqref{ec2}  we obtain the data for $\nbf{\vec\lam}{\vec\emptyset}{a}{b}{m}$,
which leads the partition function:
\beqa
& &  Z^{(S)}_{{\mathrm{N_f}} =4}(a, M_i, \ep, \es ; x,y) \CR
& & ~= 1 + \frac{ ( -a +\ep -2M_1 )( a -2 M_3)}{\ep (-2a + \ep)} x 
+ \frac{(a+\ep +\es -2M_2)(-a -2M_4)}{\ep(2a +\ep + \es)} y  \CR
& & ~~~ + \frac{ (-a + \ep -2M_1)(-a +2\ep -2M_1)(a-2M_3)(a - \ep -2M_3)}{2 {\ep}^2 (-2a + \ep)(-2a + 2\ep)} x^2 \CR
& & ~~~+ \frac{ (a +\ep +\es -2 M_2)(a + 2\ep +\es -2 M_2)(-a -2M_4)(-a - \ep -2M_4) }{2 {\ep}^2 (2a + \ep + \es)(2a + 2\ep + \es)} y^2 \CR
& & ~~~ + \frac{ (-a + \ep -2M_1) (-a + \ep +\es -2 M_2)(a -2M_3)(a - 2M_4)}{\ep\es(2a)(-2a+\ep)} xy \CR
& & ~~~+ \frac{(- a + \ep -2 M_1)(a + \ep +\es -2 M_2)( a -2M_3)(-a -2M_4) }{{\ep}^2(-2a)(2a+\es)} xy \CR
& & ~~~ + \frac{( a +\ep +\es -2M_1)( a + \ep +\es -2M_2)( -a - \es -2M_3)( -a -2M_4)}{\ep\es(2a+ \ep +\es)(-2a - \es)} xy \CR
& & ~~~ +  \cdots \cdots. \label{Nf4partition}
\eeqa
The instanton partition functions with surface operator for asymptotically free theories with $N_f \leq 3$ can be 
obtained by the decoupling limit, where the expansion parameters $x, y$ are promoted to $\Lambda_1, \Lambda_2$ 
with appropriate mass dimension. There are several choices of the set of $M_i$'s with $M_i \to \infty$. 
 In any case one of the characteristic features of the decoupling is that not only the denominators 
 but also the numerators of the $x^2$ and $y^2$ terms 
 are of factorized form. This is because there is only one fixed point with the corresponding topological number. 
 In \cite{Alday:2010vg} it is observed that up to an appropriate $U(1)$ factor the above partition function \eqref{Nf4partition}
coincides with the four-point conformal block of $SL(2)$ current algebra on the sphere with an insertion of the operator
$\mathcal K$ which was introduced in \cite{Alday:2010vg}. In sections 4 and 5 we will explicitly check that 
the partition function \eqref{Nf4partition} and its decoupling limit also agree 
with the Liouville correlation functions on the sphere with a degenerate field 
insertion; see subsection 5.5 for a summary and a rule of the correspondence.

One can check that the free energy $ F^{(S)}_{{\mathrm{N_f}} =4}(a, M_i, \ep, \es)
= \log  Z^{(S)}_{{\mathrm{N_f}} =4}(a, M_i, \ep, \es)$ has a correct pole structure,
namely the poles of $x^2$ and $y^2$ terms are $\ep^{-1}$, while that of $xy$ term
is $(\ep\es)^{-1}$. However, the explicit form is rather lengthy. We only quote the lowest terms:
\beqa
&&\lim_{\ep, \es \to 0} \ep\cdot  F^{(S)}_{{\mathrm{N_f}} =4}(a, M_i, \ep, \es)\vert_{x^2}
= - \frac{1}{16} \frac{(a+2M_1)(a -2 M_3)(3a^2 + 2a(M_1 - M_3) + 4M_1 M_3)}{a^3} \CR
&&\lim_{\ep, \es \to 0} \ep\cdot  F^{(S)}_{{\mathrm{N_f}} =4}(a, M_i, \ep, \es)\vert_{y^2}
=  \frac{1}{16} \frac{(a-2M_2)(a +2M_4)(3a^2 + 2a(M_4 - M_2) + 4M_2 M_4)}{a^3} \CR
&& \lim_{\ep, \es \to 0} \ep\es \cdot  F^{(S)}_{{\mathrm{N_f}} =4}(a, M_i, \ep, \es)\vert_{xy} \CR
&& ~~~= \frac{- a^4 + 4 (M_1 M_3 + M_1 M_4 + M_2 M_3 + M_2 M_4 - M_1 M_2 - M_3 M_4) a^2 
- 16 M_1 M_2 M_3 M_4}{2a^2}. \CR
&&
\eeqa


\section{CFT correlation functions with degenerate field insertion}

In \cite{Alday:2009fs} it was claimed that the surface operator ${\mathcal{S}} \subset \mathbb{R}^4$ 
in the supersymmetric gauge theory with eight supercharges corresponds 
to the degenerate primary operator $\Phi_{1,2}(z)$ in the Liouville CFT. An explanation  
of the correspondence from the viewpoint of the M2/M5-brane system was also given.
Since the operator $\Phi_{1,2}(z)$ which has the momentum $-\frac{1}{2b}$ satisfies the null state
condition $(b^2 L_{-1}^2 + L_{-2}) \Phi_{1,2}(z) =0$
, when it is inserted in any CFT correlation functions,
we have
\beq
b^2 \partial_z^2 \Phi_{1,2} (z) =  -  : T(z) \Phi_{1,2}(z) :,
\eeq
where $T(z)$ is the energy momentum tensor and $:~~:$ denotes the normal ordering. 
When the operator $\Phi_{1,2}(z)$ is inserted, the correlation function has an additional
dependence on the position $z$ of the degenerate operator. One of the points in \cite{Alday:2009fs} is
that this dependence appears in the subleading term of the semi-classical approximation:
\beqa
\Psi(a_i, z) &:=& \langle \Phi_{1,2}(z) V_{m_1}(z_1) \cdots V_{m_n}(z_n) \rangle_{\{a_i\}} \CR
& \sim & \exp \left(- \frac{\mathcal{F}(a_i)}{\hbar^2} + \frac{\mathcal{W}(a_i,z)}{b \hbar} + \cdots \right). 
\eeqa
Recall that the original observations of \cite{Alday:2009aq} are that
\beq
Z(a_i) := \langle V_{m_1}(z_1) \cdots V_{m_n}(z_n) \rangle_{\{a_i\}}  
\sim  \exp \left(- \frac{\mathcal{F}(a_i)}{\hbar^2} + \cdots \right)
\eeq
coincides with the Nekrasov partition function and that
\beq
\langle T(z) V_{m_1}(z_1) \cdots V_{m_n}(z_n) \rangle_{\{a_i\}}  
\sim - \frac{1}{\hbar^2} \phi^{\mathrm SW}(z) \langle V_{m_1}(z_1) \cdots V_{m_n}(z_n) \rangle_{\{a_i\}},
\eeq
where $x^2 = \phi^{\mathrm SW}(z)$ gives the Seiberg-Witten curve which is a double covering of 
the punctured Riemann sphere. 
In asymptotically free theories we should consider the correlation functions with respect to the state
introduced by Gaiotto  \cite{Gaiotto:2009ma}. It is natural to call them irregular conformal blocks, since
the differential equations for such correlation functions have irregular singularities in general. 
In the appendix of \cite{Awata:2009ur} it was noticed that the differential equation 
for irregular conformal blocks with $\Phi_{1,2}(z)$ insertion coincides with the differential equation 
for the instanton partition function with parabolic structure derived in \cite{Braverman:2004cr}. 
Based on these works we expect that the instanton partition functions computed 
in section 3 by localization formula are obtained from the one point function of $\Phi_{1,2}(z)$ with respect
to the Gaiotto state. In this section we check the correspondence for $N_f=0,1$ and $2$
(see also the next section for the discussion by degenerations from the superconformal theory with $N_f=4$). 
In section 4.4 we consider the multi-point irregular conformal blocks 
which should correspond to the instanton partition functions with multi-surface operators.

\subsection{Pure $SU(2)$}

\noindent Let us consider the (normalized) correlation function
\begin{equation}
Z_{\mbox{\scriptsize null\normalsize}}^{(0)}(z,a,\Lambda):=
\frac{\langle\Delta_-, \Lambda|\Phi_{1,2}(z)|\Delta_+, \Lambda\rangle}{\langle\Delta_a, \Lambda|\Delta_a, \Lambda\rangle}
=:\frac{\Psi^{(0)}(z,a,\Lambda)}{Z_c^{(0)}(a,\Lambda)},\quad \Delta_{\pm}:=\Delta(a \pm \frac{1}{4b}),
\label{eqn:1.1}
\end{equation}
where $\Delta_a:=\Delta(a):=(b+b^{-1})^2/4-a^2$ is the conformal dimension and $\Phi_{1,2}(z)$ 
is the degenerate primary field with the conformal dimension $h_{1,2}=-1/2-3/(4b^2)$. 
The Gaiotto state $|\Delta, \Lambda\rangle$ is the state in the Virasoro Verma module $V( \Delta, c)$ 
with the conformal dimension $\Delta$ and the central charge $c= 1 + 6(b + b^{-1})^2$, 
which is characterized by 
\begin{equation}
L_1|\Delta, \Lambda\rangle=\Lambda^2|\Delta, \Lambda\rangle,\quad L_2|\Delta, \Lambda\rangle=0.
\label{eqn:1.2}
\end{equation}
There is an ambiguity in the choice of the conformal weights $\Delta_{\pm} := \Delta(\alpha_{\pm})$ of the Gaiotto state. 
According to the fusion rule of $\Phi_{1,2}$ operator, $\langle\Delta_-, \Lambda|\Phi_{1,2}(z)|\Delta_+, \Lambda\rangle$ 
is non-vanishing if and only if $\alpha_{+} - \alpha_{-} = \pm \frac{1}{2b}$. 
The above choice $\alpha_{\pm} = a \pm \frac{1}{4b}$ is the symmetric one, which leads 
a result that is invariant under $ a \to -a$.

In \cite{Gaiotto:2009ma} it was conjectured that $Z_c^{(0)}(a,\Lambda)$ coincides with the Nekrasov partition function 
in the pure $SU(2)$ supersymmetric gauge theory, which has been proved in \cite{Hadasz:2010xp}. 
According to the appendix of \cite{Awata:2009ur}, by putting 
$\Psi^{(0)}(z,a,\Lambda)=z^{\Delta_--\Delta_+-h_{1,2}}Y^{(0)}(z,a,\Lambda)$, 
one obtains the second order differential equation\footnote{Note that the operator $\Phi_{1,2}(z)$ in this paper corresponds to $\Phi_{2,1}(z)$ in the convention of \cite{Awata:2009ur}.}:
\begin{equation}
\left[\Big(bz\frac{\partial}{\partial z}\Big)^2+2abz\frac{\partial}{\partial z}
+\Lambda^2(z+z^{-1})+\frac{\Lambda}{4}\frac{\partial}{\partial \Lambda}\right]Y^{(0)}(z,a,\Lambda)=0.
\label{eqn:1.3}
\end{equation}
Since we want to compare the instanton partition function computed in the previous section by localization theorem
with solutions to the differential equation (\ref{eqn:1.3}), we look for a solution of the form
\begin{equation}
Y^{(0)}(z,a,\Lambda)=\sum_{n=0}^{\infty} \Lambda^{2n} Y_n^{(0)}(z,a)
\label{eqn:1.5}
\end{equation}
with the initial condition $Y_0^{(0)}(z,a)=1$. 
It is convenient to introduce a mass scale $\hbar$ and scale the parameters as follows:
\begin{equation}
a~\longrightarrow~\frac{a}{\hbar},\quad \Lambda~\longrightarrow~\frac{\Lambda}{\hbar}.
\label{eqn:1.4}
\end{equation}
We also introduce the parameters $\epsilon_1=b\hbar$, $\epsilon_2=b^{-1}\hbar$ corresponding to the parameters 
of the $\Omega$ background of Nekrasov. Then we find the following differential equations for the coefficients $Y_n^{(0)}(z,a)$
in  the expansion \eqref{eqn:1.5},
\begin{equation}
\left[\Big(\epsilon_1z\frac{\partial}{\partial z}\Big)^2+2a \epsilon_1z\frac{\partial}{\partial z}
+\frac{n}{2}\epsilon_1\epsilon_2\right]Y_n^{(0)}(z,a)+(z+z^{-1})Y_{n-1}^{(0)}(z,a)=0,~ \quad n\geq 1.
\label{eqn:1.7}
\end{equation}
A power series solution to (\ref{eqn:1.7}) is given by
\begin{equation}
Y_n^{(0)}(z,a)=\sum_{k=-\infty}^{\infty}A_{n,k}^{(0)}z^k,\qquad A_{0,k}^{(0)}
=\delta_{0,k}, \quad A_{n,k}^{(0)}=-\frac{A_{n-1,k-1}^{(0)}+A_{n-1,k+1}^{(0)}}
{\epsilon_1\left(2ak+\epsilon_1k^2+\frac{1}{2}n\epsilon_2\right)},
\label{eqn:1.8}
\end{equation}
and we find the following lower order terms in the expansion \eqref{eqn:1.5},
\begin{eqnarray}
\label{eqn:1.9}
Y_1^{(0)}(z,a)&=&-\frac{1}{\epsilon_1(-2a+\epsilon_1+\frac{\epsilon_2}{2})z}-
\frac{z}{\epsilon_1(2a+\epsilon_1+\frac{\epsilon_2}{2})},\\
Y_2^{(0)}(z,a)&=&\frac{1}{2\epsilon_1^2(-2a+\epsilon_1+\frac{\epsilon_2}{2})
(-2a+2\epsilon_1+\frac{\epsilon_2}{2})z^2}+\frac{2\epsilon_1+\epsilon_2}
{\epsilon_1^2\epsilon_2(-2a+\epsilon_1+\frac{\epsilon_2}{2})(2a+\epsilon_1+\frac{\epsilon_2}{2})} \nonumber\\
\label{eqn:1.10}
&&\hspace{-1em}
+\frac{z^2}{2\epsilon_1^2(2a+\epsilon_1+\frac{\epsilon_2}{2})(2a+2\epsilon_1+\frac{\epsilon_2}{2})},\\
Y_3^{(0)}(z,a)&=&-\frac{1}{6\epsilon_1^3(-2a+\epsilon_1+\frac{\epsilon_2}{2})(-2a+2\epsilon_1
+\frac{\epsilon_2}{2})(-2a+3\epsilon_1+\frac{\epsilon_2}{2})z^3}\nonumber\\
&&\hspace{-1em}
-\frac{16\epsilon_1^2+14\epsilon_1\epsilon_2+3\epsilon_2^2-16a \epsilon_1-4a \epsilon_2}
{4\epsilon_1^3\epsilon_2(-2a+\epsilon_1+\frac{\epsilon_2}{2})(-2a+2\epsilon_1+\frac{\epsilon_2}{2})
(-2a+\epsilon_1+\frac{3\epsilon_2}{2})(2a+\epsilon_1+\frac{\epsilon_2}{2})z}+\cdots.\nonumber\\
\label{eqn:1.11}
&&
\end{eqnarray}
If we make the shift that was discussed in the last section to make the partition function invariant 
under the $SU(2)$ Weyl transformation $a \to -a$,
then \eqref{eqn:1.9} -- \eqref{eqn:1.11} completely agree to the instanton expansion \eqref{pureSU2-partition}
of the partition function of pure $SU(2)$ theory.
The free energy is defined by
\begin{equation}
F_{\mbox{\scriptsize null\normalsize}}^{(0)}(z,a,\Lambda):=\log Z_{\mbox{\scriptsize null\normalsize}}^{(0)}(z,a,\Lambda)=
\Big(\frac{a}{\epsilon_1}+\frac12+\frac{3\epsilon_2}{4\epsilon_1}\Big)\log z+\log\frac{1+\sum_{n=1}^{\infty}
 \Lambda^{2n} Y_n^{(0)}(z,a)}{Z_c^{(0)}(a,\Lambda)},
\label{eqn:1.12}
\end{equation}
and then we obtain
\begin{equation}
F_{\mbox{\scriptsize null\normalsize}}^{(0)}(z,a,\Lambda)=\frac{1}{\epsilon_1}\left\{F_{-1}^{(0)}
+\big(\epsilon_1F_{0,1}^{(0)}+\epsilon_2F_{0,2}^{(0)}\big)+{\cal O}(\epsilon^2) \right\},
\label{eqn:1.13}
\end{equation}
where the leading term is
\begin{equation}
F_{-1}^{(0)}=\log z^a-\frac{z^2-1}{2az}\Lambda^2-\frac{z^4-1}{16a^3z^2}\Lambda^4
-\frac{z^6+3z^4-3z^2-1}{48a^5z^3}\Lambda^6
-\frac{5z^8+16z^6-16z^2-5}{512a^7z^4}\Lambda^8+\cdots.
\label{eqn:1.14}
\end{equation}


\subsection{$SU(2)$ with one fundamental matter}

\noindent Next, we consider the correlation function
\begin{equation}
Z_{\mbox{\scriptsize null\normalsize}}^{(1)}(z,a,m,\Lambda):=\frac{\langle\Delta_-, \Lambda,m|\Phi_{1,2}(z)|\Delta_+, 
\Lambda\rangle}{\langle\Delta_a, \Lambda,m|\Delta_a, \Lambda\rangle}=:\frac{\Psi^{(1)}(z,a,m,\Lambda)}{Z_c^{(1)}(a,m,\Lambda)},
\label{eqn:1.17}
\end{equation}
where the Gaiotto state $|\Delta, \Lambda,m \rangle$ in the Virasoro Verma module $V(\Delta, c)$ satisfies
\begin{equation}
L_2|\Delta, \Lambda,m \rangle=-\Lambda^2|\Delta, \Lambda,m \rangle,\quad L_1|\Delta, \Lambda\rangle
=-2m \Lambda|\Delta, \Lambda,m \rangle.
\label{eqn:1.18}
\end{equation}
The denominator $Z_c^{(1)}(a,m,\Lambda)$ of \eqref{eqn:1.17}
coincides with the Nekrasov partition function of $SU(2)$ supersymmetric gauge theory 
with one fundamental matter \cite{Gaiotto:2009ma, Hadasz:2010xp}. 
By putting $\Psi^{(1)}(z,a,m,\Lambda)=z^{\Delta_--\Delta_+-h_{1,2}}Y^{(1)}(z,a,m,\Lambda)$, 
one obtains the second order differential equation,
\begin{equation}
\left[\Big(bz\frac{\partial}{\partial z}\Big)^2+\Big(2ab+\frac16\Big)z
\frac{\partial}{\partial z}-\Lambda^2(z^2-z^{-1})-2m\Lambda z
+\frac{\Lambda}{3}\frac{\partial}{\partial \Lambda}\right]Y^{(1)}(z,a,m,\Lambda)=0.
\label{eqn:1.19}
\end{equation}
By the decoupling limit of the matter $m \to \infty, -2m\Lambda^3 \to \Lambda^4, 4m^2z^3 \to \Lambda^2z^3$, 
the differential equation (\ref{eqn:1.19}) is reduced to (\ref{eqn:1.3}). After introducing the mass scale 
$\hbar$ as (\ref{eqn:1.4}) and $m \rightarrow m/\hbar$, we obtain the following solution to (\ref{eqn:1.19}),
\begin{eqnarray}
\label{eqn:1.20}
&&
Y^{(1)}(z,a,m,\Lambda)=\sum_{n=0}^{\infty} \Lambda^n Y_n^{(1)}(z,a,m),\quad Y_0^{(1)}(z,a,m)=1,\\
&&
Y_n^{(1)}(z,a,m)=\sum_{k=-\infty}^{\infty}A_{n,k}^{(1)}z^k,\nonumber\\
\label{eqn:1.22}
&&
A_{0,k}^{(1)}=\delta_{0,k}, \quad A_{n,k}^{(1)}=\frac{A_{n-2,k-2}^{(1)}-A_{n-2,k+1}^{(1)}
+2mA_{n-1,k-1}^{(1)}}{\epsilon_1\left((2a+\frac16 \epsilon_2)k + \epsilon_1k^2+\frac{1}{3}n\epsilon_2\right)}.
\end{eqnarray}
Lower order terms in the expansion \eqref{eqn:1.20} are given by
\begin{eqnarray}
\label{eqn:1.23}
Y_1^{(1)}(z,a,m)&=&\frac{2mz}{\epsilon_1(2a+\epsilon_1+\frac{\epsilon_2}{2})},\hspace{17em}\\
\label{eqn:1.24}
Y_2^{(1)}(z,a,m)&=&\frac{-1}{\epsilon_1(-2a+\epsilon_1+\frac{\epsilon_2}{2})z}
+\frac{(2\epsilon_1^2+\epsilon_1\epsilon_2+4a\epsilon_1+8m^2)z^2}{4\epsilon_1^2(2a+\epsilon_1
+\frac{\epsilon_2}{2})(2a+2\epsilon_1+\frac{\epsilon_2}{2})}, \\
Y_3^{(1)}(z,a,m)&=&\frac{-2m(2\epsilon_1+\epsilon_2)}{\epsilon_1^2\epsilon_2(-2a+\epsilon_1+\frac{\epsilon_2}{2})
(2a+\epsilon_1+\frac{\epsilon_2}{2})}\nonumber\\
\label{eqn:1.25}
&&
+\frac{m(10\epsilon_1^2+3\epsilon_1\epsilon_2+12a \epsilon_1+8m^2)z^3}
{6\epsilon_1^3(2a+\epsilon_1+\frac{\epsilon_2}{2})(2a+2\epsilon_1+\frac{\epsilon_2}{2})(2a+3\epsilon_1+\frac{\epsilon_2}{2})}.
\end{eqnarray}
Mimicking the prescription for the Nekrasov partition function \cite{Alday:2009aq}, 
we multiply $Y^{(1)}(z,a,m,\Lambda)$ by an overall factor 
$\exp(-\Lambda z/\epsilon_1)$,
\begin{equation}
{\widetilde Y}^{(1)}(z,a,m,\Lambda):=e^{-\frac{\Lambda}{\epsilon_1}z}Y^{(1)}(z,a,m,\Lambda)
=\sum_{n=0}^{\infty} \Lambda^n {\widetilde Y}_n^{(1)}(z,a,m),
\label{eqn:1.25.1}
\end{equation}
to obtain\footnote{The origin of the overall factor is made clear
in the next section where we discuss the decoupling limit at the level of differential equations.} 
\begin{eqnarray}
\label{eqn:1.25.2}
{\widetilde Y}_1^{(1)}(z,a,m)&=&\frac{(2m-2a-\epsilon_1-\frac{\epsilon_2}{2})z}
{\epsilon_1(2a+\epsilon_1+\frac{\epsilon_2}{2})},\hspace{22em}\\
\label{eqn:1.25.3}
{\widetilde Y}_2^{(1)}(z,a,m)&=&\frac{-1}{\epsilon_1(-2a+\epsilon_1+\frac{\epsilon_2}{2})z}
+\frac{(2m-2a-\epsilon_1-\frac{\epsilon_2}{2})(2m-2a-3\epsilon_1-\frac{\epsilon_2}{2})z^2}
{2\epsilon_1^2(2a+\epsilon_1+\frac{\epsilon_2}{2})(2a+2\epsilon_1+\frac{\epsilon_2}{2})}, \\
{\widetilde Y}_3^{(1)}(z,a,m)&=&-\frac{2m(2\epsilon_1+\epsilon_2)-2a\epsilon_2-\epsilon_1\epsilon_2-\frac{\epsilon_2^2}{2}}
{\epsilon_1^2\epsilon_2(-2a+\epsilon_1+\frac{\epsilon_2}{2})(2a+\epsilon_1+\frac{\epsilon_2}{2})}\nonumber\\
\label{eqn:1.25.4}
&&
+\frac{(2m-2a-\epsilon_1-\frac{\epsilon_2}{2})(2m-2a-3\epsilon_1-\frac{\epsilon_2}{2})(2m-2a-5\epsilon_1-\frac{\epsilon_2}{2})z^3}
{6\epsilon_1^3(2a+\epsilon_1+\frac{\epsilon_2}{2})(2a+2\epsilon_1+\frac{\epsilon_2}{2})(2a+3\epsilon_1+\frac{\epsilon_2}{2})}.
\end{eqnarray}
We find an agreement with the computation in the gauge theory. Namely \eqref{eqn:1.25.2} -- \eqref{eqn:1.25.4} is
consistent with the decoupling  limit of the partition function \eqref{Nf4partition}.
Especially the numerators of the coefficients of $z^2$ and $z^3$ 
are of factorized form, which is not the case in \eqref{eqn:1.23} -- \eqref{eqn:1.25}.
The multiplication of the overall factor is crucial for this factorization, which is a feature of the localization computation
formula on the gauge theory side. As in the case of pure Yang-Mills theory, the free energy is
\begin{equation}
F_{\mbox{\scriptsize null\normalsize}}^{(1)}(z,a,m,\Lambda):=\log Z_{\mbox{\scriptsize null\normalsize}}^{(1)}(z,a,m,\Lambda)
=\frac{1}{\epsilon_1}\left\{F_{-1}^{(1)}+\big(\epsilon_1F_{0,1}^{(1)}+\epsilon_2F_{0,2}^{(1)}\big)+{\cal O}(\epsilon^2) \right\}
\label{eqn:1.26}
\end{equation}
and the leading term is
\begin{eqnarray}
F_{-1}^{(1)}&=&\log z^a+\frac{mz}{a}\Lambda+\frac{(a^2-m^2)z^3+2a^2}{4a^3z}\Lambda^2-\frac{m(a^2-m^2)z^3}
{6a^5}\Lambda^3  \nonumber\\
&&
+\left\{-\frac{(a^2-m^2)(a^2-5m^2)z^4}{32a^7}+\frac{(a^2-m^2)z}{4a^5}+\frac{1}{16a^3z^2}\right\}\Lambda^4+\cdots.
\label{eqn:1.27}
\end{eqnarray}


\subsection{$SU(2)$ with two fundamental matters (first realization)}

Let us concentrate on the first realization \cite{Gaiotto:2009hg, Gaiotto:2009ma} of 
$SU(2)$ theory with two fundamental matters and consider the correlation function
\begin{equation}
Z_{\mbox{\scriptsize null\normalsize}}^{(2)}(z,a,m_i,\Lambda):=\frac{\langle\Delta_-, \Lambda,m_2|\Phi_{1,2}(z)
|\Delta_+, \Lambda,m_1\rangle}
{\langle\Delta_a, \Lambda,m_2|\Delta_a, \Lambda,m_1\rangle}
=:\frac{\Psi^{(2)}(z,a,m_i,\Lambda)}{Z_c^{(2)}(a,m_i,\Lambda)},
\label{eqn:1.30}
\end{equation}
where by multiplying an overall factor, $\exp (-2\Lambda^2/\epsilon_1\epsilon_2)\cdot Z_c^{(2)}(a,m_i,\Lambda)$ coincides 
with the Nekrasov partition function after the scaling $\Lambda \to 2\Lambda$ \cite{Gaiotto:2009ma, Hadasz:2010xp}. 
In parallel with the above computations, by putting $\Psi^{(2)}(z,a,m_i,\Lambda)
=z^{\Delta_--\Delta_+-h_{1,2}}Y^{(2)}(z,a,m_i,\Lambda)$, one obtains the second order differential equation,
\begin{equation}
\left[\Big(bz\frac{\partial}{\partial z}\Big)^2+2abz\frac{\partial}{\partial z}-\Lambda^2(z^2+z^{-2})
-2\Lambda (m_2z+m_1z^{-1})+\frac{\Lambda}{2}\frac{\partial}{\partial \Lambda}\right]Y^{(2)}(z,a,m_i,\Lambda)=0,
\label{eqn:1.31}
\end{equation}
where by the decoupling limit of the two fundamental matters $m_i \to \infty, m_i\Lambda \to -\Lambda^2/2$, 
we see that the differential equation (\ref{eqn:1.31}) is reduced to (\ref{eqn:1.3}). 
By introducing the mass scale $\hbar$ as \eqref{eqn:1.4} and $m_i \rightarrow m_i/\hbar$, 
we obtain the following solution to (\ref{eqn:1.31}),
\begin{eqnarray}
\label{eqn:1.32}
&&
Y^{(2)}(z,a,m_i,\Lambda)=\sum_{n=0}^{\infty} \Lambda^n Y_n^{(2)}(z,a,m_i),\quad Y_0^{(2)}(z,a,m_i)=1,\hspace{2em}\\
&&
Y_n^{(2)}(z,a,m_i)=\sum_{k=-\infty}^{\infty}A_{n,k}^{(2)}z^k,\nonumber\\
\label{eqn:1.34}
&&
A_{0,k}^{(2)}=\delta_{0,k},\quad A_{n,k}^{(2)}=\frac{A_{n-2,k-2}^{(2)}+A_{n-2,k+2}^{(2)}+2(m_2A_{n-1,k-1}^{(2)}+m_1A_{n-1,k+1}^{(2)})}
{\epsilon_1\left(2ak+\epsilon_1k^2+\frac{1}{2}n\epsilon_2\right)},
\end{eqnarray}
with
\begin{eqnarray}
\label{eqn:1.35}
Y_1^{(2)}(z,a,m_i)&=&\frac{2m_1}{\epsilon_1(-2a+\epsilon_1+\frac{\epsilon_2}{2})z}
+\frac{2m_2z}{\epsilon_1(2a+\epsilon_1+\frac{\epsilon_2}{2})},\hspace{13em} \\
Y_2^{(2)}(z,a,m_i)&=&\frac{2\epsilon_1^2+\epsilon_1\epsilon_2-4a\epsilon_1+8m_1^2}
{4\epsilon_1^2(-2a+\epsilon_1+\frac{\epsilon_2}{2})(-2a+2\epsilon_1+\frac{\epsilon_2}{2})z^2}\nonumber\\
\label{eqn:1.36}
&&\hspace{-3.5em}
+\frac{4m_1m_2(2\epsilon_1+\epsilon_2)}{\epsilon_1^2\epsilon_2(-2a+\epsilon_1+\frac{\epsilon_2}{2})
(2a+\epsilon_1+\frac{\epsilon_2}{2})}+\frac{(2\epsilon_1^2+\epsilon_1\epsilon_2+4a\epsilon_1+8m_2^2)z^2}
{4\epsilon_1^2(2a+\epsilon_1+\frac{\epsilon_2}{2})(2a+2\epsilon_1+\frac{\epsilon_2}{2})},\\
\label{eqn:1.37}
Y_3^{(2)}(z,a,m_i)&=&\frac{m_1(10\epsilon_1^2+3\epsilon_1\epsilon_2-12a\epsilon_1+8m_1^2)}
{6\epsilon_1^3(-2a+\epsilon_1+\frac{\epsilon_2}{2})(-2a+2\epsilon_1+\frac{\epsilon_2}{2})
(-2a+3\epsilon_1+\frac{\epsilon_2}{2})z^3}+\cdots.
\end{eqnarray}
As before, multiplying $Y^{(2)}(z,a,m_i,\Lambda)$ by an overall factor 
$\exp(-\Lambda (z+z^{-1})/\epsilon_1-2\Lambda^2/\epsilon_1\epsilon_2)$,
\begin{equation}
{\widetilde Y}^{(2)}(z,a,m_i,\Lambda):=e^{-\frac{\Lambda}{\epsilon_1}(z+z^{-1})}
e^{-\frac{2\Lambda^2}{\epsilon_1\epsilon_2}}Y^{(2)}(z,a,m_i,\Lambda)
=\sum_{n=0}^{\infty} \Lambda^n {\widetilde Y}_n^{(2)}(z,a,m_i),
\label{eqn:1.37.1}
\end{equation}
we arrive at 
\begin{eqnarray}
\label{eqn:1.37.2}
{\widetilde Y}_1^{(2)}(z,a,m_i)&=&\frac{2m_1+2a-\epsilon_1-\frac{\epsilon_2}{2}}
{\epsilon_1(-2a+\epsilon_1+\frac{\epsilon_2}{2})z}+\frac{(2m_2-2a-\epsilon_1-\frac{\epsilon_2}{2})z}
{\epsilon_1(2a+\epsilon_1+\frac{\epsilon_2}{2})},\hspace{10em} \\
{\widetilde Y}_2^{(2)}(z,a,m_i)&=&\frac{(2m_1+2a-\epsilon_1-\frac{\epsilon_2}{2})(2m_1+2a-3\epsilon_1-\frac{\epsilon_2}{2})}
{2\epsilon_1^2(-2a+\epsilon_1+\frac{\epsilon_2}{2})(-2a+2\epsilon_1+\frac{\epsilon_2}{2})z^2}\nonumber\\
\label{eqn:1.37.3}
&&
+\frac{(2m_2-2a-\epsilon_1-\frac{\epsilon_2}{2})(2m_2-2a-3\epsilon_1-\frac{\epsilon_2}{2})z^2}
{2\epsilon_1^2(2a+\epsilon_1+\frac{\epsilon_2}{2})(2a+2\epsilon_1+\frac{\epsilon_2}{2})}+\cdots,\\
{\widetilde Y}_3^{(2)}(z,a,m_i)&=&\frac{(2m_1+2a-\epsilon_1-\frac{\epsilon_2}{2})
(2m_1+2a-3\epsilon_1-\frac{\epsilon_2}{2})(2m_1+2a-5\epsilon_1-\frac{\epsilon_2}{2})}
{6\epsilon_1^3(-2a+\epsilon_1+\frac{\epsilon_2}{2})(-2a+2\epsilon_1+\frac{\epsilon_2}{2})
(-2a+3\epsilon_1+\frac{\epsilon_2}{2})z^3}+\cdots. \nonumber\\
\label{eqn:1.37.4}
&&
\end{eqnarray}
Again we find an agreement with the decoupling limit of the partition function \eqref{Nf4partition} 
computed by the localization theorem on the gauge theory side. 
The multiplication of the overall factor makes the numerator factorized. The free energy is
\begin{equation}
F_{\mbox{\scriptsize null\normalsize}}^{(2)}(z,a,m_i,\Lambda):=\log Z_{\mbox{\scriptsize null\normalsize}}^{(2)}(z,a,m_i,\Lambda)
=\frac{1}{\epsilon_1}\left\{F_{-1}^{(2)}+\big(\epsilon_1F_{0,1}^{(2)}+\epsilon_2F_{0,2}^{(2)}\big)+{\cal O}(\epsilon^2) \right\},
\label{eqn:1.38}
\end{equation}
where the leading term is
\begin{equation}
F_{-1}^{(2)}=\log z^a-\frac{m_1-m_2z^2}{az}\Lambda+\frac{m_1^2-a^2+(a^2-m_2^2)z^4}{4a^3z^2}\Lambda^2+\cdots.
\label{eqn:1.39}
\end{equation}


\subsection{Multi-point irregular conformal block}

In the above computations, we explicitly checked that a single surface operator in $SU(2)$ gauge theories
 corresponds to the degenerate primary operator $\Phi_{1,2}(z)$ on the CFT side. 
 It is natural to expect that the multi-surface operators correspond 
 to the multi-degenerate primary operators $\Phi_{1,2}(z_1), \ldots, \Phi_{1,2}(z_h)$. 
 Here we introduce the multi-point irregular conformal blocks which are to be compared with 
 the computations in the B model in sections 5 and 6 (see also Appendix B for more detail). 
 Let us consider 
\begin{equation}
Z_{\mbox{\scriptsize null\normalsize}}(z_1,\ldots,z_h):=\frac{\langle G'|\Phi_{1,2}(z_1)\cdots\Phi_{1,2}(z_h)|G'' \rangle}
{\langle G|G \rangle} =:\frac{\Psi(z_1,\ldots,z_h)}{Z_c},
\label{eqn:1.42}
\end{equation}
where $|G \rangle$ is the Gaiotto state that reproduces the Nekrasov partition function $Z_c$.
The states $|G' \rangle$ and $|G'' \rangle$ in the numerator should have shifted $a$ parameters in order to be
consistent with the fusion rule of the $\Phi_{1,2}$ operator. 
In Appendix B.1 $\Psi(z_1,z_2)$ and $\Psi(z_1,z_2,z_3)$ for $N_f=0$ theory are computed 
by solving the differential equation and in Appendix B.2 $\Psi(z_1,z_2)$ is computed for $N_f=1$ theory.

After the scaling (\ref{eqn:1.4}) we consider the self-dual case $\epsilon_1=-\epsilon_2=i \hbar$ and define the free energy
\begin{equation}
F_{\mbox{\scriptsize null\normalsize}}(z_1,\ldots,z_h)=\log Z_{\mbox{\scriptsize null\normalsize}}(z_1,\ldots,z_h)
=\sum_{k=-1}^{\infty}\hbar^k F_{\mbox{\scriptsize CFT\normalsize}}^{(k)}(z_1,\ldots,z_h),
\label{eqn:1.43}
\end{equation}
since the B model computations in sections 5 and 6 only provide the free energy with the self-dual $\Omega$ background. 
Some of the explicit computation are provided in Appendix B.


\Section{Degeneration scheme of CFT differential equations}

\subsection{Ward identities}
In general, the $N$ (or $N+2$)-points block on the Riemann sphere
\beq
\Psi(z_1,\ldots,z_N)=\langle A | {\cal O}_{h_1}(z_1)\cdots {\cal O}_{N}(z_N) | B \rangle,
\eeq
can be determined by the Ward identity 
\beq
\sum_{{\rm all} \ {\rm poles}} \langle A | {\rm Res}\Big( \xi(x)T(x)dx \Big){\cal O}_{1}(z_1)\cdots {\cal O}_{N}(z_N) | B \rangle=0,
\eeq
where $\xi=\xi(x)\frac{\partial}{\partial x}$ is any rational vector field.
For example, choosing the vector field $\xi$ as $\xi(x)=\frac{z_1 x}{z_1-x}$, one has
\beq
\begin{array}l
\displaystyle
\Big[\Big(L_0+\frac{1}{z_1} L_1+\frac{1}{z_1^2} L_2+\cdots \Big)_{0}+\Big(-z_1^2 L_{-2}-z_1 L_{-1}\Big)_{z_1}\\
\displaystyle
+\sum_{j=2}^{N}\Big(\frac{z_1z_j}{z_1-z_j}L_{-1}+\frac{z_1^2}{(z_1-z_j)^2}L_0+\frac{z_1^2}{(z_1-z_j)^3}L_1
+\cdots\Big)_{z_j}\\
\displaystyle
+\Big(z_1 L_{-1}+z_1^2 L_{-2}+z_1^3 L_{-3}+\cdots\Big)_{\infty}\Big]\Psi=0,
\end{array}
\eeq
where $(\cdots)_{z}$ means the action of $T(x)$ at $z$ defined by
\beq
(L_n)_{z}\langle \cdots {\cal O}(z) \cdots \rangle=\oint_{x=z}\frac{dx}{2\pi i}(x-z)^{n+1}
\langle \cdots T(x){\cal O}(z)\cdots \rangle.
\eeq

In the case where the operator ${\cal O}_{1}$ is the degenerate field $\Phi_{1,2}$ 
such that 
\beq
L_{-2}\Phi_{1,2}=-b^2 L_{-1}^2 \Phi_{1,2},
\eeq
and other ${\cal O}_j$ are
primaries with dimension $h_j$, then we have
\beq\label{eq:cftNpt}
\begin{array}l
\displaystyle
\Big[\Big(L_0+\frac{1}{z_1} L_1+\frac{1}{z_1^2} L_2+\cdots \Big)_{0}+\Big(b^2 z_1^2 \partial_{z_1}^2-z_1 \partial_{z_1}\Big)\\
\displaystyle
+\sum_{j=2}^{N}\Big(\frac{z_1z_j}{z_1-z_j}\partial_{z_j}+\frac{z_1^2}{(z_1-z_j)^2}h_{j}\Big)
+\Big(z_1 L_{-1}+z_1^2 L_{-2}+z_1^3 L_{-3}+\cdots\Big)_{\infty}\Big]\Psi=0.
\end{array}
\eeq

\subsection{Differential equations}

\begin{figure}[tbp]
\setlength{\unitlength}{1.1mm}
\begin{center}
\begin{picture}(50,20)(-5,-3)
\put(0,0){\line(1,0){40}}
\put(10,0){\line(0,1){10}}
\put(20,0){\line(0,1){10}}
\put(30,0){\line(0,1){10}}
\put(38,1){$m_1$}\put(39,-3){$0$}
\put(30,12){$m_2$}\put(29,-3){$t$}
\put(20,12){$h_{1,2}$}\put(19,-3){$z$}
\put(10,12){$m_3$}\put(9,-3){$1$}
\put(0,1){$m_4$}\put(-1,-3){$\infty$}
\put(-5,10){$N_f=4$:}
\end{picture}\\
$\downarrow$\\
\begin{picture}(50,20)(-5,-3)
\put(0,0){\line(1,0){40}}
\put(10,0){\line(0,1){10}}
\put(20,0){\line(0,1){10}}
\put(30,1){$| \Delta_+,\Lambda,m\rangle$}\put(30,-3){$0$}
\put(20,12){$h_{1,2}$}\put(19,-3){$z$}
\put(10,12){$m_-$}\put(9,-3){$1$}
\put(0,1){$m_+$}\put(-1,-3){$\infty$}
\put(-5,10){$N_f=3$:}
\end{picture}\\
$\swarrow$\hskip10mm$\searrow$\\
\begin{picture}(50,20)(-10,-3)
\put(0,0){\line(1,0){40}}
\put(20,0){\line(0,1){10}}
\put(30,1){$| \Delta_+,\Lambda,m_1\rangle$}\put(30,-3){$0$}
\put(20,12){$h_{1,2}$}\put(19,-3){$z$}
\put(-5,1){$\langle \Delta_-,\Lambda,m_2|$}\put(7,-3){$\infty$}
\put(-5,10){$N_f=2^{(1)}$:}
\end{picture}
\begin{picture}(50,20)(-20,-3)
\put(0,0){\line(1,0){40}}
\put(10,0){\line(0,1){10}}
\put(20,0){\line(0,1){10}}
\put(30,1){$| \Delta_+,\Lambda\rangle$}\put(30,-3){$0$}
\put(20,12){$h_{1,2}$}\put(19,-3){$z$}
\put(10,12){$m_-$}\put(9,-3){$1$}
\put(0,1){$m_+$}\put(-1,-3){$\infty$}
\put(-10,10){$N_f=2^{(2)}$:}
\end{picture}\\
$\searrow$\hskip10mm$\swarrow$\\
\begin{picture}(50,20)(-5,-3)
\put(0,0){\line(1,0){40}}
\put(20,0){\line(0,1){10}}
\put(30,1){$| \Delta_+,\Lambda\rangle$}\put(30,-3){$0$}
\put(20,12){$h_{1,2}$}\put(19,-3){$z$}
\put(-5,1){$\langle \Delta_-,\Lambda,m|$}\put(7,-3){$\infty$}
\put(-5,10){$N_f=1$:}
\end{picture}\\
$\downarrow$\\
\begin{picture}(50,20)(-5,-3)
\put(0,0){\line(1,0){40}}
\put(20,0){\line(0,1){10}}
\put(30,1){$| \Delta_+,\Lambda\rangle$}\put(30,-3){$0$}
\put(20,12){$h_{1,2}$}\put(19,-3){$z$}
\put(0,1){$\langle \Delta_-,\Lambda|$}\put(7,-3){$\infty$}
\put(-5,10){$N_f=0$:}
\end{picture}
\end{center}
\caption{Conformal blocks and their degenerations}
\label{fig:degeneration}
\end{figure}
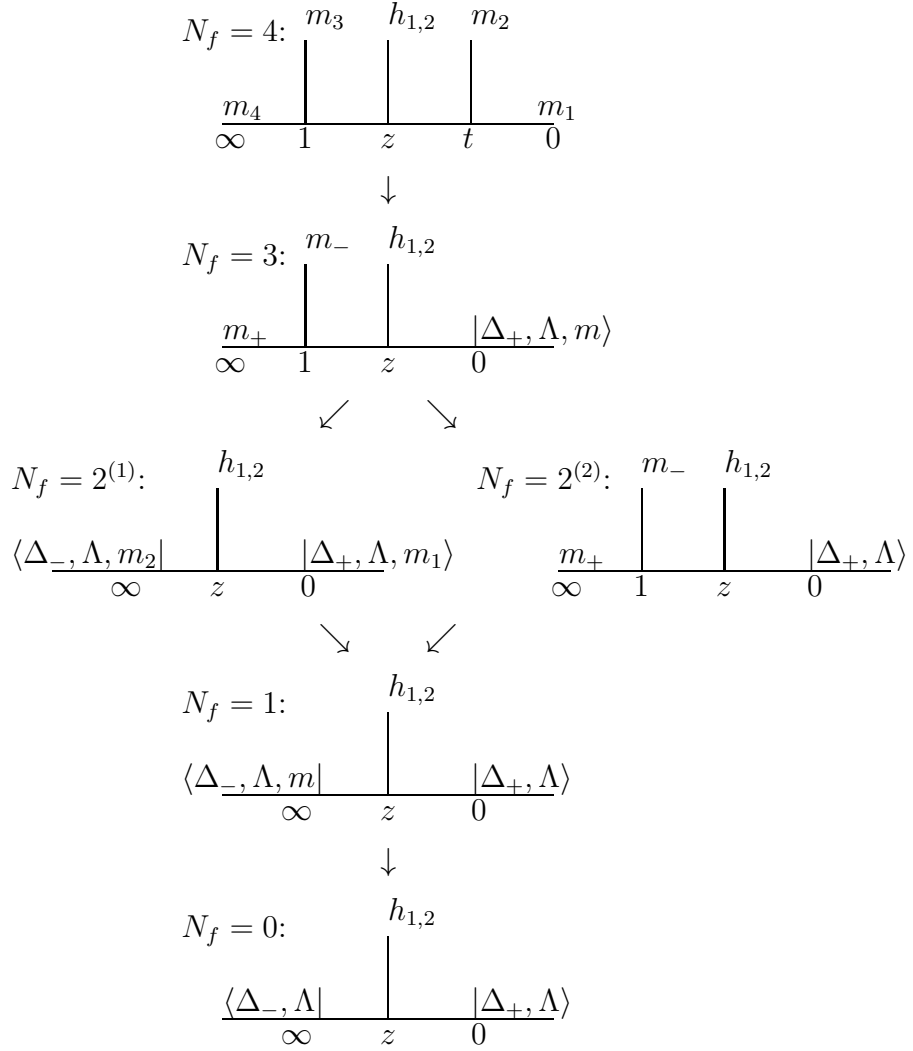

We will give a list of CFT differential equations with single $\Phi_{1,2}(z)$ operator insertion (Fig.\ref{fig:degeneration}).

\noindent $\bullet$ {$N_f=0$}: $\Psi(z,\Lambda)=\big\langle \Delta_-, \Lambda \big|\Phi_{1,2}(z)\big|\Delta_+,\Lambda\big\rangle$, 
\beq
b^2 z^2 \Psi_{zz}-\frac{3}{2} z \Psi_{z}
+\frac{\Lambda}{4}\Psi_{\Lambda}+\Big(\frac{\Delta_{-}+\Delta_{+}-h_{1,2}}{2}+\Lambda^2 (z+\frac{1}{z})\Big) \Psi=0.
\eeq

\noindent $\bullet$ {$N_f=1$}: $\Psi(z,\Lambda)=\big\langle \Delta_-, \Lambda, m \big|\Phi_{1,2}(z)\big|\Delta_+,\Lambda\big\rangle$, 
\beq
b^2 z^2 \Psi_{zz}-\frac{4}{3} z \Psi_{z}+\frac{\Lambda}{3}\Psi_{\Lambda}+\Big(\frac{\Delta_{-}+2 \Delta_{+}-h_{1,2}}{3}
+\Lambda^2 (\frac{1}{z}-z^2)-2 \Lambda m z\Big)\Psi=0.
\eeq

\noindent $\bullet$ {$N_f=2^{(1)}$ (1st realization)}:
$\Psi(z,\Lambda)=\big\langle \Delta_-, \Lambda, m_2 \big|\Phi_{1,2}(z)\big|\Delta_+,\Lambda, m_1\big\rangle$,
\beq
b^2 z^2 \Psi_{zz}-\dfrac{3}{2}z \Psi_z+\frac{\Lambda}{2} \Psi_{\Lambda}
+\Big(\frac{\Delta_{-}+\Delta_{+}-h_{1,2}}{2}
-\Lambda^2 (z^2+\frac{1}{z^2}) -2 \Lambda (\frac{m_1}{z}+m_2 z)\Big) \Psi=0.
\eeq

\noindent $\bullet$ {$N_f=2^{(2)}$ (2nd realization)}:
$\Psi(z,\Lambda)=
\big\langle \Delta_{m_+} \big|\Phi_{\Delta_{m_-}}(1) \Phi_{1,2}(z)\big|\Delta_+,\Lambda\big\rangle$, 
\beq
b^2 z(z-1) \Psi_{zz}-(2 z-1) \Psi_{z} -\frac{\Lambda}{2 z} \Psi_{\Lambda}
+\Big(\frac{\Delta_{m_-}}{z-1}+\Delta_{m_+}-\frac{\Delta_{+}}{z}-h_{1,2}+\frac{\Lambda^2(z-1)}{z^2}\Big)\Psi=0.
\eeq

\noindent $\bullet$ {$N_f=3$}: $\Psi(z,\Lambda)=
\big\langle \Delta_{m_+} \big|\Phi_{\Delta_{m_-}}(1) \Phi_{1,2}(z)\big|\Delta_+,\Lambda, m\big\rangle$, 
\beq
\begin{array}{l}
\displaystyle
b^2 z(z-1) \Psi_{zz}-(2 z-1) \Psi_{z}-\frac{\Lambda}{z}\Psi_{\Lambda} \\
\displaystyle
+\Big(\frac{\Delta_{m_-}}{z-1}+\Delta_{m_+}-\frac{\Delta_{+}}{z}-h_{1,2}-\frac{2 m \Lambda (z-1)}{z^2}-\frac{\Lambda^2 (z-1) }{z^3}\Big)\Psi=0.
\end{array}
\eeq

\noindent $\bullet$ {$N_f=4$}:
$\Psi(z,t)=
\big\langle \Delta_{m_4} \big|\Phi_{\Delta_{m_3}}(1) \Phi_{1,2}(z)
\Phi_{\Delta_{m_2}}(t)\big|\Delta_{m_1}\big\rangle$, 
\beq
\begin{array}{l}
\displaystyle
b^2 (z-1)z \Psi_{zz}-(2z-1)\Psi_{z}+\frac{(t-1)t}{(z-t)} \Psi_{t}\\
\displaystyle
+\Big(\frac{\Delta_{m_3}}{z-1}+\Delta_{m_4}-\frac{\Delta_{m_1}}{z}-\frac{\Delta_{m_2}(t^2-2 t z+z)}{(z-t)^2}-h_{1,2}\Big)\Psi
=0.
\end{array}
\eeq

For the block of the form $\Psi(z)=\langle A | {\cal O}_{1}(1)\Phi_{1,2}(z) \cdots| B \rangle$
($N_f=2^{(2)}, 3$ and $4$), a convenient choice of the vector field $\xi$ is
$\xi=\frac{x(x-1)}{x-z} \frac{\partial}{\partial x}$.


\subsection{Quantum Seiberg-Witten curves}

The CFT differential equation in previous subsection can be considered as a natural
candidate for the speculated quantized  Seiberg-Witten curve, that is an operator version of
the equation $x^2=\phi_2(z)$ (see the end of section 5 of \cite{Alday:2009aq}).
By a gauge transformation\footnote{The factors $U$ are determined by comparison with
the gauge theory (localization) results. It may be interesting to note that
the factor $U$ for $N_f=4$ case is exactly the same as the pre-factor appearing
in the integral (free field) representation of the conformal block.} $\Psi=U {\cal Z}$, it can be written
in the form ${\cal D}^{\rm SW}_{N_f}{\cal Z}=0$ which looks 
like the Seiberg-Witten curve in the standard brane set-up. 
The operator $\widehat{v}=b z \partial_z$ is a quantization of the variable $v=zx$.

\noindent $\bullet$ {$N_f=0$}: $U=z^{\Delta_--\Delta_+-h_{1,2}}$, 
\beq
{\cal D}^{\rm SW}_{0}=\frac{1}{4}\Lambda \partial_{\Lambda}+2 a \widehat {v}+\widehat {v}^2+\Lambda^2(z+\frac{1}{z}).
\eeq

\noindent $\bullet$ {$N_f=1$}: $U=e^{\frac{\Lambda}{b}z}z^{\Delta_--\Delta_+-h_{1,2}}$, 
\beq
{\cal D}^{\rm SW}_{1}=\frac{1}{3}\Lambda\partial_\Lambda
+(2 a +\frac{1}{6b}) \widehat {v}+\widehat {v}^2+2 \Lambda z \Big(\widehat {v}+a+\frac{1}{4b}+\frac{b}{2}-m\Big)+\frac{\Lambda^2}{z}.
\eeq

\noindent $\bullet$ {$N_f=2^{(1)}$}: $U=e^{2\Lambda^2+\frac{\Lambda}{b}(z+\frac{1}{z})}z^{\Delta_--\Delta_+-h_{1,2}}$, 
\beq
{\cal D}^{\rm SW}_{2^{(1)}}=\frac{1}{2}\Lambda\partial_\Lambda
+2 a \widehat {v}+\widehat {v}^2  -\frac{2 \Lambda}{z} \Big(\widehat {v}+a-\frac{1}{4b}-\frac{b}{2}+m_1\Big)
+2 \Lambda z \Big(\widehat {v}+a+\frac{1}{4b}+\frac{b}{2}-m_2\Big).
\eeq

\noindent $\bullet$ {$N_f=2^{(2)}$}: $U=(z-1)^{\frac{1+b^2-2 b m_-}{2 b^2}} z^{\Delta_--\Delta_+-h_{1,2}}$,
\beq
{\cal D}^{\rm SW}_{2^{(2)}}=\frac{1}{2}\Lambda \partial_\Lambda-\Lambda^2
+\Big(2 a+\frac{1}{2 b}\Big) \widehat {v}+\widehat {v}^2+\frac{\Lambda^2}{z}-z \Big(\widehat {v}+a+\frac{1}{4b}+\frac{b}{2}-m_--m_+\Big)
\Big(\widehat {v}+a+\frac{1}{4 b}+\frac{b}{2}-m_-+m_+\Big).
\eeq

\noindent $\bullet$ {$N_f=3$}: $U=e^{-\Lambda(b+\frac{1}{b}-2m_--\frac{1}{b z})}(z-1)^{\frac{1+b^2-2 b m_-}{2 b^2}}z^{\Delta_--\Delta_+-h_{1,2}}$, 
\beq
\begin{array}l
\displaystyle
{\cal D}^{\rm SW}_{3}=\Lambda \partial_\Lambda
+(2 a -\frac{1}{2b}-b+2 m)\Lambda+(2 a+\frac{1}{2b}+2 \Lambda) \widehat {v} +\widehat {v}^2-\frac{2\Lambda}{z}
\Big(\widehat {v}+a-\frac{1}{4 b}-\frac{b}{2}+m\Big)\\
\displaystyle
-z \Big(\widehat {v} +a+\frac{1}{4 b}+\frac{b}{2}-m_--m_+\Big)\Big(\widehat {v}+a+\frac{1}{4 b}+\frac{b}{2}-m_-+m_+\Big).
\end{array}
\eeq

\noindent $\bullet$ {$N_f=4$}:
$U=(1-t)^{-\frac{(1+b^2-2 b m_2)(1+b^2-2 b m_3)}{2 b^2}}
(1-\frac{t}{z})^{\frac{1+b^2-2 b m_2}{2 b^2}} (1-z)^{\frac{1+b^2-2 b m_3}{2 b^2}}
z^{\Delta_--\Delta_+-h_{1,2}} t^{\Delta_+-\Delta_{m_1}-\Delta_{m_2}}$,
\beq 
\begin{array}{l}
\displaystyle
{\cal D}^{\rm SW}_{4}=(1-t) t\partial_t
+(\mu_{1} \mu_{2}+\mu_{3} \mu_{4}) t+\Big((2 a+\frac{1}{2 b})(1-t)+(\mu_{1}+\mu_{2}+\mu_{3}+\mu_{4}) t\Big) \widehat {v}\\
\displaystyle
+(1+t) \widehat {v}^2
-\frac{t}{z}(\widehat {v}+\mu_1)(\widehat {v}+\mu_2)
-z(\widehat {v}+\mu_3)(\widehat {v}+\mu_4),
\end{array}
\eeq
where 
\beq
\begin{array}{ll}
\mu_1=a-\frac{1}{4 b}-\frac{b}{2}-m_1+m_2,\quad &
\mu_2=a-\frac{1}{4 b}-\frac{b}{2}+m_1+m_2,\\
\mu_3=a+\frac{1}{4 b}+\frac{b}{2}-m_3-m_4,\quad &
\mu_4=a+\frac{1}{4 b}+\frac{b}{2}-m_3+m_4.\\
\end{array}
\eeq

We remark that the equations for the (irregular) conformal blocks considered here 
are the same as the Schr\"odinger equation for quantum Painlev\'e equations \cite{nagoya}. 
The connection between CFT, iso-monodromy deformation, and Seiberg-Witten curves are natural 
because (i) CFT (KZ equation for example)
are the quantization of iso-monodromy deformation (Schlesinger system) \cite{KZ-IMD, Teschner:2010je}
and (ii) the cubic equations which determine the classical Painlev\'e Hamiltonians coincide
with the $SU(2)$ Seiberg-Witten curves \cite{KMNOY}.

  
\subsection{Degenerations}

Here, we give the degeneration scheme that connect the $N_f$ and $N_f-1$ equations.
We use the notation $\Lambda_{N_f}$ and $z_{N_f}$ for $\Lambda$, $z$ variables of $N_f=0,1,2^{(1)},2^{(2)},3$ and $4$.

\noindent $\bullet$ 
We have ${\cal D}^{\rm SW}_4 \rightarrow {\cal D}^{\rm SW}_{3}$ under the limit
\beq
m_1+m_2=m, \quad m_3=m_-, \quad m_4=m_+, \quad t m_2 =\Lambda_3, \quad m_2 \rightarrow \infty, \quad
z_4=z_3.
\eeq

\noindent $\bullet$ 
We have ${\cal D}^{\rm SW}_3 \rightarrow {\cal D}^{\rm SW}_{2^{(1)}}$ under the limit
\beq
m=m_1, \quad
m_++m_-=m_2, \quad
m_-\rightarrow \infty, \quad
\Lambda_3 m_-=\Lambda_{2^{(1)}}^2, \quad
\frac{z_3}{\Lambda_3}=\frac{z_{2^{(1)}}}{\Lambda_{2^{(1)}}}.
\eeq

\noindent $\bullet$ 
We have ${\cal D}^{\rm SW}_3 \rightarrow {\cal D}^{\rm SW}_{2^{(2)}}$ under the limit
\beq
m\rightarrow \infty, \quad
-2 \Lambda_3 m=\Lambda_{2^{(2)}}^2, \quad
z_3=z_{2^{(2)}}.
\eeq

\noindent $\bullet$ 
We have ${\cal D}^{\rm SW}_{2^{(1)}} \rightarrow {\cal D}^{\rm SW}_{1}$ under the limit
\beq
m_1 \rightarrow \infty, \quad
m_2=m, \quad
-2 \Lambda_{2^{(1)}}^2 m_1=\Lambda_1^3, \quad
z_{2^{(1)}}\Lambda_{2^{(1)}}=z_1\Lambda_1.
\eeq

\noindent $\bullet$ 
We have ${\cal D}^{\rm SW}_{2^{(2)}} \rightarrow {\cal D}^{\rm SW}_{1}$ under the limit
\beq
m_- \rightarrow \infty, \quad
m_++m_-=m, \quad
\Lambda_{2^{(2)}}^2 m_-=\Lambda_1^3, \quad
\frac{z_{2^{(2)}}}{\Lambda_{2^{(2)}}^2}=\frac{z_1}{\Lambda_1^2}.
\eeq

\noindent $\bullet$ 
We have ${\cal D}^{\rm SW}_1 \rightarrow {\cal D}^{\rm SW}_{0}$ under the limit
\beq
m \rightarrow \infty, \quad
-2 \Lambda_1^3 m=\Lambda_0^4, \quad
\frac{z_1}{\Lambda_1^2}=\frac{z_0}{\Lambda_0^2}.
\eeq

In all the cases, the degenerations of the CFT are consistent with the 
decoupling limit of the gauge theory as discussed in the case of without surface operator \cite{Marshakov:2009gn}.
By the parameter relation (\ref{eq:AGTpara}), the decoupling limit are described as follows:
\beq
\begin{array}{ll}
4 \rightarrow 3 \quad &: \quad
t=\Lambda_3 \frac{\hbar}{M_2}, \quad M_2 \rightarrow \infty,\\
3 \rightarrow 2^{(1)} \quad &: \quad
\Lambda_3=\Lambda_{2^{(1)}}^2\frac{\hbar}{-M_3},\quad 
z_3=z_{2^{(1)}}\Lambda_{2^{(1)}}\frac{\hbar}{-M_3},\quad -M_3 \rightarrow \infty,\\
3 \rightarrow 2^{(2)} \quad &: \quad
\Lambda_3=\Lambda_{2^{(2)}}^2\frac{\hbar}{4M_4},\quad M_4 \rightarrow \infty,\\
2^{(1)} \rightarrow 1 \quad &: \quad
\Lambda_{2^{(1)}}^2=\Lambda_1^3\frac{\hbar}{4M_4},\quad 
z_{2^{(1)}}=z_{1}(\Lambda_1\frac{\hbar}{4M_4})^{-1/2},\quad M_4 \rightarrow \infty,\\
2^{(2)} \rightarrow 1 \quad &: \quad
\Lambda_{2^{(2)}}^2=\Lambda_1^3\frac{\hbar}{-M_3},\quad 
z_{2^{(2)}}=z_{1}\Lambda_1\frac{\hbar}{-M_3},\quad -M_3 \rightarrow \infty\\
1 \rightarrow 0 \quad &: \quad
\Lambda_1^3=\Lambda_0^4 \frac{\hbar}{-4 M_1},\quad 
z_1=z_0(\Lambda_0 \frac{\hbar}{-4 M_1})^{2/3},\quad
-M_1 \rightarrow \infty,\\
&\hbar=(\epsilon_1\epsilon_2)^{1/2}.
\end{array}
\eeq

The degeneration relation of $N_f=4 \rightarrow N_f=3$ (and similarly 
$N_f=3 \rightarrow N_f=2^{(1)}$) has simple interpretation in operator level
as follows. Consider the product
\beq
\big| \psi \big\rangle=z^{\Delta_{m_1}+\Delta_{m_2}-\Delta_a}\Phi_{\Delta_{m_1}}(z) \big| \Delta_{m_2} \big\rangle.
\eeq
By the definition of the primary filed, it satisfies
\beq
L_n \big| \psi \big\rangle=
z^n\big(z \partial_z+\Delta_a+n\Delta_{m_1}-\Delta_{m_2}\big)\big| \psi \big\rangle. \quad (n>0)
\eeq
Here, we will put
\beq\label{eq:C-op-deg}
z=\epsilon \Lambda, \quad m_1=-\frac{1}{\epsilon}-m, \quad
m_2=\frac{1}{\epsilon}.
\eeq
Then, under the limit $\epsilon \rightarrow 0$, we have
$\Delta_a+n\Delta_{m_1}-\Delta_{m_2}=\frac{1-n}{\epsilon^2}+\frac{-2 mn}{\epsilon}+O(\epsilon^0)$ and
\beq
L_1 \big| \psi \big\rangle=-2 m \Lambda \big| \psi \big\rangle, \quad
L_2 \big| \psi \big\rangle=- \Lambda^2 \big| \psi \big\rangle, \quad
L_n \big| \psi \big\rangle=0, \quad (n\geq 3).
\eeq
Thus, the Gaiotto state $\big| \Delta_a,\Lambda, m \big\rangle$ can be obtained as a degeneration
limit (\ref{eq:C-op-deg}) of two primaries.


\subsection{Solutions}

The equation ${\cal D}^{\rm SW}_{4} {\cal Z}(z,t)=0$ has the following series solution 
\beq
{\cal Z}(z,t)=\sum_{n=0}^{\infty}{\cal Z}_n(z)t^n, \quad {\cal Z}_n(z)=\sum_{k=-n}^{\infty} c_{n,k}z^k.
\eeq
First few terms are as follows
\beq
\begin{array}{l}
\displaystyle
{\cal Z}_0=1 
 + \frac{2 \mu_{3} \mu_{4}}{2 b^2+4 a b+1} z
 + \frac{2 \mu_{3} (b+\mu_{3}) \mu_{4} (b+\mu_{4})}{\left(2 b^2+4 a b+1\right) \left(4 b^2+4 a b+1\right)} z^2+{\cal O}(z^3),\\
 \displaystyle
{\cal Z}_1=
 \frac{2 \mu_{1} \mu_{2}}{2 b^2-4 a b+1} \frac{1}{z}
 -\mu_{1} \mu_{2}-\mu_{3} \mu_{4} +\frac{2 \mu_{1}\mu_{2} (b-\mu_{3}) (b-\mu_{4}) }{2 b^2-4 a b+1}+\frac{2 (b+\mu_{1}) (b+\mu_{2}) \mu_{3} \mu_{4}}{2 b^2+4 a b+1}+{\cal O}(z),\\
 \displaystyle
{\cal Z}_2= \frac{2 (b-\mu_{1}) \mu_{1} (b-\mu_{2}) \mu_{2}}{\left(2 b^2-4 a b+1\right) \left(4 b^2-4 a b+1\right)}  \frac{1}{z^2} +{\cal O}(\frac{1}{z}),
\end{array}
\eeq
These agree with the localization results \eqref{Nf4partition} 
by the following correspondence\footnote{We have checked this up to the order 7 in $x$ and $y$ variables.} :
\beq\label{eq:AGTpara}
\begin{array}{l}
\displaystyle
a \rightarrow \frac{-a-\epsilon_{2}/4}{\sqrt{\epsilon_{1}\epsilon_{2}}},\quad
b \rightarrow \sqrt{\frac{\epsilon_{1}}{\epsilon_{2}}},\quad
z \rightarrow -x, \quad
\frac{t}{z} \rightarrow -y,\\
\displaystyle
\mu_1 \rightarrow \frac{-a-\epsilon_{1}-\epsilon_{2}+2 M_{2}}{\sqrt{\epsilon_{1}\epsilon_{2}}},\quad
\mu_2 \rightarrow \frac{-a-2M_{4}}{\sqrt{\epsilon_{1}\epsilon_{2}}},\quad
\mu_3 \rightarrow \frac{-a+\epsilon_{1}-2 M_{1}}{\sqrt{\epsilon_{1}\epsilon_{2}}},\quad
\mu_4 \rightarrow \frac{-a+2 M_{3}}{\sqrt{\epsilon_{1}\epsilon_{2}}}.
\end{array}
\eeq

The coefficients of the border terms $z^n$ and $(\frac{t}{z})^n$ are always factorized. 
From the CFT point of view, this can be understood by fusion (not degeneration) of
primary operators. 
More precisely, for $t\rightarrow 0$, then $\Phi_{\Delta(m_2)}(t)$ and $|\Delta(m_1)\rangle$
are fused and we have
\beq
{\cal Z}(z,0)={}_2F_{1}(\frac{\mu_3}{b},\frac{\mu_4}{b},\frac{2a}{b}+1+\frac{1}{2b^2},z).
\eeq
Similarly, for $z \rightarrow 0$ (with $u=\frac{t}{z}$ fixed), then $\langle \Delta(m_4)|$ and
$\Phi_{\Delta(m_2)}(1)$ are fused and
\beq
{\cal Z}(z,uz)={}_2F_{1}(-\frac{\mu_1}{b},-\frac{\mu_2}{b},-\frac{2a}{b}+1+\frac{1}{2b^2},u).
\eeq

For the degenerate cases $N_f\leq 3$, one can also solve the 
differential equations ${\cal D}^{\rm SW}_{N_f}{\cal Z}=0$ in series expansion.
Alternatively, such solutions can be obtained through the limiting procedure starting form $N_f=4$ case.
Since the limit can be taken term by term with respect to the variables $z$ and $\Lambda$ (or $t$), 
we will illustrate the procedure on simplest examples.

The first example is the term $z$ in $N_f=4$ solution and its degenerations:
{\large
\beq
\begin{array}{c}
\frac{\left(2 b^2+4 a b-4 m_{3} b-4 m_{4} b+1\right) \left(2 b^2+4 a b-4 m_{3} b+4 m_{4} b+1\right) z}{8 b^2 \left(2 b^2+4 a b+1\right)} \\
\downarrow\\
\frac{\left(2 b^2+4 a b-4 m_{-} b-4 m_{+} b+1\right) \left(2 b^2+4 a b-4 m_{-} b+4 m_{+} b+1\right) z}{8 b^2 \left(2 b^2+4 a b+1\right)} \\
\swarrow \hskip20mm \searrow\\
\hskip 20mm -\frac{\Lambda \left(2 b^2+4 a b-4 m_{2} b+1\right) z}{b \left(2 b^2+4 a b+1\right)} \hskip10mm
\frac{\left(2 b^2+4 a b-4 m_{-} b-4 m_{+} b+1\right) \left(2 b^2+4 a b-4 m_{-} b+4 m_{+} b+1\right) z}{8 b^2 \left(2 b^2+4 a b+1\right)}\\
\searrow \hskip20mm \swarrow\\
-\frac{\Lambda \left(2 b^2+4 a b-4 m b+1\right) z}{b \left(2 b^2+4 a b+1\right)} \\
\downarrow\\
-\frac{2 \Lambda^2 z}{2 b^2+4 a b+1}. \nonumber
\end{array}
\eeq}
Here, the arrows are the same meaning as Fig.\ref{fig:degeneration}. This degeneration corresponds
to the decoupling of the fundamental matter attached to the vertical brane at $z=0$.
The next example represents the similar decoupling process around the brane at $z=\infty$:
{\large\beq
\begin{array}{c}
\frac{\left(2 b^2-4 a b-4 m_{1} b-4 m_{2} b+1\right) \left(2 b^2-4 a b+4 m_{1} b-4 m_{2} b+1\right) t}{8 b^2 \left(2 b^2-4 a b+1\right) z}\\
\downarrow\\
-\frac{\left(2 b^2-4 a b-4 m b+1\right) \Lambda}{b \left(2 b^2-4 a b+1\right) z}\\
\swarrow \hskip20mm \searrow\\
-\frac{\Lambda \left(2 b^2-4 a b-4 m_{1} b+1\right)}{b \left(2 b^2-4 a b+1\right) z} \hskip10mm
\frac{2 \Lambda^2}{\left(-2 b^2+4 a b-1\right) z} \\
\searrow \hskip20mm \swarrow\\
\frac{2 \Lambda^2}{\left(-2 b^2+4 a b-1\right) z} \\
\downarrow\\
\frac{2 \Lambda^2}{\left(-2 b^2+4 a b-1\right) z}. \nonumber
\end{array}
\eeq}


\section{B model computations via Seiberg-Witten curve}

In \cite{Alday:2009aq} it is argued that the Seiberg-Witten curve arises 
in the \textquotedblleft semiclassical limit\textquotedblright\hspace{0.25em}
$\epsilon_{1,2}\ll a_i,m_i$ of the expectation value of the energy momentum tensor. 
For example, by taking the limit $\hbar\to 0$ one finds the following Seiberg-Witten curves 
$x^2=\phi_n^{\mbox{\scriptsize SW\normalsize}}(z)$ \cite{Gaiotto:2009ma},
\begin{eqnarray}
\mbox{$SU(2), N_f=0$}:~&&
\langle\Delta_a, \Lambda|T(z)|\Delta_a, \Lambda\rangle~\longrightarrow~
-\frac{1}{\hbar^2}\phi_0^{\mbox{\scriptsize SW\normalsize}}(z) \langle\Delta_a, \Lambda|\Delta_a, \Lambda\rangle \hspace{8.2em}\nonumber\\
\label{eqn:2.1}
&&
\hspace{-6.6em}\phi_0^{\mbox{\scriptsize SW\normalsize}}(z)=M_0(z)^2\sigma_0(z),~\sigma_0(z):=-z\big(z^2-\frac{u}
{\Lambda^2}z+1\big),~M_0(z):=\frac{\Lambda}{z^2},\\
\mbox{$SU(2), N_f=1$}:~&&
\langle\Delta_a, \Lambda,m|T(z)|\Delta_a, \Lambda\rangle~\longrightarrow~
-\frac{1}{\hbar^2}\phi_1^{\mbox{\scriptsize SW\normalsize}}(z) \langle\Delta_a, \Lambda,m|\Delta_a, \Lambda\rangle \nonumber\\
\label{eqn:2.2}
&&
\hspace{-6.6em}\phi_1^{\mbox{\scriptsize SW\normalsize}}(z)=M_1(z)^2\sigma_1(z),~\sigma_1(z):=z\big(z^3+\frac{2m}
{\Lambda}z^2+\frac{u}{\Lambda^2}z-1\big),~M_1(z):=\frac{\Lambda}{z^2},\\
\mbox{$SU(2), N_f=2$}:~&&
\langle\Delta_a, \Lambda,m_2|T(z)|\Delta_a, \Lambda,m_1\rangle~\longrightarrow~
-\frac{1}{\hbar^2}\phi_2^{\mbox{\scriptsize SW\normalsize}}(z) \langle\Delta_a, \Lambda,m_2|\Delta_a, \Lambda,m_1\rangle \nonumber\\
\label{eqn:2.3}
&&
\hspace{-6.6em}\phi_2^{\mbox{\scriptsize SW\normalsize}}(z)=M_2(z)^2\sigma_2(z),~\sigma_2(z):=z^4+\frac{2m_2}{\Lambda}z^3+\frac{u}
{\Lambda^2}z^2+\frac{2m_1}{\Lambda}z+1,~M_2(z):=\frac{\Lambda}{z^2},
\end{eqnarray}
where in this section the subscript $n$ of $\phi_n^{\mbox{\scriptsize SW\normalsize}}(z)=M_n(z)^2\sigma_n(z)$ 
stands for the number of flavors. The Coulomb moduli parameter $u=a^2+{\cal O}(\Lambda)$ 
in each ${\cal N}=2$ supersymmetric gauge theory is determined from the period
\begin{equation}
a(u)=\oint_A \lambda_{\mbox{\scriptsize SW\normalsize}}(z),\quad \lambda_{\mbox{\scriptsize SW\normalsize}}(z):=x(z)dz,
\label{eqn:2.5}
\end{equation}
where $A$ is the $A$-cycle on the Seiberg-Witten curve $x^2=\phi_n^{\mbox{\scriptsize SW\normalsize}}(z)$. 
Using the discussion in \cite{Alday:2009fs}, 
one can find that the leading term (disk amplitude) $F_{-1}$ of the free energy $F_{\mbox{\scriptsize null\normalsize}}$ 
in section 4 is related to the Seiberg-Witten curve by
\begin{equation}
\left(\frac{\partial F_{-1}(z)}{\partial z}\right)^2=\phi_n^{\mbox{\scriptsize SW\normalsize}}(z),\quad \longrightarrow \quad F_{-1}(z)
=\pm \int^z \lambda_{\mbox{\scriptsize SW\normalsize}}(z').
\label{eqn:2.6}
\end{equation}
Note that in this computation we do not know how to determine the constant of integration for $F_{-1}(z)$. 
Actually for (\ref{eqn:2.1}) -- (\ref{eqn:2.3}), 
we can check that the right hand side of (\ref{eqn:2.6}) agrees with the computations (\ref{eqn:1.14}), (\ref{eqn:1.27}) 
and (\ref{eqn:1.39}) in section 4 for the first few orders in $\Lambda$ except constant terms in the insertion point $z$ 
of the degenerate operator.

In \cite{Eynard:2007kz}, Eynard and Orantin defined the free energies on arbitrary complex plane curves
 by the topological recursion which has its origin to the loop equation in matrix models. 
 In \cite{Kozcaz:2010af}, it was claimed that the correlation functions in the CFT for ${\cal N}=2$ superconformal 
 quiver gauge theories can be related to the free energies defined by the topological recursion on the Seiberg-Witten curves 
 obtained from the energy momentum tensor of the CFT as (\ref{eqn:2.1}) -- (\ref{eqn:2.3}). 
 In this section we generalize their claim to asymptotically free theories. 
 Following the construction of Eynard and Orantin, let us define the free energies 
 ${\cal F}_{\mbox{\scriptsize SW\normalsize}}^{(g,h)}(z_1,\ldots,z_h)$, $g,h \in {\IZ}_{\ge 0},~h \ge 1$ 
on the Seiberg-Witten curve ${\cal C}_{\mbox{\scriptsize SW\normalsize}}$: 
$x^2=\phi_n^{\mbox{\scriptsize SW\normalsize}}(z)=M_n(z)^2\sigma_n(z)$ by
\begin{eqnarray}
\label{eqn:2.7}
&&{\cal F}_{\mbox{\scriptsize SW\normalsize}}^{(g,h)}(z_1,\ldots,z_h):=\int^{z_1} \cdots \int^{z_h} W^{(g,h)}(z_1',\ldots,z_h'),\\
&&W^{(0,1)}(z):=\lambda_{\mbox{\scriptsize SW\normalsize}}(z),\quad W^{(0,2)}(z_1,z_2):=B(z_1,z_2)-\frac{dz_1dz_2}{(z_1-z_2)^2}, \nonumber\\
&&W^{(g,h)}(z_1,\ldots,z_h):=\widetilde{W}^{(g,h)}(z_1,\ldots,z_h)\quad \mbox{for}\quad (g,h)\neq (0,1),(0,2), \nonumber
\end{eqnarray}
where ${\cal F}_{\mbox{\scriptsize SW\normalsize}}^{(0,1)}(z)$ is nothing but the disk amplitude (\ref{eqn:2.6}). 
The multilinear meromorphic differentials $W^{(g,h)}(z_1,\ldots,z_h)$ 
are defined on ${\cal C}_{\mbox{\scriptsize SW\normalsize}}$ by the topological recursion relation
\begin{eqnarray}
&&
\widetilde{W}^{(0,1)}(z):=0,\quad \widetilde{W}^{(0,2)}(z_1,z_2):=B(z_1,z_2),\hspace{19em}\nonumber\\
&&
dE_{q,\bar{q}}(z):=\frac12\int_q^{\bar{q}}B(z,\xi),~\mbox{near a branch point}~q_i, \nonumber\\
&&\widetilde{W}^{(g,h+1)}(z,z_1,\ldots,z_h):=
\sum_{q_i \in{\cal C}_{\mbox{\scriptsize SW\normalsize}}}\mathop{\mbox{Res}}_{q=q_i} 
\frac{dE_{q,\bar{q}}(z)}{\lambda_{\mbox{\scriptsize SW\normalsize}}(q)-\lambda_{\mbox{\scriptsize SW\normalsize}}
({\bar q})}\biggl\{\widetilde{W}^{(g-1,h+2)}(q,\bar{q},z_1,\ldots,z_h) \nonumber\\
&&\hspace{13em} +\sum_{\ell=0}^g \sum_{J \subset H}\widetilde{W}^{(g-\ell,|J|+1)}(q,z_J)
\widetilde{W}^{(\ell,|H|-|J|+1)}(\bar{q},z_{H \backslash J})\biggr\},
\label{eqn:2.8}
\end{eqnarray}
where $q_i$ are the branch points on ${\cal C}_{\mbox{\scriptsize SW\normalsize}}$, 
$H=\{1,\ldots,h\},~J=\{i_1,\ldots,i_j\}\subset H$ and $z_J=\{z_{i_1},\ldots,z_{i_j}\}$. 
$q$ and ${\bar q}$ denote the positions on the upper and the lower sheet, respectively. 
The Bergman kernel $B(z_1,z_2)$ is given by the Akemann's formula \cite{Akemann:1996,Bouchard:2008gu},
\begin{eqnarray}
\label{eqn:2.9}
&&B(z_1,z_2)=
\frac{dz_1dz_2}{2(z_1-z_2)^2}\bigg(\frac{2f(z_1,z_2)+G(k)(z_1-z_2)^2}{2\sqrt{\sigma_n(z_1)\sigma_n(z_2)}}+1\bigg),\hspace{12.5em}\\
\label{eqn:2.10}
&&f(z_1,z_2):=z_1^2z_2^2+\frac12 z_1z_2(z_1+z_2)S_1+\frac16(z_1^2+4z_1z_2+z_2^2)S_2+\frac12(z_1+z_2)S_3+S_4,\\
\label{eqn:2.11}
&&G(k):=-\frac{1}{3}S_2+(q_1q_2+q_3q_4)-\frac{E(k)}{K(k)}(q_1-q_3)(q_2-q_4),\\
&&K(k)=\int_0^1\frac{dt}{\sqrt{(1-t^2)(1-k^2t^2)}},~ E(k)=\int_0^1dt \sqrt{\frac{1-k^2t^2}{1-t^2}},~ k^2=\frac{(q_1-q_2)(q_3-q_4)}{(q_1-q_3)(q_2-q_4)},\nonumber\\
\label{eqn:2.12}
&&
\end{eqnarray}
where $S_k$ is the coefficient of $z^{4-k}$ in $\sigma_n(z)$. 
$K(k)$ (resp. $E(k)$) is the complete elliptic integral of the first (resp. the second) kind with the modulus $k$.

For ${\cal N}=2$ superconformal quiver gauge theories with the self-dual constraint $\epsilon_1=-\epsilon_2=i \hbar$, 
the claim of \cite{Kozcaz:2010af} may be summarized as
\begin{equation}
F_{\mbox{\scriptsize CFT\normalsize}}^{(k)}(z_1,\ldots,z_{\widetilde h})=\sum_{2-2g-h=-k} 
\frac{1}{h!}\sum_{i_1,\ldots,i_h=1}^{\widetilde h} {\cal F}_{\mbox{\scriptsize SW\normalsize}}^{(g,h)}(z_{i_1},\ldots,z_{i_h}),
\label{eqn:2.12.5}
\end{equation}
where the left hand side is the ${\widetilde h}$-points free energy defined by (\ref{eqn:1.43}) on the CFT side
with internal channels chosen so that the result is symmetric in variables $z_1, \ldots, z_{\widetilde h}$.
Under this constraint  on the internal channel, the left hand side is essentially fixed.
On the other hand, there exist ambiguities of the constants of integration in \eqref{eqn:2.7}. 
Thus we will make a more modest proposal by keeping only universal terms on the right hand side of (\ref{eqn:2.12.5}),
which are independent of these ambiguities. 
For both the superconformal and the asymptotically free theories
we expect that at least a part of the relation (\ref{eqn:2.12.5}) is valid,
\begin{equation}
{F}_{\mbox{\scriptsize CFT\normalsize}}^{(h-2)}(z_1,\ldots,z_h)={\cal F}_{\mbox{\scriptsize SW\normalsize}}^{(0,h)}(z_1,\ldots,z_h),
\label{eqn:2.13}
\end{equation}
where ${\cal F}_{\mbox{\scriptsize SW\normalsize}}^{(0,h)}(z_1,\ldots,z_h)$ is the summation of all the universal terms which are
of the form $a_{n_1 n_2 \ldots n_h} z_1^{n_1} z_2^{n_2} \cdots z_h^{n_h}~(n_i \in {\mathbb Z})$ 
with the condition $\prod_{i=1}^h n_i \neq 0$. 
In the rest of this section we will explicitly check the relation (\ref{eqn:2.13}) for $N_f=0$ and $N_f=1$.

\subsection{Pure $SU(2)$}

Here we compute the free energies on the Seiberg-Witten curve (\ref{eqn:2.1}) corresponding 
to pure $SU(2)$ supersymmetric gauge theory. The period (\ref{eqn:2.5}) is obtained from the complete elliptic integral as follows:
\begin{equation}
\frac{da(u)}{du}=\oint_A \frac{\partial \lambda_{\mbox{\scriptsize SW\normalsize}}(z)}{\partial u}
=\frac{1}{2 \Lambda}\oint_A \frac{dz}{\sqrt{\sigma_0(z)}}=\frac{1}{\pi \Lambda \sqrt{q_3-q_1}}K(k),\quad k^2=\frac{q_1-q_2}{q_1-q_3},
\label{eqn:2.15}
\end{equation}
 where $q_1=0,~q_2=(u-\sqrt{u^2-4\Lambda^4})/2\Lambda^2$
and $q_3=(u+\sqrt{u^2-4\Lambda^4})/2\Lambda^2$ are the branch points of the curve (\ref{eqn:2.1}). 
Thus one obtains
\begin{equation}
u(a)=a^2+\frac{\Lambda^4}{2a^2}+\frac{5\Lambda^8}{32a^6}+\frac{9\Lambda^{12}}{64a^{10}}
+\frac{1469\Lambda^{16}}{8192a^{14}}+\frac{4471\Lambda^{20}}{16384a^{18}}+{\cal O}(\Lambda^{24}).
\label{eqn:2.16}
\end{equation}
To compute the annulus amplitude ${\cal F}_{\mbox{\scriptsize SW\normalsize}}^{(0,2)}(z_1,z_2)$ 
on the curve (\ref{eqn:2.1}), taking the limit $q_4\to \infty$ in (\ref{eqn:2.9}), 
one obtains the Bergman kernel on the curve by replacing $f(z_1,z_2)$ and $G(k)$ with
\begin{eqnarray}
&&
{\widetilde f}(z_1,z_2)=-\frac12 z_1z_2(z_1+z_2)-\frac16(z_1^2+4z_1z_2+z_2^2){\widetilde S}_1
-\frac12(z_1+z_2){\widetilde S}_2-{\widetilde S}_3,\nonumber\\
&&
{\widetilde G}(k)=\frac{1}{3}{\widetilde S}_1+q_3+\frac{E(k)}{K(k)}(q_1-q_3),
\label{eqn:2.17}
\end{eqnarray}
where ${\widetilde S}_k$ is the coefficient of $z^{3-k}$ in $\sigma_0(z)$. 
Thus we obtain the annulus amplitude
\begin{eqnarray}
&&
{\cal F}_{\mbox{\scriptsize SW\normalsize}}^{(0,2)}(z_1,z_2)=\frac{z_1^2z_2^2+1}{16a^4z_1z_2}\Lambda^4
+\frac{(z_1+z_2)(z_1^3z_2^3+1)}{32a^6z_1^2z_2^2}\Lambda^6 \nonumber\\
&&\hspace{3em}
+\frac{10(z_1^2+z_2^2)(z_1^4z_2^4+1)+9z_1z_2(z_1^4z_2^4+1)
+32z_1^2z_2^2(z_1^2z_2^2+1)-4z_1^2z_2^2(z_1^2+z_2^2)}{512a^8z_1^3z_2^3}\Lambda^8\nonumber\\
&&\hspace{3em}
+{\cal O}(\Lambda^{10}).
\label{eqn:2.18}
\end{eqnarray}
We can see that the amplitude agrees with (\ref{N02}) up to $\Lambda^8$. 
Hence the relation (\ref{eqn:2.13}) is correct as was expected.

Higher topology amplitudes are iteratively computed by the recursion (\ref{eqn:2.8}) 
and the multilinear meromorphic differentials $W^{(g,h)}(z_1,\ldots,z_h)$ can be expanded by the kernel differentials \cite{Bouchard:2007ys},
\begin{equation}
\chi_i^{(n)}(z):=\mathop{\mbox{Res}}_{q=q_i}\left(-\frac{2dE_{q,\bar{q}}(z)}{\lambda_{\mbox{\scriptsize SW\normalsize}}(q)
-\lambda_{\mbox{\scriptsize SW\normalsize}}({\bar q})}\frac{dq}{(q-q_i)^n}\right).
\label{eqn:2.19}
\end{equation}
For example, $W^{(0,3)}(z_1,z_2,z_3)$ is written as
\begin{eqnarray}
W^{(0,3)}(z_1,z_2,z_3)&=&\sum_{q_i}\mathop{\mbox{Res}}_{q=q_i}\frac{2dE_{q,\bar{q}}(z)}{\lambda_{\mbox{\scriptsize SW\normalsize}}(q)
-\lambda_{\mbox{\scriptsize SW\normalsize}}({\bar q})}B(z_2,q)B(z_3,{\bar{q}})\hspace{2em}\nonumber\\
\label{eqn:2.20}
&=&\frac{1}{2}\sum_{q_i}M_n(q_i)^2\sigma_n'(q_i)\chi_i^{(1)}(z_1)\chi_i^{(1)}(z_2)\chi_i^{(1)}(z_3),\\
\label{eqn:2.21}
&&\hspace{-8em}
\chi_i^{(1)}(z)=\frac{dz}{2M_n(q_i)\sigma_n'(q_i)\sqrt{\sigma_n(z)}}\bigg(G(k)+\frac{2f(z,q_i)}{(z-q_i)^2}\bigg).
\end{eqnarray}
Thus we obtain the three-holed amplitude ${\cal F}_{\mbox{\scriptsize SW\normalsize}}^{(0,3)}(z_1,z_2,z_3)$ 
on the Seiberg-Witten curve (\ref{eqn:2.1}),
\begin{eqnarray}
&&
{\cal F}_{\mbox{\scriptsize SW\normalsize}}^{(0,3)}(z_1,z_2,z_3)=
\frac{z_1^2z_2^2z_3^2-1}{16a^7z_1z_2z_3}\Lambda^6+\frac{3\big(z_1^3z_2^3z_3^3(z_1+z_2+z_3)-(z_1z_2+z_2z_3+z_3z_1)\big)}
{64a^9z_1^2z_2^2z_3^2}\Lambda^8\hspace{2em} \nonumber\\
&&
+\bigg\{\frac{z_1^2+z_2^2+z_3^2-(z_1^2z_2^2+z_2^2z_3^2+z_3^2z_1^2)}{128a^{11}z_1z_2z_3}
+\frac{5\big(z_1^4z_2^4z_3^4(z_1^2+z_2^2+z_3^2)-(z_1^2z_2^2+z_2^2z_3^2+z_3^2z_1^2)\big)}{128a^{11}z_1^3z_2^3z_3^3}\nonumber\\
&&
+\frac{9\big(z_1^3z_2^3z_3^3(z_1z_2+z_2z_3+z_3z_1)-(z_1+z_2+z_3)\big)}{256a^{11}z_1^2z_2^2z_3^2}
+\frac{9(z_1^2z_2^2z_3^2-1)}{64a^{11}z_1z_2z_3}\bigg\}\Lambda^{10}+{\cal O}(\Lambda^{12}),
\label{eqn:2.22}
\end{eqnarray}
and in (\ref{N03}) we checked the relation (\ref{eqn:2.13}) up to $\Lambda^6$.

\subsection{$SU(2)$ with one fundamental matter}

We compute the annulus amplitude ${\cal F}_{\mbox{\scriptsize SW\normalsize}}^{(0,2)}(z_1,z_2)$ 
on the Seiberg-Witten curve (\ref{eqn:2.2}) corresponding 
to $SU(2)$ supersymmetric gauge theory with one fundamental matter. 
The period (\ref{eqn:2.5}) is computed from
\begin{equation}
\frac{da(u)}{du}=\frac{1}{\pi \Lambda \sqrt{(q_1-q_3)(q_2-q_4)}}K(k),\quad k^2
=\frac{(q_1-q_2)(q_3-q_4)}{(q_1-q_3)(q_2-q_4)},
\label{eqn:2.23}
\end{equation}
where $q_{1,2}=-(m \pm \sqrt{m^2-u})/\Lambda+\Lambda^2/(2(m^2\pm m\sqrt{m^2-u}-u))+{\cal O}(\Lambda^5), q_3=0$ and
$q_4=\Lambda^2/u+{\cal O}(\Lambda^5)$. One finds 
\begin{equation}
u(a)=a^2-\frac{m\Lambda^3}{a^2}-\frac{(3a^2-5m^2)\Lambda^6}{8a^6}+{\cal O}(\Lambda^9).
\label{eqn:2.24}
\end{equation}
Then, from (\ref{eqn:2.9}) we obtain the annulus amplitude
\begin{eqnarray}
&&
{\cal F}_{\mbox{\scriptsize SW\normalsize}}^{(0,2)}(z_1,z_2)=-\frac{(a^2-m^2)z_1z_2}{4a^4}\Lambda^2
+\frac{m(a^2-m^2)z_1z_2(z_1+z_2)}{4a^6z_1z_2}\Lambda^3 \hspace{10em}\nonumber\\
&&\hspace{2em}
+\frac{2(a^2-m^2)(a^2-5m^2)(z_1^2+z_2^2)z_1^2z_2^2+(a^2-m^2)(a^2-9m^2)z_1^3z_2^3+2a^4}{32a^8z_1z_2}\Lambda^4
+{\cal O}(\Lambda^5).\nonumber\\
&&
\label{eqn:2.25}
\end{eqnarray}
As before this agrees with ${F}_{\mbox{\scriptsize CFT\normalsize}}^{(0)}(z_1,z_2)$ 
up to $\Lambda^4$ (see (\ref{N12})). Hence the relation (\ref{eqn:2.13}) also holds in this case.



\section{Geometric engineering and open topological B model}

Hereafter we consider the open topological string on toric Calabi-Yau threefolds (local A model) 
which is expected to realize a surface operator in $\mathcal{N}=2$ $SU(2)$ gauge theories in four dimensions. 
In this section we compute the topological open string amplitude by combining 
the local mirror symmetry with the  conjecture of remodeling the B model \cite{Marino:2006hs, Bouchard:2007ys}, by which 
we have the equality between the local A model amplitudes 
and the free energies ${\cal F}^{(g,h)}(x_1,\ldots,x_h)$, $g,h \in {\IZ}_{\ge 0},~h \ge 1$ 
on the mirror curve
\begin{equation}
{\cal C}=\left\{x,y \in {\IC}^*~|~H(x,y)=0\right\}~\subset {\IC}^* \times {\IC}^*,
\label{eqn:3.2}
\end{equation}
computed by the topological recursion relation of Eynard and Orantin we employed in section 5. 
The free energies are defined by (\ref{eqn:2.7}) and (\ref{eqn:2.8}) under the replacement
\begin{equation}
\lambda_{\mbox{\scriptsize SW\normalsize}}(z),\quad \longrightarrow \quad \omega(x):=\log y(x)\frac{dx}{x}.
\label{eqn:3.3}
\end{equation}

\subsection{Toric brane on local Hirzebruch surface $K_{{\bf F}_0}$: pure $SU(2)$}

\begin{figure}[tbp]
 \begin{center}
  \includegraphics[width=50mm]{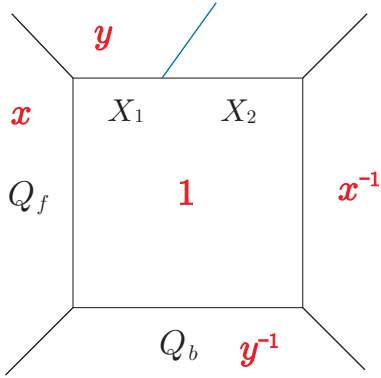}
 \end{center}
 \caption{Local Hirzebruch surface $K_{{\bf F}_0}$}
 \label{fig:3.1}
\end{figure}

The pure $SU(2)$ gauge theory is realized by the Hirzebruch surface 
${\bf F}_0={\bf P}^1_{f} \times {\bf P}^1_{b}$, and 
we insert a toric brane on the base ${\bf P}^1_b$ as the blue line in Fig.\ref{fig:3.1}. 
The K\"ahler parameters $Q_f$, $Q_b$ of ${\bf P}^1_{f}$, ${\bf P}^1_{b}$ 
and the parameters on the gauge theory side are related by \cite{Katz:1996fh},
\begin{equation}
Q_f=e^{-2\beta a},\quad Q_b=X_1X_2=\beta^4\Lambda^4,\quad X_1 : = \beta^2\Lambda^2w^{-1},\quad X_2 := \beta^2\Lambda^2w,
\label{eqn:3.1}
\end{equation}
where $\beta$ is a scale parameter which corresponds to the radius of the fifth dimension in the gauge theory
and $X_1, (X_2)$ represents the distance between the toric brane and the trivalent vertex in the web diagram
as indicated in Fig.\ref{fig:3.1}. 
The charge vectors of $K_{{\bf F}_0}$ are given by
\begin{equation}
\ell_b=(-2,1,0,1,0),\quad \ell_f=(-2,0,1,0,1).
\label{eqn:3.3.1}
\end{equation}
By taking the local coordinate patch as Fig.\ref{fig:3.1}, the mirror curve which describes the moduli of the toric brane is obtained as
\begin{equation}
xy^2+(x^2+x+z_b)y+z_fx=0,\quad \sigma(x):=(x^2+x+z_b)^2-4z_fx^2,
\label{eqn:3.4}
\end{equation}
where $z_{f}, z_{b}$ are the moduli parameters of complex structure of the mirror Calabi-Yau threefold. 
The closed and open mirror maps are given by \cite{Aganagic:2000gs,Aganagic:2001nx,Lerche:2001cw},
\begin{eqnarray}
&&
\log Q_b=\log z_b+2\sum_{m,n\ge 0, (m,n)\neq (0,0)}^{\infty}\frac{(2m+2n-1)!}{m!^2n!^2}z_b^mz_f^n,\nonumber\\
&&
\log \frac{Q_f}{Q_b}=\log \frac{z_f}{z_b},\quad X=\left(\frac{Q_b}{z_b}\right)^{\frac12}x,
\label{eqn:3.5}
\end{eqnarray}
where $X=X(x)$ is the open string moduli on the A model side. 
The disk amplitude is computed in a similar manner to \cite{Aganagic:2001nx},
\begin{eqnarray}
{\cal F}^{(0,1)}(Q_f,\Lambda,w)&=&\int^{x}\omega(x')=\int^{x} \log \bigg\{\frac{(x'^2+x'+z_b)
+\sqrt{\sigma(x')}}{2x'}\bigg\}\frac{dx'}{x'}\hspace{7.5em} \nonumber\\
&&\hspace{-5em}\simeq
-\frac{w^2-1}{(1-Q_f)w}(\beta\Lambda)^2+\frac{(1+Q_f)(w^4-1)}{4(1-Q_f)^3w^2}(\beta\Lambda)^4
-\frac{(1+4Q_f+Q_f^2)(w^6-1)}{9(1-Q_f)^5w^3}(\beta\Lambda)^6\nonumber\\
&&\hspace{-5em}
-\frac{2(w^2-1)}{(1-Q_f)^5w}(\beta\Lambda)^6
+\frac{(1+9Q_f+9Q_f^2+Q_f^3)(w^8-1)}{16(1-Q_f)^7w^4}(\beta\Lambda)^8\nonumber\\
&&\hspace{-5em}
+\frac{2(1+Q_f)(w^4-1)}{(1-Q_f)^7w^2}(\beta\Lambda)^8+{\cal O}(\Lambda^{10}),
\label{eqn:3.6}
\end{eqnarray}
where in the second equality, since the toric brane is inserted on the base ${\bf P}^1_{b}$, 
we expanded the integrand around the midpoint $x'=z_b^{1/2}$ and took away a logarithmic term from the final result. 
When we compute the annulus and the three-holed amplitudes ${\cal F}^{(0,2)},  {\cal F}^{(0,3)}$ in the following,
we will use a similar prescription as above. Using the relation (\ref{eqn:3.1}) of geometric engineering
and taking the limit $\beta \to 0$, we find a matching of (\ref{eqn:3.6}) up to $\Lambda^8$ with the leading term (\ref{eqn:1.14}) of the
free energy obtained from the CFT one point function of $\Phi_{1,2}$.

We can compute the annulus amplitude ${\cal F}^{(0,2)}(x,y)$ using (\ref{eqn:2.9}), 
where $G(k)$ can be rewritten in terms of the period $T_b=-\log Q_b$ as was shown in \cite{Manabe:2009sf},
\begin{equation}
G(k)=-\frac{1}{12}\Delta_0(z_b,z_f){\widetilde z}_b \frac{\partial}{\partial {\widetilde z}_b} 
\Big\{12\log {\widetilde z}_b\frac{\partial}{\partial {\widetilde z}_b} T_b
+4\log {\widetilde z}_b +\log \Delta_0({\widetilde z}_b,{\widetilde z}_b{\widetilde z}_f)\Big\},
\label{eqn:3.7}
\end{equation}
where ${\widetilde z}_b=z_b, {\widetilde z}_f=z_f/z_b$, and $\Delta_0(z_b,z_f):=1-8(z_b+z_f)+16(z_b-z_f)^2$ 
is a component of the discriminant of the mirror curve (\ref{eqn:3.4}). Thus we obtain the annulus amplitude
\begin{eqnarray}
{\cal F}^{(0,2)}(Q_f,\Lambda,w_i)&=&\int^{x}\int^{y}B(x',y')-\frac{dx'dy'}{(x'-y')^2} \nonumber\\
&&\hspace{-8em}\simeq
\frac{Q_f(w_1^2w_2^2+1)}{(1-Q_f)^4w_1w_2}(\beta\Lambda)^4-\frac{Q_f(1+Q_f)(w_1+w_2)(w_1^3w_2^3+1)}
{(1-Q_f)^6w_1^2w_2^2}(\beta\Lambda)^6 \nonumber\\
&&\hspace{-8em}
+\frac{Q_f(1+3Q_f+Q_f^2)(w_1^2+w_2^2)(w_1^4w_2^4+1)}{(1-Q_f)^8w_1^3w_2^3}(\beta\Lambda)^8
+\frac{Q_f(2+5Q_f+2Q_f^2)(w_1^4w_2^4+1)}{2(1-Q_f)^8w_1^2w_2^2}(\beta\Lambda)^8\nonumber\\
&&\hspace{-8em}
+\frac{2Q_f(1+6Q_f+Q_f^2)(w_1^2w_2^2+1)}{(1-Q_f)^8w_1w_2}(\beta\Lambda)^8
-\frac{2Q_f^2(w_1^2+w_2^2)}{(1-Q_f)^8w_1w_2}(\beta\Lambda)^8+{\cal O}(\Lambda^{10}),
\label{eqn:3.8}
\end{eqnarray}
where in the second equality, we used a similar prescription to the case of the disk amplitude.
The annulus amplitude with two arguments gives the contribution from a geometry 
where two toric branes are inserted on the base ${\bf P}^1_{b}$. 
$X:=\beta^2\Lambda^2w_1^{-1}$ and $Y:=\beta^2\Lambda^2w_2^{-1}$ represent 
the positions of the first and the second toric brane, respectively. 
Using the relation  (\ref{eqn:3.1}) and taking the limit $\beta \to 0$, we find that (\ref{eqn:3.8}) coincides with (\ref{eqn:2.18}) and (\ref{N02}).

Higher topology amplitudes can be also computed by the topological recursion (\ref{eqn:2.8}). 
As an example let us compute the three-holed amplitude 
${\cal F}^{(0,3)}(x,y,z)$ using (\ref{eqn:2.20}), where the moment function $M(x)$ is defined by
\begin{eqnarray}
&&\frac12\big(\omega(x)-\omega({\bar x})\big)=\frac{dx}{x}\tanh^{-1}\bigg[\frac{\sqrt{\sigma(x)}}{x^2+x+z_b}\bigg]
=:M(x)\sqrt{\sigma(x)}dx,\hspace{6em}\nonumber\\
\longrightarrow\quad&&
M(x)=\frac{1}{x\sqrt{\sigma(x)}}\tanh^{-1}\bigg[\frac{\sqrt{\sigma(x)}}{x^2+x+z_b}\bigg]
=\frac{1}{x}\sum_{n=0}^{\infty}\frac{1}{2n+1}\frac{\sigma(x)^n}{(x^2+x+z_b)^{2n+1}}.
\label{eqn:3.9}
\end{eqnarray}
Thus we obtain the three-holed amplitude
\begin{eqnarray}
&&{\cal F}^{(0,3)}(Q_f,\Lambda,w_i)=\int^{x}\int^{y}\int^{z}W^{(0,3)}(x',y',z') \nonumber\\
&&\simeq
\frac{(Q_f+6Q_f^2+Q_f^3)(w_1^2w_2^2w_3^2-1)}{(1-Q_f)^7w_1w_2w_3}(\beta\Lambda)^6\nonumber\\
&&
+\frac{(Q_f+11Q_f^2+11Q_f^3+Q_f^4)\big(w_1^3w_2^3w_3^3(w_1+w_2+w_3)-(w_1w_2+w_2w_3+w_3w_1)\big)}{(1-Q_f)^9w_1^2w_2^2w_3^2}(\beta\Lambda)^8 \nonumber\\
&&
+\bigg\{\frac{2(Q_f^2+6Q_f^3+Q_f^4)\big(w_1^2+w_2^2+w_3^2-(w_1^2w_2^2+w_2^2w_3^2+w_3^2w_1^2)\big)}{(1-Q_f)^{11}w_1w_2w_3}\nonumber\\
&&
+\frac{(Q_f+17Q_f^2+36Q_f^3+17Q_f^4+Q_f^5)\big(w_1^3w_2^3w_3^3(w_1w_2+w_2w_3+w_3w_1)-(w_1+w_2+w_3)\big)}{(1-Q_f)^{11}w_1^2w_2^2w_3^2} \nonumber\\
&&
+\frac{(Q_f+18Q_f^2+42Q_f^3+18Q_f^4+Q_f^5)\big(w_1^4w_2^4w_3^4(w_1^2+w_2^2+w_3^2)-(w_1^2w_2^2+w_2^2w_3^2+w_3^2w_1^2)\big)}{(1-Q_f)^{11}w_1^3w_2^3w_3^3}\nonumber\\
&&
+\frac{2(Q_f+28Q_f^2+86Q_f^3+28Q_f^4+Q_f^5)(w_1^2w_2^2w_3^2-1)}{(1-Q_f)^{11}w_1w_2w_3}\bigg\}(\beta\Lambda)^{10}+{\cal O}(\Lambda^{12}).
\label{eqn:3.10}
\end{eqnarray}
The three-holed amplitude with three arguments gives the leading contribution 
when three toric branes are inserted on the base ${\bf P}^1_{b}$.
$X:=\beta^2\Lambda^2w_1^{-1}$, $Y:=\beta^2\Lambda^2w_2^{-1}$ and $Z:=\beta^2\Lambda^2w_3^{-1}$ represent the position of each toric brane. 
Using (\ref{eqn:3.1}) and taking the limit $\beta \to 0$, we find that (\ref{eqn:3.10}) agrees with (\ref{eqn:2.22}) and (\ref{N03}).

\subsection{Toric brane on local del Pezzo surface $K_{{dP}_2}$: $SU(2)$ with one fundamental matter}

\begin{figure}[tbp]
 \begin{center}
  \includegraphics[width=50mm]{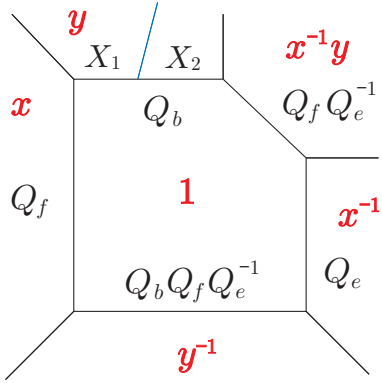}
 \end{center}
 \caption{Local del Pezzo surface $K_{{dP}_2}$}
 \label{fig:3.2}
\end{figure}

By the geometric engineering, one can also introduce fundamental matters. The $SU(2)$ theory 
with one fundamental matter is realized by the del Pezzo surface ${dP}_2$, 
which is obtained by a blow up at a torus fixed point on the Hirzebruch surface ${\bf F}_0$, . 
Let us insert a toric brane on the base ${\bf P}^1_{b}$ as the blue line in Fig.\ref{fig:3.2}. 
As in (\ref{eqn:3.1}), the K\"ahler parameters $Q_f$, (resp. $Q_b$, $Q_e$) 
of the fiber ${\bf P}^1_{f}$, (resp. base ${\bf P}^1_{b}$, the exceptional curve ${\bf P}^1_{e}$), 
and the distance between the toric brane and the vertices in the web diagram $X_1, (X_2)$ 
can be related to the parameters on the gauge theory side as \cite{Katz:1996fh,Eguchi:2003sj},
\begin{equation}
Q_f=e^{-2\beta a},\quad Q_e=e^{-\beta (a-m)}, \quad Q_b=X_1X_2=2\beta^3\Lambda^3,
\quad X_1 : = \beta^2\Lambda^2 w^{-1},\quad X_2 : = 2\beta \Lambda w.
\label{eqn:3.11}
\end{equation}
The charge vectors of $K_{{dP}_2}$ are given by
\begin{equation}
\ell_b=(-1,1,0,0,1,-1),\quad \ell_f=(-2,0,1,0,0,1),\quad \ell_e=(-1,0,1,-1,1,0).
\label{eqn:3.11.1}
\end{equation}
By taking the local coordinate patch as Fig.\ref{fig:3.2}, we obtain the mirror curve which describes the moduli of the toric brane,
\begin{eqnarray}
&&
(x+z_b)y^2+(x^2+x+z_bz_fz_e^{-1})y+z_fx=0,\nonumber\\
&&
\sigma(x):=(x^2+x+z_bz_fz_e^{-1})^2-4z_fx(x+z_b),
\label{eqn:3.12}
\end{eqnarray}
where $z_f, z_b$, and $z_e$ are the moduli parameters of complex structure the mirror Calabi-Yau threefold. 
The closed and open mirror maps are given by \cite{Lerche:2001cw},
\begin{eqnarray}
&&
\log Q_b=\log z_b+\sum_{m,n,k \ge 0, (m,n,k)\neq (0,0,0)}^{\infty}(-1)^m\frac{(3m+2n+2k-1)!}{m!n!k!(m+k)!(m+n)!}z_b^{m+k}z_f^{m+n+k}z_e^{-k},\nonumber\\
&&
\log \frac{Q_f}{Q_b^2}=\log \frac{z_f}{z_b^2},\quad \log \frac{Q_e}{Q_b}=\log \frac{z_e}{z_b},\quad X=\frac{Q_b}{z_b}x,
\label{eqn:3.13}
\end{eqnarray}
where $X=X(x)$ is the open string moduli on the A model side. 
The disk amplitude is computed as
\begin{eqnarray}
&&
{\cal F}^{(0,1)}(Q_f,Q_e,\Lambda,w)=\int^{x}\omega(x')
=\int^{x} \log \bigg\{\frac{(x'^2+x'+z_bz_fz_e^{-1})+\sqrt{\sigma(x')}}{2(x'+z_b)}\bigg\}\frac{dx'}{x'} \nonumber\\
&&
\simeq
-\frac{2(Q_f-Q_e)w}{(1-Q_f)Q_e}\beta\Lambda+\frac{1}{(1-Q_f)w}(\beta\Lambda)^2
+\frac{(Q_f-Q_e)(Q_f^2+Q_f+Q_e-3Q_fQ_e)w^2}{(1-Q_f)^3Q_e^2}(\beta\Lambda)^2 \nonumber\\
&&
-\frac{8(Q_f-Q_e)\big(Q_f^4+4Q_f^3+Q_f^2-(8Q_f^3+5Q_f^2-Q_f)Q_e+(10Q_f^2-5Q_f+1)Q_e^2\big)w^3}
{9(1-Q_f)^5Q_e^3}(\beta\Lambda)^3 \nonumber\\
&&
+\bigg\{\frac{(Q_f-Q_e)}{(1-Q_f)^7Q_e^4}\Big((Q_f+1)(Q_f^2+8Q_f+1)Q_f^3-(15Q_f^3+39Q_f^2+7Q_f-1)Q_f^2Q_e\nonumber\\
&&\hspace{7em}
+(45Q_f^3+21Q_f^2-7Q_f+1)Q_fQ_e^2-(35Q_f^3-21Q_f^2+7Q_f-1)Q_e^3\Big)w^4 \nonumber\\
&&\hspace{1em}
-\frac{8Q_f^2(Q_f-Q_e)(1-Q_e)w}{(1-Q_f)^5Q_e^2}-\frac{1+Q_f}{4(1-Q_f)^3w^2}\bigg\}(\beta\Lambda)^4+{\cal O}(\Lambda^{5}),
\label{eqn:3.14}
\end{eqnarray}
where in the second equality we expanded the integrand around the midpoint $x'=z_b^{1/2}$ and removed a logarithmic term. 
We see that this result is consistent with (\ref{eqn:3.6}) in the limit $Q_e \to 0, Q_f/Q_e \to 1$.
Using the relation (\ref{eqn:3.11}) and taking the limit $\beta \to 0$, 
we find that (\ref{eqn:3.14}) agrees with (\ref{eqn:1.27}) except the coefficient of $w\Lambda$. 
The difference of the coefficient of $w\Lambda$ is nothing but the overall factor $\exp (-\Lambda w/\hbar)$ 
which compensates the difference between the instanton partition function with surface operator and 
the correlation function with the degenerate primary field insertion. Therefore (\ref{eqn:3.14}) agrees with the computation on the gauge theory side.

From (\ref{eqn:2.9}) we can also compute the annulus amplitude ${\cal F}^{(0,2)}(x,y)$, 
where $G(k)$ can be rewritten in terms of the period $T_b=-\log Q_b$ as \cite{Manabe:2009sf},
\begin{eqnarray}
&&
G(k)=-\frac{\Delta_0(z_b,z_f,z_e)}{24z_e^2C(z_b,z_f,z_e)}
{\widetilde z}_b \frac{\partial}{\partial {\widetilde z}_b} \Big\{12\log {\widetilde z}_b
\frac{\partial}{\partial {\widetilde z}_b} T_b
+7\log {\widetilde z}_b +\log \Delta_0({\widetilde z}_b,{\widetilde z}_b^2{\widetilde z}_f,{\widetilde z}_b{\widetilde z}_e)\Big\}, \nonumber\\
&&
\Delta_0(z_b,z_f,z_e):=16z_b^3z_f^2z_e-\big(27z_e^4-12(3z_e-2z_f)z_e^2+8(z_e^2+2z_fz_e-2z_f^2)\big)z_b^2z_f \nonumber\\
&&\hspace{4em}
+\big(12(3z_e-2z_f)z_fz_e^2-(z_e^3+46z_fz_e^2-64z_f^2z_e+32z_f^3)+(z_e^2+8z_fz_e-8z_f^2)\big)z_bz_e \nonumber\\
&&\hspace{4em}
+(4z_f-1)^2(z_e^2-z_e+z_f)z_e^2, \nonumber\\
&&
C(z_b,z_f,z_e):=-\frac{9}{8}z_bz_e^2+(z_b+z_e)z_e-\frac{1}{8}\big(7z_e+4z_f(z_b+z_e)\big)+z_f,
\label{eqn:3.15}
\end{eqnarray}
where ${\widetilde z}_b=z_b, {\widetilde z}_f=z_f/z_b^2, {\widetilde z}_e=z_e/z_b$, and $\Delta_0(z_b,z_f,z_e)$ 
is a component of the discriminant of the mirror curve (\ref{eqn:3.12}). Thus we obtain the annulus amplitude
\begin{eqnarray}
&&
{\cal F}^{(0,2)}(Q_f,Q_e,\Lambda,w_i)=\int^{x}\int^{y}B(x',y')-\frac{dx'dy'}{(x'-y')^2} \nonumber\\
&&\simeq
\frac{4Q_f^2(Q_f-Q_e)(1-Q_e)w_1w_2}{(1-Q_f)^4Q_e^2}(\beta\Lambda)^2\nonumber\\
&&
-\frac{8Q_f^3(Q_f-Q_e)(1-Q_e)(Q_f+1-2Q_e)w_1w_2(w_1+w_2)}{(1-Q_f)^6Q_e^3}(\beta\Lambda)^3 \nonumber\\
&&
+\bigg\{\frac{16Q_f^4(Q_f-Q_e)(1-Q_e)\big(Q_f^2+3Q_f+1+5(Q_e-Q_f-1)Q_e\big)(w_1^2+w_2^2)w_1w_2}{(1-Q_f)^8Q_e^4}
\nonumber\\
&&\hspace{1em}
+\frac{8Q_f^4(Q_f-Q_e)(1-Q_e)(2Q_f+1-3Q_e)(Q_f+2-3Q_e)w_1^2w_2^2}{(1-Q_f)^8Q_e^4}
\nonumber\\
&&\hspace{1em}
+\frac{Q_f}{(1-Q_f)^4w_1w_2}\bigg\}(\beta\Lambda)^4+{\cal O}(\Lambda^5),
\label{eqn:3.16}
\end{eqnarray}
where in the second equality we expanded the Bergman kernel around the point $x'=z_b^{1/2}, y'=z_b^{1/2}$
and removed a logarithmic term. $X:=\beta^2 \Lambda^2 w_1^{-1}$, and $Y:=\beta^2 \Lambda^2 w_2^{-1}$ 
represent the position of two toric branes. Using the geometric engineering (\ref{eqn:3.11}) 
and taking the limit $\beta \to 0$, we see that (\ref{eqn:3.16}) agrees with (\ref{eqn:2.25}), and (\ref{N12}).


\Section{Vortex counting and open topological A model}

In a recent paper \cite{Dimofte:2010tz} the instanton partition function with surface
operator has been worked out from the viewpoint of the coupling of four dimensional gauge theory 
with a two dimensional theory on the surface. It was argued that in the decoupling limit
$\Lambda_{\mathrm{inst}} \to 0$, where only zero instanton sector of the four dimensional theory
survives, the partition function reduces to the vortex counting in the two dimensional theory.
By the localization computation on the affine Laumon space which we used in section 3,
the vortex counting of \cite{Dimofte:2010tz} can be derived in the following way.
Recall our identification of the instanton number $k$ and the monopole number $\mathfrak{m}$:
\beq
k = k_1, \qquad \mathfrak{m} = k_2 - k_1,
\eeq
where $k_1$ and $k_2$ are given by \eqref{topnumber} in terms of a pair of Young diagrams $(\lam_1, \lam_2)$.
Thus when we look at the zero instanton sector we have to set $k_1=0$ and the monopole
number is restricted to take non-negative integer\footnote{As we will see the next subsection, 
for $k>0$ the negative monopole number is allowed.}. 
This means that $\lambda_1$ has to be trivial and $\lambda_2$ has
only a single row, whose length gives the monopole number. Let us first consider 
pure $SU(2)$ theory for simplicity. Then almost all terms in the diagonal part of the equivariant 
character vanish. The remaining terms are
\beqa
{\rm ch}_{\vec\lam,\vec\lam}(a,a)
&=& e^{a_1 - a_2 + \ep +\es} \frac{e^{\ep\lam_{2,1}} -1}{e^{\ep} -1} 
- e^{\ep}  \frac{(e^{\ep\lam_{2,1}} -1)(e^{-\ep\lam_{2,1}}-1)}{e^{\ep} -1} 
- e^{\ep}  \frac{(e^{-\ep\lam_{2,1}}-1)}{e^{\ep} -1}  \CR
&=& \left( e^{a_1 - a_2 + \ep +\es} + e^{\ep} \right)
 \frac{e^{\ep\mathfrak{m}} -1}{e^{\ep} -1} 
=  \left( e^{a_1 - a_2 +\es} + 1\right) \sum_{k=1}^{\mathfrak{m}} e^{k\ep}.
\eeqa
Hence the zero instanton part of the partition function \eqref{Zpure} is
\beq
Z_{\mathrm{monopole}} (a, \ep, \es ; z)=1 +  \sum_{\mathfrak{m}=1}^\infty
\frac{1}{\mathfrak{m}!} \left( \frac{z}{\ep} \right)^{\mathfrak{m}} \prod_{k=1}^{\mathfrak{m}}
\frac{1}{2a + \es + k \ep}, \label{vortex}
\eeq
where we have replaced the parameter $\Lambda_2$ in \eqref{Zpure} to $z$.
We see that with the choice of equivariant parameters  $\ep = \hbar, \es \to 0$ the partition function \eqref{vortex}
agrees with the generating function of vortex counting, eq. (3.24) in \cite{Dimofte:2010tz}, 
where it was argued that the $K$ theory version of (3.24) coincides
with a refined open topological string amplitude in the limit where the K\"ahler parameter
$Q_{b}$ of the base ${\bf P}^1_{b}$ vanishes.

Let us look at a similar vortex counting in zero instanton background in $N_f =4$ theory.
The results for $1 \leq {N}_f  \leq 3$ theories can be obtained by the decoupling limit. 
The contributions from $(\emptyset, (\mathfrak{m}) )$
gives the partition function
\beq
Z_{\mathrm{monopole}}^{N_f =4}(a, M_i, \ep, \es ; y) = 1 +
\sum_{\mathfrak{m}=1}^\infty y^{\mathfrak{m}}
\frac{ \nbf{\vec\emptyset}{(\emptyset, (\mathfrak{m}) )}{a_1}{a}{m_1} 
\cdot \nbf{(\emptyset, (\mathfrak{m}) )}{\vec\emptyset}{a}{a_2}{m_2}}
{n_{\mathrm{vec}}[(\emptyset, (\mathfrak{m}) )](\vec{a})}.
\eeq
Since only the non-vanishing component of $\vec\lam$ is $\lam_{2,1} = \mathfrak{m}$, we have
\beqa
 \nbf{(\emptyset, (\mathfrak{m}) )}{\vec\emptyset}{a}{b}{m}
&=& \prod_{k=0}^{\mathfrak{m}-1} (- a +  b - k \ep -m), \\
 \nbf{\vec\emptyset}{(\emptyset, (\mathfrak{m}) )}{a}{b}{m} 
&=& \prod_{k=1}^\mathfrak{m} (a + b + k \ep + \es -m).
\eeqa
Combined with the previous computation of 
$n_{\mathrm{vec}}[(\emptyset, (\mathfrak{m}) )](\vec{a})$ in pure $SU(2)$ theory,
this gives
\beq
Z_{\mathrm{monopole}}^{N_f =4}(a, M_i, \ep, \es ; y) = 1 +
\sum_{\mathfrak{m}=1}^\infty (-y)^{\mathfrak{m}}
\frac{ \prod_{k=1}^{\mathfrak{m}}(a -2M_2 + k \ep + \es) (a + 2M_4 + (k-1) \ep)}
{\mathfrak{m}! \ep^{\mathfrak{m}}
 \prod_{k=1}^{\mathfrak{m}} (2a + k \ep +\es)} \label{matter-monopole}
\eeq
After the identification $z \equiv (-y), 2M_2 \equiv m_1 - \frac{1}{2} \ep + \frac{1}{2} \es,
2M_4 \equiv -m_2 + \frac{3}{2} \ep + \frac{1}{2} \es$, \eqref{matter-monopole} 
agrees to (4.26) in \cite{Dimofte:2010tz} up to $U(1)$ factor (an appropriate power of $(1-z)$),
which is also related to the hypergeometric series.


%
\subsection{Localization on one instanton sector}
We generalize the vortex counting to one instanton sector of the four
dimensional gauge theory. For one instanton sector, the pair of partitions
$\vec{\lambda}:=(\lambda_1,\lambda_2)$ satisfies
\begin{eqnarray}
k_1=1,\quad k_2=\mathfrak{m}+1.
\end{eqnarray}
There are four choices for $\vec{\lambda}$ as follows:
\begin{eqnarray}
&&(A)\quad \vec{\lambda}_{\mathfrak{m}A}=((1),(\mathfrak{m}+1)),\quad \mathfrak{m}\ge -1,  \qquad
(B)\quad \vec{\lambda}_{\mathfrak{m}B}=(\emptyset,(\mathfrak{m}+1,1)),\quad \mathfrak{m}\ge 0,
\nonumber \\
&&(C)\quad \vec{\lambda}_{\mathfrak{m}C}=(\emptyset,(\mathfrak{m},1,1)), \quad \mathfrak{m}\ge 1, \qquad
(D)\quad \vec{\lambda}_{\mathfrak{m}D}=((1,1),(\mathfrak{m})), \quad \mathfrak{m}\ge 0.
\label{fixed}
\end{eqnarray}
For $N_f=4$ theory, by evaluating the three characters ${\rm Tr}_{{\rm Ext}(\vec{\lambda},\vec{\lambda})}[g]$,
${\rm Tr}_{{\rm Ext}(\vec{\lambda},\vec{\emptyset})}[g]$ with $a_1=a$, $a_2=-a$,  $b_1=M_3-M_4$, $b_2=M_4-M_3$, $m_2=M_3+M_4$ and 
${\rm Tr}_{{\rm Ext}(\vec{\emptyset},\vec{\lambda})}[g]$  with $a_1=M_1-M_2$, $a_2=M_2-M_1$, $b_1=a$, $b_2=-a$, $m_1=M_1+M_2$
we can read off the partition function in one instanton sector for each
partition.
\begin{eqnarray}
&&\hspace*{-0.5cm}
Z_1^{(A)}(a,M_i,\epsilon_1,\epsilon_2;x,y)
=\sum_{\mathfrak{m}=-1}^{\infty}xy^{\mathfrak{m}+1}
\frac{n_f^S[\vec{\emptyset},\vec{\lambda}_{\mathfrak{m}A}](a_1,a;m_1)\cdot 
n_f^S[\vec{\lambda}_{\mathfrak{m}A},\vec{\emptyset}](a_1,a;m_2)}
{n_{\rm vec}[\vec{\lambda}_{\mathfrak{m}A}](\vec{a})},
 \\
&&\quad
n_f^S(\vec{\lambda}_{\mathfrak{m}A},\vec{\emptyset})
=(a-2M_3)\prod_{k=0}^{\mathfrak{m}}(-a-2M_4-k\epsilon_1),
\nonumber \\
&&\quad
n_f^S(\vec{\emptyset},\vec{\lambda}_{\mathfrak{m}A})
=(-a-2M_1+\epsilon_1)\prod_{k=1}^{\mathfrak{m}+1}(a-2M_2+k\epsilon_1+\epsilon_2),
\nonumber \\
&&\quad
n_{\rm vec}[\vec{\lambda}_{\mathfrak{m}A}](\vec{a})=
(-2a-\mathfrak{m}\epsilon_1)\epsilon_1^{\mathfrak{m}+2}(\mathfrak{m}+1)!\prod_{k=0}^{\mathfrak{m}}(2a+k\epsilon_1+\epsilon_2),
\nonumber
\end{eqnarray}
\begin{eqnarray}
&&\hspace*{-0.5cm}
Z_1^{(B)}(a,M_i,\epsilon_1,\epsilon_2;x,y)
=\sum_{\mathfrak{m}=0}^{\infty}xy^{\mathfrak{m}+1}
\frac{n_f^S[\vec{\emptyset},\vec{\lambda}_{\mathfrak{m}B}](a_1,a;m_1)\cdot 
n_f^S[\vec{\lambda}_{\mathfrak{m}B},\vec{\emptyset}](a_1,a;m_2)}
{n_{\rm vec}[\vec{\lambda}_{\mathfrak{m}B}](\vec{a})},
 \\
&&\quad
n_f^S(\vec{\lambda}_{\mathfrak{m}B},\vec{\emptyset})
=(-a-2M_3-\epsilon_2)\prod_{k=0}^{\mathfrak{m}}(-a-2M_4-k\epsilon_1),
\nonumber \\
&&\quad
n_f^S(\vec{\emptyset},\vec{\lambda}_{\mathfrak{m}B})
=(a-2M_1+\epsilon_1+\epsilon_2)
\prod_{k=1}^{\mathfrak{m}+1}(a-2M_2+k\epsilon_1+\epsilon_2),
\nonumber \\
&&\quad
n_{\rm vec}[\vec{\lambda}_{\mathfrak{m}B}](\vec{a})=
(\mathfrak{m}\epsilon_1-\epsilon_2)\epsilon_1^{\mathfrak{m}+1}
\mathfrak{m}!\prod_{k=0}^{\mathfrak{m}+1}(2a+k\epsilon_1+\epsilon_2),
\nonumber 
\end{eqnarray}
\begin{eqnarray}
&&\hspace*{-0.5cm} 
Z_1^{(C)}(a,M_i,\epsilon_1,\epsilon_2;x,y)
=\sum_{\mathfrak{m}=1}^{\infty}xy^{\mathfrak{m}+1}
\frac{n_f^S[\vec{\emptyset},\vec{\lambda}_{\mathfrak{m}C}](a_1,a;m_1)\cdot 
n_f^S[\vec{\lambda}_{\mathfrak{m}C},\vec{\emptyset}](a_1,a;m_2)}
{n_{\rm vec}[\vec{\lambda}_{\mathfrak{m}C}](\vec{a})},
 \\
&&\quad
n_f^S(\vec{\lambda}_{\mathfrak{m}C},\vec{\emptyset})
=(-a-2M_3-\epsilon_2)(-a-2M_4-\epsilon_2)
\prod_{k=0}^{\mathfrak{m}-1}(-a-2M_4-k\epsilon_1)
\nonumber \\
&&\quad
n_f^S(\vec{\emptyset},\vec{\lambda}_{\mathfrak{m}C})
=(a-2M_1+\epsilon_1+\epsilon_2) (a-2M_2+\epsilon_1+2\epsilon_2)
\prod_{k=1}^{\mathfrak{m}}(a-2M_2+k\epsilon_1+\epsilon_2),
\nonumber \\
&&\quad 
n_{\rm vec}[\vec{\lambda}_{\mathfrak{m}C}](\vec{a})
=(2a+\epsilon_1+2\epsilon_2)
(-\mathfrak{m} \epsilon_1+\epsilon_2)\epsilon_2
\epsilon_1^{\mathfrak{m}}(\mathfrak{m}-1)!\prod_{k=0}^{\mathfrak{m}}(2a+k\epsilon_1+\epsilon_2),
\nonumber 
\end{eqnarray}
\begin{eqnarray}
&&\hspace*{-0.5cm}
Z_1^{(D)}(a,M_i,\epsilon_1,\epsilon_2;x,y)
=\sum_{\mathfrak{m}=0}^{\infty}xy^{\mathfrak{m}+1}
\frac{n_f^S[\vec{\emptyset},\vec{\lambda}_{\mathfrak{m}D}](a_1,a;m_1)\cdot 
n_f^S[\vec{\lambda}_{\mathfrak{m}D},\vec{\emptyset}](a_1,a;m_2)}
{n_{\rm vec}[\vec{\lambda}_{\mathfrak{m}D}](\vec{a})},
 \\
&&\quad
n_f^S(\vec{\lambda}_{\mathfrak{m}D},\vec{\emptyset})
=(a-2M_3)(a-2M_4)\prod_{k=0}^{\mathfrak{m}-1}(-a-2M_4-k\epsilon_1),
\nonumber \\
&&\quad
n_f^S(\vec{\emptyset},\vec{\lambda}_{\mathfrak{m}D})
=(a+2M_1-\epsilon_1)(a+2M_2-\epsilon_1-\epsilon_2)
\prod_{k=1}^{\mathfrak{m}}(a-2M_2+k\epsilon_1+\epsilon_2),
\nonumber \\
&&\quad
n_{\rm vec}[\vec{\lambda}_{\mathfrak{m}D}](\vec{a})=(-2a+\epsilon_1)(2a+\mathfrak{m}\epsilon_1)
\epsilon_2\epsilon_1^{\mathfrak{m}+1}\mathfrak{m}!
\prod_{k=0}^{\mathfrak{m}-1}(2a+k\epsilon_1+\epsilon_2).\nonumber
\end{eqnarray}

Taking a decoupling limit
\begin{eqnarray}
M_2,M_3,M_4\to \infty,
\quad \Lambda_1=-2M_3x, \quad
\Lambda_2=4M_2M_4y,
\end{eqnarray}
one obtains 
the partition function for $N_f=1$ theory
\begin{eqnarray}
&&
Z_1^{(A)}(a,M_1,\epsilon_1,\epsilon_2;\Lambda_1,\Lambda_2)
\nonumber \\
&&\quad
=\sum_{\mathfrak{m}=-1}^{\infty}\Lambda_1\Lambda_2^{\mathfrak{m}+1}
\frac{a+2M_1-\epsilon_1}
{(2a+\mathfrak{m}\epsilon_1)\epsilon_1^{\mathfrak{m}+2}(\mathfrak{m}+1)!\prod_{k=0}^{\mathfrak{m}}(2a+k\epsilon_1+\epsilon_2)},
\nonumber \\
&&
Z_1^{(B)}(a,M_1,\epsilon_1,\epsilon_2;\Lambda_1,\Lambda_2)
\nonumber \\
&&\quad
=\sum_{\mathfrak{m}=0}^{\infty}\Lambda_1\Lambda_2^{\mathfrak{m}+1}
\frac{a-2M_1+\epsilon_1+\epsilon_2}
{(\mathfrak{m}\epsilon_1-\epsilon_2)\epsilon_1^{\mathfrak{m}+1}
\mathfrak{m}!\prod_{k=0}^{\mathfrak{m}+1}(2a+k\epsilon_1+\epsilon_2)},
\nonumber\\
&&
Z_1^{(C)}(a,M_1,\epsilon_1,\epsilon_2;\Lambda_1,\Lambda_2)
\nonumber \\
&&\quad=\sum_{\mathfrak{m}=1}^{\infty}\Lambda_1\Lambda_2^{\mathfrak{m}+1}
\frac{a-2M_1+\epsilon_1+\epsilon_2}
{(2a+\epsilon_1+2\epsilon_2)
(-\mathfrak{m} \epsilon_1+\epsilon_2)\epsilon_2
\epsilon_1^{\mathfrak{m}}(\mathfrak{m}-1)!\prod_{k=0}^{\mathfrak{m}}(2a+k\epsilon_1+\epsilon_2)},
\nonumber\\
&&
Z_1^{(D)}(a,M_1,\epsilon_1,\epsilon_2;\Lambda_1,\Lambda_2)
\nonumber \\
&&\quad=\sum_{\mathfrak{m}=0}^{\infty}\Lambda_1\Lambda_2^{\mathfrak{m}+1}
\frac{a+2M_1-\epsilon_1}
{(2a-\epsilon_1)(2a+\mathfrak{m}\epsilon_1)\epsilon_2
\epsilon_1^{\mathfrak{m}+1}\mathfrak{m}!\prod_{k=0}^{\mathfrak{m}-1}(2a+k\epsilon_1+\epsilon_2)}.
\label{1-instanton_Nf=1}
\end{eqnarray}
The decoupling limit
\begin{eqnarray}
M_1,M_2,M_3,M_4\to \infty,
\quad \Lambda_1=4M_1M_3x, \quad
\Lambda_2=4M_2M_4y,
\end{eqnarray}
gives the one instanton partition function for $N_f=0$ theory
\begin{eqnarray}
&&Z_1^{(A)}(a,\epsilon_1,\epsilon_2;\Lambda_1,\Lambda_2)
\nonumber \\
&&\quad=-\sum_{\mathfrak{m}=-1}^{\infty}\Lambda_1\Lambda_2^{\mathfrak{m}+1}
\frac{1}
{(2a+\mathfrak{m}\epsilon_1)\epsilon_1^{\mathfrak{m}+2}(\mathfrak{m}+1)!\prod_{k=0}^{\mathfrak{m}}(2a+k\epsilon_1+\epsilon_2)},
\nonumber\\
&&Z_1^{(B)}(a,\epsilon_1,\epsilon_2;\Lambda_1,\Lambda_2)
\nonumber \\
&&\quad=\sum_{\mathfrak{m}=0}^{\infty}\Lambda_1\Lambda_2^{\mathfrak{m}+1}
\frac{1}
{(\mathfrak{m}\epsilon_1-\epsilon_2)\epsilon_1^{\mathfrak{m}+1}
\mathfrak{m}!\prod_{k=0}^{\mathfrak{m}+1}(2a+k\epsilon_1+\epsilon_2)},
\nonumber\\
&&Z_1^{(C)}(a,\epsilon_1,\epsilon_2;\Lambda_1,\Lambda_2)
\nonumber \\
&&\quad=\sum_{\mathfrak{m}=1}^{\infty}\Lambda_1\Lambda_2^{\mathfrak{m}+1}
\frac{1}
{(-\mathfrak{m} \epsilon_1+\epsilon_2)
(2a+\epsilon_1+2\epsilon_2)
\epsilon_2\epsilon_1^{\mathfrak{m}}(\mathfrak{m}-1)!
\prod_{k=0}^{\mathfrak{m}}(2a+k\epsilon_1+\epsilon_2)},
\nonumber\\
&&Z_1^{(D)}(a,\epsilon_1,\epsilon_2;\Lambda_1,\Lambda_2)
\nonumber \\
&&\quad=\sum_{\mathfrak{m}=0}^{\infty}\Lambda_1\Lambda_2^{\mathfrak{m}+1}
\frac{1}
{(2a-\epsilon_1)(2a+\mathfrak{m}\epsilon_1)(-\epsilon_2)
\epsilon_1^{\mathfrak{m}+1}\mathfrak{m}!
\prod_{k=0}^{\mathfrak{m}-1}(2a+k\epsilon_1+\epsilon_2)}.
\label{1-instanton_Nf=0}
\end{eqnarray}


\subsection{Topological vertex computation}
Now we discuss the one instanton partition function for the four
dimensional gauge theory from the A model 
via geometric engineering.
\begin{figure}[tbp]
 \begin{center}
   \includegraphics[width=50mm,clip]{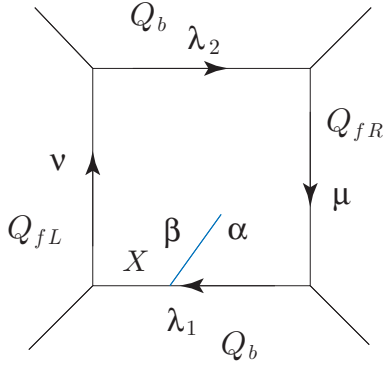}
  \caption{D-brane inserted on an inner leg of local Hirzebruch surface}
 \end{center}
\end{figure}
The pure gauge theory is engineered by the A model on
the local Hirzebruch surface
${\bf F}_0={\bf P}^1_{b}\times{\bf P}^1_{f}$
with a toric brane.
The K\"ahler parameters for the base ${\bf P}^1_{b}$ and the 
fiber ${\bf P}^1_{f}$ correspond to $Q_b=\beta^4\Lambda^4$ and
$Q_f=e^{-2\beta a}$, respectively.
To realize the surface operator in four dimensional theory, 
the toric brane should be inserted on the inner leg which denotes
the base ${\bf P}^1_{b}$ in the toric diagram
\cite{Gukovtalk,Kozcaz:2010af,Dimofte:2010tz}, 
and we choose the open string moduli $X=\beta^2\Lambda^2w$.

The topological vertex computes the open BPS invariants \cite{Aganagic:2003db,Halmagyi:2005vk}.
\begin{eqnarray}
Z^{\rm open}_{\rm BPS}(X,Q_b,Q_{fL},Q_{fR};q)
=\frac{Z^{\rm D\hbox{-}brane}(X,Q_b,Q_{fL},Q_{fR};q)}{Z^{\rm closed}_{\rm
BPS}(Q_b,Q_{fL},Q_{fR};q)}.
\end{eqnarray}
Each factor is given by the representation sums
\begin{eqnarray}
&&Z^{\rm closed}_{\rm BPS}(Q_b,Q_{fL},Q_{fR};q)
\nonumber \\
&&
=\sum_{\mu,\nu,\lambda_1,\lambda_2}
C_{\lambda_1\mu^t\emptyset}C_{\nu\lambda_1^{t}\emptyset}
C_{\lambda_2\nu^{t}\emptyset}C_{\mu\lambda_2^t\emptyset}
q^{(\kappa_{\nu}+\kappa_{\lambda_2}+\kappa_{\mu}+\kappa_{\lambda_1})/2}
Q_{fL}^{|\nu|}Q_{fR}^{|\mu|}Q_b^{|\lambda_1|+|\lambda_2|}, 
\\
&&Z^{\rm D\hbox{-}brane}(X,Q_b,Q_{fL},Q_{fR};q)
\nonumber \\
&&
=\sum_{\mu,\nu,\lambda_1,\lambda_2,\alpha,\beta}
C_{(\lambda_1\otimes\alpha)\mu^t\emptyset}C_{\nu(\lambda_1^t\otimes\beta)\emptyset}
C_{\lambda_2\nu^{t}\emptyset}C_{\mu\lambda_2^t\emptyset}
q^{(\kappa_{\nu}+\kappa_{\lambda_2}+\kappa_{\mu}
+(p-1)\kappa_{\lambda_1^t\otimes\beta}+p\kappa_{\lambda_1\otimes\alpha})/2}
\nonumber\\
&&\quad\quad\times
Q_{fL}^{|\nu|}Q_{fR}^{|\mu|}Q_b^{|\lambda_1|+|\lambda_2|+|\beta|}
(-1)^{|\lambda_1|+(p-1)|\lambda_1^t\otimes\beta|+p|\lambda_1\otimes\alpha|}
~{\rm Tr}_{\alpha}V~{\rm Tr}_{\beta}V^{-1},
\label{open_vertex_hirz}
\end{eqnarray}
where $V=X$ and $p$ denotes the framing of the toric brane.
For the tensor product representation $\alpha\otimes \beta$, 
$\kappa_{\alpha\otimes\beta}=\kappa_{\alpha}+\kappa_{\beta}$ and 
$C_{(\alpha\otimes\beta)\mu\nu}=\sum_{\gamma}c^{\gamma}_{\alpha\beta}C_{\gamma\mu\nu}$
where $c^{\gamma}_{\alpha\beta}$ is Littlewood-Richardson coefficient \cite{Macdonald}.
In the following we choose the framing $p=-1$.
In order to compare with the four dimensional gauge theory in detail, 
we have set the K\"ahler parameter for the fiber ${\bf P}^1_{f}$
in the left/right side in the toric diagram independently
as $Q_{fL}=e^{-2\beta a_L}$ and $Q_{fR}=e^{-2\beta a_R}$.

The one instanton part of the topological string amplitude is the first order in $Q_b^1$. 
For the closed string partition function $Z^{\rm closed}_{\rm BPS}(Q_b,Q_{fL},Q_{fR};q)$, 
we only need to consider the terms with $\lambda_1=\lambda_2=\emptyset$.
For such choice of the partitions, 
one finds the closed string partition function
\begin{eqnarray}
&&Z^{\rm closed}_{\rm BPS\;0}(Q_b,Q_{fL},Q_{fR};q)
=M(Q_{fL};q)M(Q_{fR};q), \\
&& M(Q;q)=\prod_{n=1}^{\infty}(1-Qq^n)^{-n}.
\end{eqnarray}

On the other hand, the D-brane partition function $Z^{\rm
D\hbox{-}brane}(X,Q_b,Q_{fL},Q_{fR};q)$ in one instanton sector comes from the 
following three choices of the partitions:
\begin{eqnarray}
&&(1)\; (\beta,\lambda_1,\lambda_2)=(\Box,\emptyset,\emptyset),
\quad
(2)\; (\beta,\lambda_1,\lambda_2)=(\emptyset,\emptyset,\Box),
 \quad
(3)\; (\beta,\lambda_1,\lambda_2)=(\emptyset,\Box,\emptyset).
\nonumber \\
&&\label{3_classes}
\end{eqnarray}
At this point we should point out a crucial difference from the case of 
the geometric engineering of the Nekrasov partition function in terms of 
closed topological string. In the case of the Nekrasov partition function for
$SU(N)$ gauge theory, the fixed points on the instanton moduli space are in one to one
correspondence with the assignments of the Young diagrams 
on $N$ parallel inner edges representing the base ${\bf P}^1_b$ of ALE fibration of
type $A_{N-1}$. However, in the present case even at one instanton level
the one to one correspondence is lost. In fact we found four fixed points \eqref{fixed} on 
the affine Laumon space with instanton number one,  while \eqref{3_classes}
gives only three configurations. The lack of one to one correspondence makes the
problem of matching the instanton partition function with surface operator to
open topological string amplitudes highly non-trivial.

Let us compute the open BPS partition function for each choice of partitions.
For case (1), the open BPS partition function 
$Z^{\rm open\;(1)}_{\rm BPS\;1}(X,Q_b,Q_{fL},Q_{fR};q)$ becomes
\begin{eqnarray}
\hspace*{-0.3cm}
&&Z^{\rm open\;(1)}_{\rm BPS\;1}(X,Q_b,Q_{fL},Q_{fR};q)
\nonumber \\
&&
=\frac{1}{Z^{\rm closed}_{\rm BPS\;0}(Q_b,Q_{fL},Q_{fR};q)}
\nonumber \\
&&\quad\times
Q_b\sum_{\alpha}
s_{\alpha}(q^{-\rho})s_{\alpha}(X)s_{\Box}(q^{\rho})q^{-\kappa_{\Box}/2}
\sum_{\mu}s_{\mu}(q^{\rho+\alpha})s_{\mu}(q^{\rho})Q_{fR}^{|\mu|}
\nonumber \\
&&\quad\times
\sum_{\nu}s_{\nu}(q^{\rho+\Box})s_{\nu}(q^{\rho})Q_{fL}^{|\nu|}s_{\Box}(X^{-1})
\nonumber \\
&&
=Q_b\frac{q^{1/2}}{q-1}\frac{1}{1-Q_{fL}}\sum_{\mathfrak{m}^{\prime}=0}^{\infty}
\frac{X^{-1}(Xq^{1/2})^{\mathfrak{m}^{\prime}}}{\prod_{k=1}^{\mathfrak{m}^{\prime}}(1-q^k)(1-Q_{fR}q^{k-1})}.
\end{eqnarray}
In the computation, we used the following relations.
\begin{eqnarray}
&&s_{(\mathfrak{m})}(q^{-\rho})=\frac{q^{\mathfrak{m}/2}}{\prod_{k=1}^{\mathfrak{m}}(1-q^k)},\quad
{\rm Tr}_{(\mathfrak{m})}X=s_{(\mathfrak{m})}(X)=X^{\mathfrak{m}},
\\
&&\sum_{\mu}s_{\mu}(q^{\rho+(\mathfrak{m})})s_{\mu}(q^{\rho})Q^{|\mu|}
=M(Q;q)\prod_{k=1}^{\mathfrak{m}}\frac{1}{1-Qq^{k-1}}.
\end{eqnarray}

For case (2), we obtain
\begin{eqnarray}
&&Z^{\rm open\;(2)}_{\rm BPS\;1}(X,Q_b,Q_{fL},Q_{fR};q)
\nonumber \\
&&=\frac{1}{Z^{\rm closed}_{\rm BPS\;0}(Q_b,Q_{fL},Q_{fR};q)}
\nonumber \\
&&\quad\quad\times
Q_b\sum_{\alpha}
s_{\alpha}(q^{-\rho})s_{\alpha}(X)s_{\Box}(q^{\rho})^2
\sum_{\mu}s_{\mu}(q^{\rho+\alpha})s_{\mu}(q^{\rho+\Box})Q_{fR}^{|\mu|}
\sum_{\nu}s_{\nu}(q^{\rho})s_{\nu}(q^{\rho+\Box})Q_{fL}^{|\nu|}
\nonumber \\
&&
=Q_b\frac{q}{(q-1)^2}\frac{1}{1-Q_{fL}}
\sum_{{\mathfrak{m}}=0}^{\infty}\frac{1}{(1-Q_{fR}q^{-1})(1-Q_{fR}q^{{\mathfrak{m}}})}
\frac{(Xq^{1/2})^{\mathfrak{m}}}{\prod_{k=1}^{\mathfrak{m}}(1-q^k)\prod_{k=1}^{\mathfrak{m}-1}(1-Q_{fR}q^{k-1})}.
\nonumber \\
&&
\label{case2_hirz}
\end{eqnarray}
To derive this result, we applied a relation 
\begin{eqnarray}
\hspace*{-0.2cm}
\sum_{\mu}s_{\mu}(q^{\rho+(\mathfrak{m})})s_{\mu}(q^{\rho+\Box})Q^{|\mu|}
=M(Q;q)\frac{1}{(1-Qq^{-1})(1-Qq^{\mathfrak{m}})}
\frac{1}{\prod_{k=1}^{\mathfrak{m}-1}(1-Qq^{k-1})}.
\end{eqnarray}
This is found from the Cauchy formula (\ref{s1}).

For case (3), we have to consider the topological vertex with a tensor
product representation $C_{(\alpha\otimes\Box)\mu^t\emptyset}$
seriously. 
For the tensor product representation $\alpha\otimes\beta$,
the Schur function obeys \cite{Macdonald}
\begin{eqnarray}
s_{\alpha\otimes \beta}=s_{\alpha}s_{\beta}=\sum_{\gamma}c_{\alpha\beta}^{\gamma}s_{\gamma}.
\end{eqnarray}
Then the topological vertex with a tensor product representation 
is computed as
\begin{eqnarray}
&&C_{\emptyset(\alpha\otimes\beta)\mu}
=
\sum_\gamma c_{\alpha\beta}^\gamma  C_{\emptyset\gamma\mu}
=
\sum_\gamma c_{\alpha\beta}^\gamma  s_\gamma(q^{\rho}) s_{\mu^t}(q^{\rho+\gamma}) q^{\kappa_\mu/2}
=
\sum_\gamma c_{\alpha\beta}^\gamma  s_\gamma(q^{\rho+\mu^t}) s_{\mu^t}(q^{\rho}) q^{\kappa_\mu/2}
\nonumber \\
&&
=
s_\alpha(q^{\rho+\mu^t}) s_\beta(q^{\rho+\mu^t}) s_{\mu^t}(q^{\rho}) q^{\kappa_\mu/2}
=
s_\alpha(q^\rho) s_\beta(q^\rho) s_{\mu^t}(q^{\rho+\alpha}) 
{ s_{\mu^t}(q^{\rho+\beta}) \over s_{\mu^t}(q^\rho) } 
q^{\kappa_\mu/2}.
\end{eqnarray}
Applying this expression to (\ref{open_vertex_hirz}), we find 
that the partition function in this case coincides with (\ref{case2_hirz}):
\begin{equation}
Z^{\rm open\;(3)}_{\rm BPS\;1}(X,Q_b,Q_{fL},Q_{fR};q)=Z^{\rm open\;(2)}_{\rm BPS\;1}(X,Q_b,Q_{fL},Q_{fR};q).
\end{equation}

Summing these three contributions, we find the partition function in one
instanton sector.
\begin{eqnarray}
&&Z^{\rm open}_{\rm BPS\;1}(X,Q_b,Q_{fL},Q_{fR};q)
\nonumber \\
&&=Q_b\frac{q^{1/2}}{q-1}\frac{1}{1-Q_{fL}}\sum_{\mathfrak{m}^{\prime}=0}^{\infty}
\frac{X^{-1}(Xq^{1/2})^{\mathfrak{m}^{\prime}}}{\prod_{k=1}^{\mathfrak{m}^{\prime}}(1-q^k)(1-Q_{fR}q^{k-1})}
\nonumber \\
&&\quad
+2Q_b\frac{q}{(q-1)^2}\frac{1}{1-Q_{fL}}
\sum_{\mathfrak{m}=0}^{\infty}
\frac{1}{(1-Q_{fR}q^{-1})(1-Q_{fR}q^{\mathfrak{m}})}
\frac{(Xq^{1/2})^{\mathfrak{m}}}
{\prod_{k=1}^{\mathfrak{m}}(1-q^k)\prod_{k=1}^{\mathfrak{m}-1}(1-Q_{fR}q^{k-1})}.
\nonumber \\
&&
\end{eqnarray}
In the four dimensional limit $\beta\to 0$, the open BPS
partition function become 
\begin{eqnarray}
&&Z^{\rm open(4D)}_{\rm BPS\;1}(w,\Lambda,a_L,a_R;\hbar)
\nonumber \\
&&=-\Lambda^4\sum_{\mathfrak{m}^{\prime}=0}^{\infty} 
(\Lambda^2w)^{\mathfrak{m}^{\prime}-1}\frac{1}{(2a_L)\hbar^{\mathfrak{m}^{\prime}+1} 
\mathfrak{m}^{\prime}!\prod_{k=1}^{\mathfrak{m}^{\prime}}(2a_R+(k-1)\hbar)}
\nonumber \\
&&\quad
+2\Lambda^4\sum_{\mathfrak{m}=1}^{\infty}(\Lambda^2w)^\mathfrak{m}\frac{1}{(2a_L)(2a_R-\hbar)(2a_R+\mathfrak{m}\hbar)\hbar^{\mathfrak{m}+2} \mathfrak{m}!
\prod_{k=1}^{\mathfrak{m}-1}(2a_R+(k-1)\hbar)},
\end{eqnarray}
where $q=e^{-\beta\hbar}$.

On the other hand, in the self-dual case $\epsilon_1=-\epsilon_2=\hbar$,
the one instanton partition function for 
the gauge theory (\ref{1-instanton_Nf=0}) 
yields
\begin{eqnarray}
&&Z_{\rm 1\hbox{-}inst}^{\rm (4D)}(a,\hbar,-\hbar;\Lambda^2w,\Lambda^2w^{-1})
\nonumber \\
&&
=-\Lambda^4\sum_{\mathfrak{m}=-1}^{\infty}(\Lambda^2w)^{\mathfrak{m}}
\frac{1}{\hbar^{\mathfrak{m}+2} (\mathfrak{m}+1)!
\prod_{k=0}^{\mathfrak{m}+1}(2a+(k-1)\hbar)}
\nonumber \\
&&\quad
+2\Lambda^4\sum_{\mathfrak{m}=0}^{\infty}(\Lambda^2w)^{\mathfrak{m}}
\frac{1}{(2a-\hbar)^2(2a+\mathfrak{m}\hbar)
\hbar^{\mathfrak{m}+2} \mathfrak{m}!
\prod_{k=1}^{\mathfrak{m}-1}(2a+(k-1)\hbar)}.
\end{eqnarray}
Choosing $a_L$ and $a_R$ by
\begin{eqnarray}
a_L=a-\hbar/2,\quad a_R=a,
\end{eqnarray}
we find a coincidence between the one instanton partition
function for gauge theory and four dimensional limit of the partition
function for the open BPS states in the A model.

\subsubsection{Geometric engineering of $N_f=1$ theory}
\begin{figure}[tbp]
 \begin{center}
  \includegraphics[width=50mm,clip]{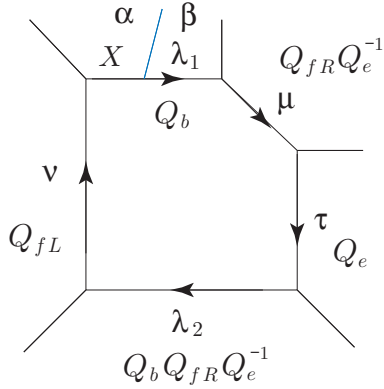}
 \end{center}
 \caption{D-brane inserted on an inner leg of local del Pezzo surface}
 \label{del_Pezzo_vertex}
\end{figure}
Geometrically the four dimensional gauge theory with $N_f=1$ flavor
is engineered by the A model on local del Pezzo surface $dP_2$
with K\"ahler parameters
\begin{eqnarray}
Q_{f}=e^{-2\beta a},\quad Q_e=e^{-\beta(a-m)},\quad
 Q_b=2\beta^3\Lambda^3.
\end{eqnarray}
So as to realize the surface operator, we introduce toric D-brane
as Fig.\ref{del_Pezzo_vertex}, and 
the open string moduli is also identified by
\begin{eqnarray}
X=\beta^2\Lambda^2w.
\end{eqnarray}

The topological vertex computes the open BPS invariants on local
del Pezzo surface.
\begin{eqnarray}
&&Z^{\rm open}_{\rm BPS}(X,Q_b,Q_e,Q_{fL},Q_{fR};q)
=\frac{Z^{\rm D\hbox{-}brane}(X,Q_b,Q_e,Q_{fL},Q_{fR};q)}
{Z^{\rm closed}_{\rm BPS}(Q_b,Q_e,Q_{fL},Q_{fR};q)},
\\
&&Z^{\rm closed}_{\rm BPS}(Q_b,Q_e,Q_{fL},Q_{fR};q)
\nonumber \\
&&=\sum_{\lambda_1,\lambda_2,\mu,\nu,\tau}
C_{\lambda_1\nu^t\emptyset}
C_{\mu\lambda_1^t\emptyset}
C_{\tau\mu^t\emptyset}C_{\lambda_2\tau^t\phi}C_{\nu\lambda_2^t\emptyset}
\nonumber \\
&&\quad\quad\times q^{(\kappa_{\nu}+\kappa_{\lambda_2})/2}
(-Q_b)^{|\lambda_1|}(-Q_{fR}Q_e^{-1})^{|\mu|}
(-Q_e)^{|\tau|}(Q_bQ_{fR}Q_e^{-1})^{|\lambda_2|}
Q_{fL}^{|\nu|},
\\
&&Z^{\rm D\hbox{-}brane}_{\rm BPS}(X,Q_b,Q_e,Q_{fL},Q_{fR};q)
\nonumber \\
&&=
\sum_{\lambda_1,\lambda_2,\mu,\nu,\tau,\alpha,\beta}
C_{(\lambda_1\otimes\alpha)\nu^t\emptyset}
C_{\mu(\lambda_1^t\otimes\beta)\emptyset}
C_{\tau\mu^t\emptyset}C_{\lambda_2\tau^t\phi}C_{\nu\lambda_2^t\emptyset}
\nonumber \\
&&\quad\quad\times q^{(p\kappa_{\lambda_1\otimes\alpha}
+p\kappa_{\lambda_1^t\otimes\beta}
+\kappa_{\nu}+\kappa_{\lambda_2}
)/2}
(-Q_b)^{|\lambda_1|}(-Q_{fR}Q_e^{-1})^{|\mu|}
(-Q_e)^{|\tau|}(Q_bQ_{fR}Q_e^{-1})^{|\lambda_2|}
Q_{fL}^{|\nu|}Q_b^{|\beta|}
\nonumber \\
&&\quad\quad\times
(-1)^{p|\alpha|+p|\beta|}
~{\rm Tr}_{\alpha}V~{\rm Tr}_{\beta}V^{-1}.
\end{eqnarray}
For later convenience, we have changed 
the K\"ahler parameter $Q_f$ as in the local Hirzebruch case.
\begin{eqnarray}
Q_{fL}=e^{-2\beta a_L},\quad Q_{fR}=e^{-2\beta a_R},
\quad Q_e=e^{-\beta(a_R-m)}.
\end{eqnarray}
In the following we choose the framing $p=-1$.

The one instanton sector for four dimensional theory comes from 
a part of the above representation sums which satisfies 
$|\beta|+|\lambda_1|+|\lambda_2|=1$ for the D-brane partition function 
and $|\lambda_1|+|\lambda_2|=0$ for closed string partition function.
We find the closed string partition function 
\begin{eqnarray}
&&Z^{\rm closed}_{\rm BPS\;0}(Q_b,Q_e,Q_{fL},Q_{fR};q)
=\frac{M(Q_{fL};q)M(Q_{fR};q)}{M(Q_e;q)M(Q_{fR}Q_e^{-1};q)}.
\end{eqnarray}

The computation of the D-brane partition function for the one instanton sector
is classified into three cases (\ref{3_classes}). 
Each partition function is computed in the same way as local Hirzebruch surface.
\begin{eqnarray}
&&Z^{\rm open\;(1)}_{\rm BPS\;1}(X,Q_b,Q_e,Q_{fL},Q_{fR};q)
\nonumber \\
&&=(-Q_b)\frac{q^{1/2}}{q-1}
\sum_{\mathfrak{m}^{\prime}=0}^{\infty}\frac{(1-Q_{fR}Q_e^{-1})X^{-1}(Xq^{1/2})^{\mathfrak{m}^{\prime}}}{(1-Q_{fR})\prod_{k=1}^{\mathfrak{m}^{\prime}}(1-q^k)(1-Q_{fL}q^{k-1})},
\\
&&Z^{\rm open\;(2)}_{\rm BPS\;1}(X,Q_b,Q_e,Q_{fL},Q_{fR};q)
\nonumber \\
&&=(Q_bQ_fQ_e^{-1})
\left(\frac{q^{1/2}}{q-1}\right)^2
\nonumber \\
&&\times
\sum_{\mathfrak{m}=0}^{\infty}\frac{(1-Q_e)(Xq^{1/2})^{\mathfrak{m}}}{(1-Q_{fR})(1-Q_{fL}q^{-1})(1-Q_{fL}q^{\mathfrak{m}})\prod_{k=1}^{\mathfrak{m}}(1-q^k)\prod_{k=1}^{\mathfrak{m}-1}(1-Q_{fL}q^{k-1})},
\\
&&Z^{\rm open\;(3)}_{\rm BPS\;1}(X,Q_b,Q_e,Q_{fL},Q_{fR};q)
\nonumber \\
&&=(-Q_b)
\left(\frac{q^{1/2}}{q-1}\right)^2
\nonumber \\
&&\times
\sum_{\mathfrak{m}=0}^{\infty}\frac{(1-Q_{fR}Q_e^{-1})(Xq^{1/2})^{\mathfrak{m}}}
{(1-Q_{fR})(1-Q_{fL}q^{-1})(1-Q_{fL}q^{\mathfrak{m}})
\prod_{k=1}^{\mathfrak{m}}(1-q^k)\prod_{k=1}^{\mathfrak{m}-1}(1-Q_{fL}q^{k-1})}.
\end{eqnarray}
Summing all these contributions, one finds
\begin{eqnarray}
&&Z^{\rm open}_{\rm BPS\;1}(X,Q_b,Q_e,Q_{fL},Q_{fR};q)
\nonumber \\
&&=Q_b\frac{q^{1/2}}{q-1}
\sum_{\mathfrak{m}^{\prime}=0}^{\infty}\frac{(Q_{fR}Q_e^{-1}-1)X^{-1}(Xq^{1/2})}{(1-Q_{fR})\prod_{k=1}^{\mathfrak{m}^{\prime}}(1-q^k)(1-Q_{fL}q^{k-1})}
\nonumber \\
&&+Q_b\left(\frac{q^{1/2}}{q-1}\right)^2
\sum_{\mathfrak{m}=0}^{\infty}
\frac{(2Q_{fR}Q_e^{-1}-Q_{fR}-1)(Xq^{1/2})^{\mathfrak{m}}}
{(1-Q_{fR})(1-Q_{fL}q^{-1})(1-Q_{fL}q^{\mathfrak{m}})
\prod_{k=1}^{\mathfrak{m}}(1-q^k)
\prod_{k=1}^{\mathfrak{m}-1}(1-Q_{fL}q^{k-1})}.
\nonumber \\
&&
\label{1-instanton_Nf1}
\end{eqnarray}
In the four dimensional limit ($\beta\to 0$), 
this partition function yields
\begin{eqnarray}
&&Z^{\rm open(4D)}_{\rm BPS\;1}(w,\Lambda,a_L,a_R,M;\hbar)
\nonumber \\
&&=2\Lambda^3\sum_{\mathfrak{m}^{\prime}=0}^{\infty} 
(\Lambda^2 w)^{\mathfrak{m}^{\prime}-1}
\frac{a_R+m}
{(2a_R)\hbar^{\mathfrak{m}^{\prime}+1}\mathfrak{m^{\prime}}!
\prod_{k=1}^{\mathfrak{m}^{\prime}}(2a_L+(k-1)\hbar)}
\nonumber \\
&&\quad
+2\Lambda^3
\sum_{\mathfrak{m}=1}^{\infty}(\Lambda^2 w)^\mathfrak{m} \frac{-2m}{(2a_R)(2a_L-\hbar)
(2a_L+\mathfrak{m}\hbar)\hbar^{\mathfrak{m}+2} \mathfrak{m}!
\prod_{k=1}^{\mathfrak{m}-1}(2a_L+(k-1)\hbar)}.
\label{4d_1-inst_Nf1}
\end{eqnarray}

In the self-dual case $\epsilon_1=-\epsilon_2=\hbar$, 
the one instanton partition function  (\ref{1-instanton_Nf=1}) for the gauge theory is reduced to
\begin{eqnarray}
&&Z_{\rm 1\hbox{-}inst}^{\rm (4D)}(a,\hbar,-\hbar,M_1;\Lambda^2w,2\Lambda^3w^{-1})
\nonumber \\
&&
=2\Lambda^3\sum_{\mathfrak{m}=-1}^{\infty}(\Lambda^2 w)^{\mathfrak{m}}
\frac{a+2M_1-\hbar}
{(2a-\hbar)(2a+\mathfrak{m}\hbar)\hbar^{\mathfrak{m}+2}(\mathfrak{m}+1)!\prod_{k=1}^{\mathfrak{m}}(2a+k\hbar)}
\nonumber \\
&&\quad
+2\Lambda^3\sum_{\mathfrak{m}=0}^{\infty}(\Lambda^2 w)^{\mathfrak{m}}
\frac{-(4M_1-\hbar)}{(2a-\hbar)^2(2a+\mathfrak{m}\hbar)\hbar^{\mathfrak{m}+2}\mathfrak{m}!\prod_{k=1}^{\mathfrak{m}-1}(2a+(k-1)\hbar)}.
\end{eqnarray}
This result coincides with the topological vertex
computation (\ref{4d_1-inst_Nf1}) under the following shifts of parameters:
\begin{eqnarray}
a_L=a,\quad a_R=a-\hbar/2,\quad m=2M_1-\hbar/2.
\end{eqnarray}

%
\newpage 

\section*{Acknowledgments}


We would like to thank S.~Hirano, H.~Itoyama, H.~Nagoya and S.Yanagida for enlightening discussions. 
We also thank T.~Dimofte, M.~Mari\~no and Y.~Tachikawa for helpful correspondences. 
Special thanks are due to K.~Nagao for his clear explanation of the work of Braverman and Etingof. 
This work is partially supported by the Grant-in-Aid for Nagoya
University Global COE Program, "Quest for Fundamental Principles in the
Universe: from Particles to the Solar System and the Cosmos", from the Ministry
of Education, Culture, Sports, Science and Technology of Japan.
The work of H.A and H.K. is supported in part by Daiko Foundation. 
The work is also supported in part by Grant-in-Aid for Young Scientists (B)
[\# 21740179] (H.F.), Grant-in-Aid for Scientific Research
[\# 22224001] (H.K.) and Grant-in-Aid for Scientific Research [\#21340036] (Y.Y.)
from the Japan Ministry of Education, Culture, Sports, Science and Technology.

\vskip 10mm

\section*{Note added}
After this paper was submitted to arXiv, there appeared a new article \cite{Kozcaz:2010yp}, 
where the results in \cite{Alday:2010vg} are extended to affine $sl(N)$ case.

\newpage


%
\Section*{Appendix A : Equivariant Character of the Affine Laumon space}
\renewcommand{\theequation}{A.\arabic{equation}}\setcounter{equation}{0}
\renewcommand{\thesubsection}{A.\arabic{subsection}}\setcounter{subsection}{0}


\newcommand{\ha}{\frac 12}

%
\newcommand\FigSgn{%
%
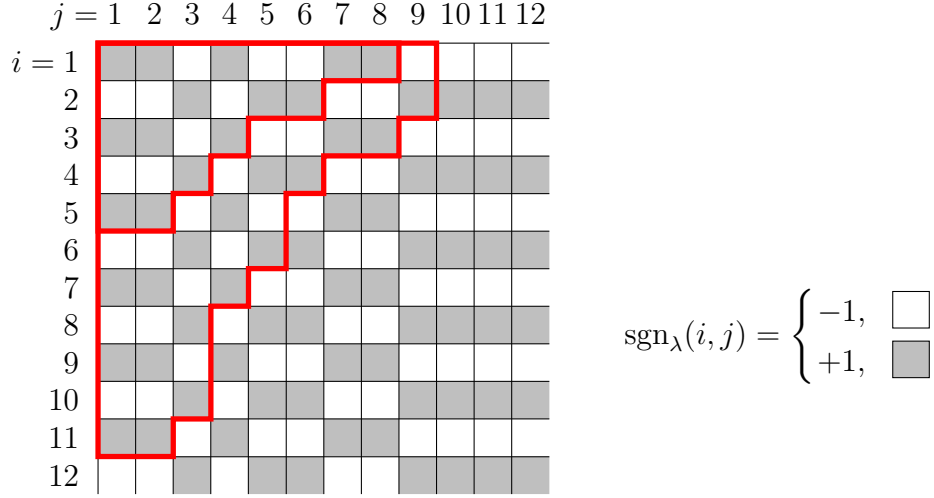
\begin{figure}[tbp]
\psset{unit=5mm}
\begin{center}
\begin{pspicture}(-3,-11)(22,1)
%
%
\multirput(0,0)(0,-2){6}{
\psframe*[linecolor=lightgray](0,0)(2,-1)%
\psframe*[linecolor=lightgray](3,0)(4,-1)%
\psframe*[linecolor=lightgray](6,0)(8,-1)%
}%
\multirput(0,0)(0,-2){6}{
\psframe*[linecolor=lightgray](2,-1)(3,-2)%
\psframe*[linecolor=lightgray](4,-1)(6,-2)%
\psframe*[linecolor=lightgray](8,-1)(12,-2)%
}%
%
\multirput(0,0)(0,-1){12}{\psline[linewidth=.4pt](0,0)(12,0)}%
\multirput(0,0)(1,0){12}{\psline[linewidth=.4pt](0,0)(0,-12)}%
\put(-3,-1){\makebox(2.5,1)[r]{$i=1$}}%
\put(-3,-2){\makebox(2.5,1)[r]{$2$}}%
\put(-3,-3){\makebox(2.5,1)[r]{$3$}}%
\put(-3,-4){\makebox(2.5,1)[r]{$4$}}%
\put(-3,-5){\makebox(2.5,1)[r]{$5$}}%
\put(-3,-6){\makebox(2.5,1)[r]{$6$}}%
\put(-3,-7){\makebox(2.5,1)[r]{$7$}}%
\put(-3,-8){\makebox(2.5,1)[r]{$8$}}%
\put(-3,-9){\makebox(2.5,1)[r]{$9$}}%
\put(-3,-10){\makebox(2.5,1)[r]{$10$}}%
\put(-3,-11){\makebox(2.5,1)[r]{$11$}}%
\put(-3,-12){\makebox(2.5,1)[r]{$12$}}%
\put(-0.8,0.35){\makebox(1,1)[b]{$j=1$}}%
\put(1,0.5){\makebox(1,1)[b]{$2$}}%
\put(2,0.5){\makebox(1,1)[b]{$3$}}%
\put(3,0.5){\makebox(1,1)[b]{$4$}}%
\put(4,0.5){\makebox(1,1)[b]{$5$}}%
\put(5,0.5){\makebox(1,1)[b]{$6$}}%
\put(6,0.5){\makebox(1,1)[b]{$7$}}%
\put(7,0.5){\makebox(1,1)[b]{$8$}}%
\put(8,0.5){\makebox(1,1)[b]{$9$}}%
\put(9,0.5){\makebox(1,1)[b]{$10$}}%
\put(10,0.5){\makebox(1,1)[b]{$11$}}%
\put(11,0.5){\makebox(1,1)[b]{$12$}}%
%
%
\pspolygon[linewidth=2pt,linecolor=red]%
(8,-0)(8,-1)%
(6,-1)(6,-2)%
(4,-2)(4,-3)%
(3,-3)(3,-4)%
(2,-4)(2,-5)%
(0,-5)(0,-0)%
\pspolygon[linewidth=2pt,linecolor=red]%
(9,-0)(9,-2)%
(8,-2)(8,-3)%
(6,-3)(6,-4)%
(5,-4)(5,-6)%
(4,-6)(4,-7)%
(3,-7)(3,-10)%
(2,-10)(2,-11)%
(0,-11)(0,-0)%
\put(14,-8){
$
{\rm sgn}_\lambda(i,j)
=
\left\{
\begin{array}{ll}
-1, \ \ &
\put(0,-.2){\makebox(1,1)[lb]{
\psframe[linewidth=.4pt](0,1)(1,0)%
}}
\\
+1, \ \ &
\put(0,-.2){\makebox(1,1)[lb]{
\psframe[linewidth=.4pt,fillstyle=solid,fillcolor=lightgray](0,1)(1,0)%
}}
\end{array}
\right.
$
}
%
%
\end{pspicture}
\end{center}
\caption{
Example of 
${\rm sgn}_\lambda(i,j)$ for 
$\lambda=(9,9,8,6,5,5,4,3,3,3,2)$,
which equals that for 
$\lambda=(8,6,4,3,2)$.
The black and white boxes denote
${\rm sgn}_\lambda(i,j)=1$ and $-1$, respectively.  
}
\label{fig:SGN}
\end{figure}
}

The fixed points of the toric action on the affine $U(2)$ Laumon space 
are isolated and labeled by a pair of partitions $\vec{\lam} := (\lam_1, \lam_2)$.
We denote by $\lambda_{k,i}$ the $i$-th component of the partition 
$\lam_k = (\lambda_{k,1},\lambda_{k,2},\cdots,\lambda_{k,N})$. 
The equivariant character 
$
{\rm ch}_{\vec\lam,\vec\mu}(\vec a,\vec b)
:=
\mathrm{Tr}_{\mathrm{Ext}(\vec\lam, \vec\mu)}
[\mathrm{diag.} (\ep,\es ; \vec{a}, \vec{b})] 
$
in \cite{Feigin} computes the contribution of a bifundamental multiplet,
from which those of an adjoint and an (anti-)fundamental 
multiplet are derived. Hence the relevant gauge group is $U(2) \times U(2)$
in the following. We need a second pair of partitions $\vec{\mu} := (\mu_1, \mu_2)$ 
and the Coulomb moduli  parameters $\vec{a} := (a_1, a_2), \vec{b} := (b_1, b_2)$
to write down the formula of the equivariant character.
With the convention $a_k \equiv a_{k+2}$ and $\lambda_{k,i} \equiv \lambda_{k+2,i}$,
the equivariant character at a fixed point of the toric action is%
\footnote{
In the $SU(2)$ case, $a_k := (-1)^{k-1}a$ and $b_k:= (-1)^{k-1}b$.
}
\beqa
{\rm ch}_{\vec\lam,\vec\mu}(\vec a,\vec b)
&:=&
\sum_{k, \ell \geq1} e^{a_{k} - b_{\ell+1}} 
e^{\ep + \es(\floor{ \frac{\ell}{2}- \frac{1}{2}  } - \floor{ \frac{k}{2} -1 })}
\frac{(e^{\ep\mu_{\ell+1,\ell}} -1)(e^{-\ep\lam_{k,k}} -1)}{e^{\ep} -1} 
\cr
&+&
\sum_{k, \ell \geq1} e^{a_{k+1} - b_{\ell}} 
e^{\ep + \es(\floor{ \frac{\ell}{2}-1 } - \floor{ \frac{k}{2} -\frac{3}{2} })}
\frac{(e^{\ep\mu_{\ell,\ell}} -1)(e^{-\ep\lam_{k+1,k}} -1)}{e^{\ep} -1} 
\cr
&+&
\sum_{\ell \geq 1} e^{a_1 - b_{\ell+1}} 
e^{\ep + \es(\floor{ \frac{\ell}{2} - \frac{1}{2} }+1)}
\frac{e^{\ep\mu_{\ell+1,\ell}} -1}{e^{\ep} -1} 
\cr
&+&
\sum_{\ell \geq 1} e^{a_2 - b_{\ell}} 
e^{\ep + \es(\floor{ \frac{\ell}{2} - 1 }+1)}
\frac{e^{\ep\mu_{\ell,\ell}} -1}{e^{\ep} -1}
\cr
&-&
\sum_{k, \ell \geq1} e^{a_{k} - b_{\ell}} 
e^{\ep + \es(\floor{ \frac{\ell}{2}- 1  } - \floor{ \frac{k}{2} -1 })}
\frac{(e^{\ep\mu_{\ell,\ell}} -1)(e^{-\ep\lam_{k,k}} -1)}{e^{\ep} -1}
\cr
&-&
\sum_{k, \ell \geq1} e^{a_{k+1} - b_{\ell+1}} 
e^{\ep + \es(\floor{ \frac{\ell}{2}- \frac{3}{2}  } - \floor{ \frac{k}{2} - \frac{3}{2} })}
\frac{(e^{\ep\mu_{\ell+1,\ell}} -1)(e^{-\ep\lam_{k+1,k}} -1)}{e^{\ep} -1}
\cr
&-&
\sum_{k \geq 1} e^{a_{k} - b_{1}} 
e^{\ep + \es(-1- \floor{ \frac{k}{2}-1 })}
\frac{e^{-\ep\lam_{k,k}} -1}{e^{\ep} -1}
\cr
&-&
\sum_{k \geq 1} e^{a_{k+1} - b_{2}} 
e^{\ep + \es(-1- \floor{ \frac{k}{2}- \frac{3}{2} })}
\frac{e^{-\ep\lam_{k+1,k}} -1}{e^{\ep} -1}. \label{A1}
\eeqa
Here the floor function $\floor k$ denotes
the largest integer not greater than $k$.
We have rewritten the original formula by Feigin et. al. (\cite{Feigin}. Prop.4.15) 
to arrive at \eqref{A1}. 


We can rewrite this character as a Laurent polynomial in
$e^{a_i}$, $e^{b_i}$ and $e^{\epsilon_i}$ 
with non-negative integer coefficients as follows:
\\
{\bf Proposition.}~{\it 
\beqa
{\rm ch}_{\vec\lam,\vec\mu}(\vec a,\vec b)
&=&
\sum_{k\geq1 \atop \ell\geq0} 
e^{a_{k} - b_{\ell+1}} 
e^{\es(\floor{ \frac{\ell+1}{2} } - \floor{ \frac{k}{2} })}
\sum_{i=
\min(1,1+\mu_{\ell+1,\ell+1}-\lam_{k,k})
}^{
\min(0,  \mu_{\ell+1,\ell  }-\lam_{k,k})
}
e^{i\ep} 
\cr
&+&
\sum_{k\geq1 \atop \ell\geq0} 
e^{a_{k-1} - b_{\ell}} 
e^{\es(\floor{ \frac{\ell}{2} } - \floor{ \frac{k-1}{2} })}
\sum_{i=
\min(1,1+\mu_{\ell,\ell+1}-\lam_{k-1,k})
}^{
\min(0,  \mu_{\ell,\ell  }-\lam_{k-1,k})
}
e^{i\ep} 
\cr
&+&
\sum_{k\geq0 \atop \ell\geq1} 
e^{a_{k} - b_{\ell}} 
e^{\es(\floor{ \frac{\ell}{2} } - \floor{ \frac{k}{2} })}
\sum_{i=
\max(1,1+\mu_{\ell,\ell}-\lam_{k,k  })
}^{
\max(0,  \mu_{\ell,\ell}-\lam_{k,k+1})
}
e^{i\ep} 
\cr
&+&
\sum_{k\geq0 \atop \ell\geq1} 
e^{a_{k+1} - b_{\ell+1}} 
e^{\es(\floor{ \frac{\ell+1}{2} } - \floor{ \frac{k+1}{2} })}
\sum_{i=
\max(1,1+\mu_{\ell+1,\ell}-\lam_{k+1,k  })
}^{
\max(0,  \mu_{\ell+1,\ell}-\lam_{k+1,k+1})
}
e^{i\ep} 
\label{eq:propCh4}%
\eeqa
with $\lam_{k,0}=\mu_{k,0}:=\infty$. 
}

{\noindent{\it Proof.\hskip10pt}}
Since 
\beq
q\frac{(q^N-1)(q^{-M}-1)}{q-1}
=
\left[
\sum_{i=1-M}^{\min(0,N-M)}
-
\sum_{i=\max(1,1+N-M)}^N 
\right]
q^{i}
\eeq
for any $M,N=0,1,2,3,\cdots$, 
the character ${\rm ch}_{\vec\lam,\vec\mu}(\vec a,\vec b)$ reduces to
\beqa
&&
\sum_{k, \ell \geq1} 
e^{a_{k} - b_{\ell+1}} 
e^{\es(\floor{ \frac{\ell+1}{2} } - \floor{ \frac{k}{2} })}
\left[
\sum_{i=
1-\lam_{k,k}
}^{
\min(0,  \mu_{\ell+1,\ell}-\lam_{k,k})}
-
\sum_{i=
\max(1,1+\mu_{\ell+1,\ell}-\lam_{k,k})
}^{      \mu_{\ell+1,\ell}
}
\right]
e^{i\ep} 
\label{eq:proofCh1}%
\\
&+&
\sum_{k, \ell \geq1} 
e^{a_{k-1} - b_{\ell}} 
e^{\es(\floor{ \frac{\ell}{2} } - \floor{ \frac{k-1}{2} })}
\left[
\sum_{i=
1-\lam_{k-1,k}
}^{
\min(0,  \mu_{\ell,\ell}-\lam_{k-1,k})}
-
\sum_{i=
\max(1,1+\mu_{\ell,\ell}-\lam_{k-1,k})
}^{      \mu_{\ell,\ell}
}
\right]
e^{i\ep} 
\\
&+&
\sum_{\ell \geq 1} 
e^{a_1 - b_{\ell+1}} 
e^{\es\floor{ \frac{\ell+1}{2} } }
\sum_{i=1}^{\mu_{\ell+1,\ell}}
e^{i\ep} 
+
\sum_{\ell \geq 1} 
e^{a_0 - b_{\ell}} 
e^{\es\floor{ \frac{\ell}{2} } }
\sum_{i=1}^{\mu_{\ell,\ell}}
e^{i\ep} 
\\
&+&
\sum_{k, \ell \geq1} 
e^{a_{k} - b_{\ell}} 
e^{\es(\floor{ \frac{\ell}{2} } - \floor{ \frac{k}{2} })}
\left[
\sum_{i=
\max(1,1+\mu_{\ell,\ell}-\lam_{k,k})
}^{      \mu_{\ell,\ell}
}
-
\sum_{i=
1-\lam_{k,k}
}^{
\min(0,  \mu_{\ell,\ell}-\lam_{k,k})}
\right]
e^{i\ep} 
\label{eq:proofCh5}%
\\
&+&
\sum_{k, \ell \geq1} 
e^{a_{k+1} - b_{\ell+1}} 
e^{\es(\floor{ \frac{\ell+1}{2} } - \floor{ \frac{k+1}{2} })}
\left[
\sum_{i=
\max(1,1+\mu_{\ell+1,\ell}-\lam_{k+1,k})
}^{      \mu_{\ell+1,\ell}
}
\!\!-\!\!
\sum_{i=
1-\lam_{k+1,k}
}^{
\min(0,  \mu_{\ell+1,\ell}-\lam_{k+1,k})}
\right]
e^{i\ep}
~~~~
\\
&+&
\sum_{k \geq 1} 
e^{a_{k} - b_{1}} 
e^{-\es\floor{ \frac{k}{2} }}
\sum_{i=1-\lam_{k,k}}^0
e^{i\ep} 
\label{eq:proofCh7}%
+
\sum_{k \geq 1} 
e^{a_{k+1} - b_{2}} 
e^{\es(1 - \floor{ \frac{k+1}{2} })}
\sum_{i=1-\lam_{k+1,k}}^0
e^{i\ep}. 
\eeqa
Adding the second term of (\ref{eq:proofCh5}) with $\ell\geq2$
and
the first term of (\ref{eq:proofCh1})
yields
\beq
\sum_{k,\ell\geq1} 
e^{a_{k} - b_{\ell+1}} 
e^{\es(\floor{ \frac{\ell+1}{2} } - \floor{ \frac{k}{2} })}
\sum_{i=
\min(1,1+\mu_{\ell+1,\ell+1}-\lam_{k,k})
}^{
\min(0,  \mu_{\ell+1,\ell  }-\lam_{k,k})
}
e^{i\ep}. 
\label{eq:proofCh5plus1}%
\eeq
On the other hand, adding the second term of (\ref{eq:proofCh5}) with $\ell=1$
and the first term of 
(\ref{eq:proofCh7})
yields
\beq
\sum_{k\geq1} 
e^{a_{k} - b_{1}} 
e^{-\es\floor{ \frac{k}{2} }}
\sum_{i=\min(1,1+\mu_{1,1}-\lam_{k,k})}^0
e^{i\ep}.
\label{eq:proofCh5plus7}%
\eeq
Combining (\ref{eq:proofCh5plus1}) with (\ref{eq:proofCh5plus7}) 
gives the first term of (\ref{eq:propCh4}).
In the same manner we can get other terms. 
\hfill\fbox{}


Let us introduce the following signature 
(Fig. \ref{fig:SGN})
\beq
{\rm sgn}_\lambda(i,j)
:=
\left\{
\begin{array}{l}
-1,
\\
\\
+1,
\end{array}
\right.
\quad
\begin{array}{l}
{\rm if }~~
\left\{
\begin{array}{ll}
\lambda_{2n+1}<j\leq \lambda_{2n}
~~&{\rm and}~~
i=2m-1
~~{\rm or}
\\
\lambda_{2n+2}<j\leq \lambda_{2n+1}
~&{\rm and}~~
i=2m,
\end{array}
\right.
\\
{\rm if }~~
\left\{
\begin{array}{ll}
\lambda_{2n+1}<j\leq \lambda_{2n}
~~&{\rm and}~~
i=2m
~~\hskip21pt{\rm or}
\\
\lambda_{2n+2}<j\leq \lambda_{2n+1}
~&{\rm and}~~
i=2m-1
\end{array}
\right.
\end{array}
\eeq
with $n=0,1,2,\cdots$ and $m=1,2,3,\cdots$.
Here $\lambda_0:=\infty$.
Note that if $\lambda_{k+1} = \lambda_{k+2}$ then 
${\rm sgn}_\lambda(i,j)={\rm sgn}_{\lambda^{\rm red}}(i,j)$ 
with 
$\lambda^{\rm red}:=(\lambda_1,\cdots,\lambda_{k},\lambda_{k+3},\cdots)$.
%
\FigSgn
%
%
Then we can represent the character 
as a summation over some squares in the Young diagrams
as follows:
\\
{\bf Proposition.}~{\it 
\beqa
{\rm ch}_{\vec\lambda,\vec\mu}(\vec a,\vec b)
&=&
\sum_{I,J=1}^2
{\rm ch}_{\lambda_I,\mu_J}^{I,J}(a_I,b_J),
\cr
{\rm ch}_{\lambda,\mu}^{I,J}(a,b)
&:=&
{\rm exp}\left\{
a-b
+\ha\ep
+\left(\ha+J-I\right){\frac\es 2}
\right\}
\cr
&\times&
\left\{
\sum_{(i,j)\in\lambda \atop {\rm sgn}_\mu(i,j) = (-1)^{I+J+1} }
{\rm exp}\left\{
-\left(\lambda_i-j+\ha\right)\ep 
+\left(\mu'_j-i+\ha\right){\frac\es 2}
\right\}
\right.
\cr
&&\hskip6pt
+
\left.
\sum_{(i,j)\in\mu \atop {\rm sgn}_\lambda(i,j) = (-1)^{I+J} }
{\rm exp}\left\{
~~\left(\mu_i-j+\ha\right)\ep 
-\left(\lambda'_j-i+\ha\right){\frac\es 2}
\right\}
\right\}.~~~
\label{eq:PropByYang}%
\eeqa
Moreover ${\rm ch}_{\vec\lambda,\vec\mu}(\vec a,\vec b)$
is symmetric under the replacement
$(a_1+{\frac\es 4},b_1+{\frac\es 4}) \leftrightarrow 
(a_2-{\frac\es 4},b_2-{\frac\es 4})$ and
$(\lambda_1,\mu_1) \leftrightarrow (\lambda_2,\mu_2)$.
}

{\noindent{\it Proof.\hskip10pt}}
Let
${\rm ch}_{\lambda_I,\mu_J}^{I,J}(a_I,b_J)$
be a part of 
${\rm ch}_{\vec\lambda,\vec\mu}(\vec a,\vec b)$,
which contains $e^{a_I-b_J}$,
i.e.,
\beqa
{\rm ch}_{\lam,\mu}^{1,1}(a,b)
&:=&
\sum_{k\geq1 \atop \ell\geq0} 
e^{a - b} 
e^{\es(\ell - k + 1)}
\sum_{i=
\min(1,1+\mu_{2\ell+1}-\lam_{2k-1})
}^{
\min(0,  \mu_{2\ell  }-\lam_{2k-1})
}
e^{i\ep} 
\cr
&+&
\sum_{k,\ell\geq1}
e^{a - b} 
e^{\es(\ell - k )}
\sum_{i=
\min(1,1+\mu_{2\ell  }-\lam_{2k})
}^{
\min(0,  \mu_{2\ell-1}-\lam_{2k})
}
e^{i\ep} 
\cr
&+&
\sum_{k,\ell\geq1}
e^{a - b} 
e^{\es(\ell - k )}
\sum_{i=
\max(1,1+\mu_{2\ell-1}-\lam_{2k-1})
}^{
\max(0,  \mu_{2\ell-1}-\lam_{2k  })
}
e^{i\ep} 
\cr
&+&
\sum_{k\geq0 \atop \ell\geq1} 
e^{a - b} 
e^{\es(\ell - k )}
\sum_{i=
\max(1,1+\mu_{2\ell}-\lam_{2k  })
}^{
\max(0,  \mu_{2\ell}-\lam_{2k+1})
}
e^{i\ep}, 
\eeqa
\beqa
{\rm ch}_{\lam,\mu}^{2,1}(a,b)
&:=&
\sum_{k\geq1 \atop \ell\geq0} 
e^{a - b} 
e^{\es(\ell - k  )}
\sum_{i=
\min(1,1+\mu_{2\ell+1}-\lam_{2k})
}^{
\min(0,  \mu_{2\ell  }-\lam_{2k})
}
e^{i\ep} 
\cr
&+&
\sum_{k,\ell\geq1}
e^{a - b} 
e^{\es(\ell - k )}
\sum_{i=
\min(1,1+\mu_{2\ell  }-\lam_{2k-1})
}^{
\min(0,  \mu_{2\ell-1}-\lam_{2k-1})
}
e^{i\ep} 
\cr
&+&
\sum_{k\geq0 \atop \ell\geq1}
e^{a - b} 
e^{\es(\ell - k -1)}
\sum_{i=
\max(1,1+\mu_{2\ell-1}-\lam_{2k  })
}^{
\max(0,  \mu_{2\ell-1}-\lam_{2k+1})
}
e^{i\ep}
\cr
&+&
\sum_{k,\ell\geq1}
e^{a - b} 
e^{\es(\ell - k)}
\sum_{i=
\max(1,1+\mu_{2\ell}-\lam_{2k-1})
}^{
\max(0,  \mu_{2\ell}-\lam_{2k  })
}
e^{i\ep} 
\eeqa
and
$
{\rm ch}_{\lam,\mu}^{2,2}(a,b)
:=
{\rm ch}_{\lam,\mu}^{1,1}(a,b)
$
and 
$
{\rm ch}_{\lam,\mu}^{1,2}(a,b)
:=
{\rm ch}_{\lam,\mu}^{2,1}(a,b) e^{\es}
$.
Then we obtain
\beqa
{\rm ch}_{\lam,\mu}^{1,1}(a,b)
&=&
\sum_{(i,j)\in\lambda \atop {\rm sgn}_\mu(i,j) = -1 }
{\rm exp}\left\{ a-b 
-\left(\lambda_i-j \right)\ep 
+\left(\mu'_j-i+1\right){\frac\es 2}
\right\}
\cr
&+&
\sum_{(i,j)\in\mu \atop {\rm sgn}_\lambda(i,j) = 1 }
{\rm exp}\left\{ a-b 
+\left(\mu_i-j+1 \right)\ep 
-\left(\lambda'_j-i\right){\frac\es 2}
\right\},
\cr
{\rm ch}_{\lam,\mu}^{2,1}(a,b)
&=&
\sum_{(i,j)\in\lambda \atop {\rm sgn}_\mu(i,j) = 1 }
{\rm exp}\left\{ a-b 
-\left(\lambda_i-j \right)\ep 
+\left(\mu'_j-i\right){\frac\es 2}
\right\}
\cr
&+&
\sum_{(i,j)\in\mu \atop {\rm sgn}_\lambda(i,j) = -1 }
{\rm exp}\left\{ a-b 
+\left(\mu_i-j+1 \right)\ep 
-\left(\lambda'_j-i+1\right){\frac\es 2}
\right\}
,
\eeqa
which proves (\ref{eq:PropByYang}).
Since
\beqa
{\rm ch}_{\lam_2,\mu_2}^{2,2}(a_2+{\frac\es 4},b_2+{\frac\es 4})
&=&
{\rm ch}_{\lam_2,\mu_2}^{1,1}(a_2-{\frac\es 4},b_2-{\frac\es 4}),
\cr
{\rm ch}_{\lam_1,\mu_2}^{1,2}(a_1-{\frac\es 4},b_2+{\frac\es 4})
&=&
{\rm ch}_{\lam_1,\mu_2}^{2,1}(a_1+{\frac\es 4},b_2-{\frac\es 4}),
\eeqa
the proposition follows.
\hfill\fbox{}


Especially when 
$\vec\mu=\vec\lambda$, 
$\vec\mu=\vec\emptyset$ or
$\vec\lambda=\vec\emptyset$,
we can also represent the character 
as a summation over all squares in the Young diagrams:
\\
{\bf Corollary.}~{\it 
\beqa
{\rm ch}_{\vec\lambda,\vec\lambda}(\vec a,\vec b)
&=&
\sum_{I,J=1}^2
\widetilde{\rm ch}_{\lambda_I,\lambda_J}^{I,J}(\vec a,\vec b),
\label{eq:CorLamLam}%
\cr
\widetilde{\rm ch}_{\lambda,\mu}^{I,J}(\vec a,\vec b)
&:=&
\sum_{(i,j)\in\lambda}
{\rm exp}\left\{
a_{I+( 1-{\rm sgn}_\mu(i,j)^{I-J})/2  }
-
b_{J+( 1-{\rm sgn}_\mu(i,j)^{I-J})/2  }
+\ha\left(\ep+{\frac\es 2}\right)
\right\}
\cr
&&\hskip-60pt
\times
{\rm exp}\left\{
\left(
\left(\lambda_i-j+\ha\right)\ep 
-\left(\mu'_j-i+\ha+J-I\right){\frac\es 2}
\right)
(-1)^{I+J}
{\rm sgn}_\mu(i,j)
\right\},
\eeqa
\beqa
{\rm ch}_{\vec\lambda,\vec\emptyset}(\vec a,\vec b)
&=&
\sum_{I=1}^2
e^{a_I}
\left[
e^{-b_I}
\sum_{ (i,j)\in\lambda_I \atop i:{\rm odd} }
+
e^{-b_{I+1}+ {\frac\es 2}(-1)^{I+1} }
\sum_{ (i,j)\in\lambda_I \atop i:{\rm even} }
\right]
e^{ 
\left(1-j\right)\ep 
+\left(1-i\right){\frac\es 2}
}, 
\label{eq:CorLamEmpty}%
\\
{\rm ch}_{\vec\emptyset,\vec\mu}(\vec a,\vec b)
&=&
\sum_{J=1}^2
e^{-b_J}
\left[
e^{a_J}
\sum_{ (i,j)\in\mu_J \atop i:{\rm even} }
+
e^{a_{J+1}+ {\frac\es 2}(-1)^J }
\sum_{ (i,j)\in\mu_J \atop i:{\rm odd} }
\right]
e^{ 
j\ep 
+i{\frac\es 2}
}. 
\label{eq:CorEmptyLam}%
\eeqa
}

{\noindent{\it Proof.\hskip10pt}}
Let
$
\widetilde{\rm ch}_{\lambda,\lambda}^{I,I}(\vec a,\vec b)
:=
{\rm ch}_{\lambda,\lambda}^{I,I}(a_I,b_I).
$
For $I\neq J$, let
$
\widetilde{\rm ch}_{\lambda,\mu}^{I,J}(\vec a,\vec b)
$
be the combination of the terms of 
$
{\rm ch}_{\lambda,\mu}^{I,J}(a_I,b_J)
$
with negative powers in $e^{\ep}$
and those of 
$
{\rm ch}_{\mu,\lambda}^{J,I}(a_J,b_I)
$
with positive powers. 
Then we get (\ref{eq:CorLamLam}).
When $\vec\mu=\vec\emptyset$,
since 
${\rm sgn}_\emptyset(i,j) = (-1)^i$,
if $i$ is odd or even number, 
then $I=J$ or $I\neq J$, respectively.
Thus
$
{\rm ch}_{\lambda,\emptyset}^{1,1}(a_1,b_1)
+
{\rm ch}_{\lambda,\emptyset}^{1,2}(a_1,b_2)
$
and 
$
{\rm ch}_{\lambda,\emptyset}^{2,2}(a_2,b_2)
+
{\rm ch}_{\lambda,\emptyset}^{2,1}(a_2,b_1)
$
give the $I=1$ and $2$ part of (\ref{eq:CorLamEmpty}), respectively.
On the other hand, when $\vec\lambda=\vec\emptyset$,
if $i$ is even or odd number, 
then $I= J$ or $I\neq J$, respectively,
and in the same manner we obtain (\ref{eq:CorEmptyLam}).
\hfill\fbox{}


\Section*{Appendix B : Multi points insertion of degenerate operators}
\renewcommand{\theequation}{B.\arabic{equation}}\setcounter{equation}{0}
\renewcommand{\thesubsection}{B.\arabic{subsection}}\setcounter{subsection}{0}

\subsection{$N_f=0$ case}

In $N_f=0$ case, we put $\langle A |=\langle \Delta', \Lambda |$ and $| B \rangle=|\Delta, \Lambda \rangle$
in eq.(\ref{eq:cftNpt}), then we have
\beq
\begin{array}l
\displaystyle
\Big[\frac{\Lambda}{4}\partial_{\Lambda}+
\frac{\Delta+\Delta'-N h_{1,2}}{2}
+\Lambda^2(z_1+\frac{1}{z_1})+b^2 z_1^2 \partial_{z_1}^2-\frac{3}{2}z_1\partial_{z_1}\\
\displaystyle 
+\sum_{j=2}^N\big\{(\frac{z_1z_j}{z_1-z_j}-\frac{z_j}{2})\partial_{z_j}+\frac{z_1^2h_{1,2}}{(z_1-z_j)^2}\big\}\Big] \Psi=0.
\end{array}
\eeq
We set the dimensions of initial state as $\Delta=\Delta(a_0)$, then the dimensions of intermediate and final states $\Delta(a_i)$ are 
restricted by the fusion rule as $a_{i+1}=a_i\pm \dfrac{1}{2b}$ and $\Delta'=\Delta(a_N)$. 
For each choice of the intermediate channels (called fusion path), one has a series solution of the form
\beq
\Psi=\prod_{i=1}^{N}z_i^{\Delta(a_{i})-\Delta(a_{i-1})-h_{1,2}}
\prod_{1\leq i<j \leq N}(1-\frac{z_j}{z_i})^{-\frac{1}{2 b^2}}Y(z), \quad
Y(z)=\sum_{n=0}^{\infty} Y_n(z)\Lambda^{2n}.
\eeq

\noindent $\bullet$ The case $N=2$: For the simplest fusion path $a_{i}=a+ \frac{i}{2b}$, we have
\beq
\begin{array}l
\displaystyle
Y_0=1, \quad
Y_1=\frac{1}{-b^2-2ab-1}(\frac{1}{z_2}+\frac{1}{z_1})+\frac{1}{-b^2+2ab+1}(z_1+z_2), \\[6mm]
\displaystyle
Y_2=c_{1}
+c_{2}(\frac{1}{z_2^2}+\frac{1}{z_1^2})
+c_{3}(\frac{z_1}{z_2}+\frac{z_2}{z_1})
+c_{4}z_1z_2
+\frac{c_{5}}{z_1z_2}
+c_{6}(z_1^2+z_2^2),\\
c_{1}=\frac{2(b^2+1)}{(b^2-2ab-1)(b^2+2ab+1)},\quad
c_{2}=\frac{1}{2(b^2+2ab+1)(2b^2+2ab+1)},\\
c_{3}=-\frac{1}{(-b^2+2ab+1)(b^2+2ab+1)},\quad
c_{4}=\frac{2(a-b)}{(2a-b)(-2b^2+2ab+1)(-b^2+2ab+1)},\\
c_{5}=\frac{2(b^2+ab+1)}{(b^2+2ab+1)(b^2+2ab+2)(2b^2+2ab+1)},\quad
c_{6}=\frac{1}{2(-2b^2+2ab+1)(-b^2+2ab+1)},\\[6mm]
\displaystyle
Y_3=c_{1}(\frac{1}{z_2^3}+\frac{1}{z_1^3})
+c_{2}(\frac{1}{z_2}+\frac{1}{z_1})
+c_{3}(z_1+z_2)
+c_{4}(z_1^3+z_2^3)\\
\displaystyle
+c_{5}(\frac{1}{z_2^2z_1}+\frac{1}{z_2z_1^2})
+c_{6}(z_2z_1^2+z_2^2z_1)
+c_{7}(\frac{z_1}{z_2^2}+\frac{z_2}{z_1^2})
+c_{8}(\frac{z_1^2}{z_2}+\frac{z_2^2}{z_1}),\\
c_{1}=-\frac{1}{6(b^2+2ab+1)(2b^2+2ab+1)(3b^2+2ab+1)},\quad
c_{2}=\frac{8b^4+8ab^3+13b^2+6ab+6}{2(-b^2+2ab+1)(b^2+2ab+1)(b^2+2ab+2)(2b^2+2ab+1)},\\
c_{3}=-\frac{-8b^3+8ab^2-5b+6a}{2(2a-b)(-2b^2+2ab+1)(-b^2+2ab+1)(b^2+2ab+1)},\quad
c_{4}=\frac{1}{6(-3b^2+2ab+1)(-2b^2+2ab+1)(-b^2+2ab+1)},\\
c_{5}=-\frac{3b^2+2ab+2}{2(b^2+2ab+1)(b^2+2ab+2)(2b^2+2ab+1)(3b^2+2ab+1)},\quad
c_{6}=\frac{2a-3b}{2(2a-b)(-3b^2+2ab+1)(-2b^2+2ab+1)(-b^2+2ab+1)},\\
c_{7}=\frac{1}{2(-b^2+2ab+1)(b^2+2ab+1)(2b^2+2ab+1)},\quad
c_{8}=-\frac{1}{2(-2b^2+2ab+1)(-b^2+2ab+1)(b^2+2ab+1)}.
\end{array}
\eeq
Then the free energy is given as
\beq
\log Y(z_1,z_2)=g(z_1)+g(z_2)+g(z_1,z_2),
\eeq
where
\beq
\begin{array}l
g(z_1)=\Lambda^2 (\frac{z_1}{2 a b-b^2+1}-\frac{1}{z_1 (2 a b+b^2+1)})\\
+\Lambda^4 (\frac{b^2 z_1^2}{2 (2 a b-2 b^2+1) (2 a b-b^2+1)^2}
-\frac{b^2}{2 z_1^2 (2 a b+b^2+1)^2 (2 a b+2 b^2+1)}
-\frac{b^2}{(2 a b-b^2+1) (2 a b+b^2+1)})\\
+\Lambda^6 (\frac{2 b^4 z_1^3}{3 (2 a b-3 b^2+1) (2 a b-2 b^2+1) (2 a b-b^2+1)^3}
-\frac{2 b^4}{z_1 (2 a b-b^2+1) (2 a b+b^2+1)^2 (2 a b+b^2+2) (2 a b+2 b^2+1)}\\
-\frac{2 b^4}{3 z_1^3 (2 a b+b^2+1)^3 (2 a b+2 b^2+1) (2 a b+3 b^2+1)}
+\frac{2 b^3 z_1}{(2a-b) (2 a b-2 b^2+1) (2 a b-b^2+1)^2 (2 a b+b^2+1)})+{\cal O}(\Lambda^8),
\end{array}
\eeq
and
\beq
\begin{array}l
g(z_1,z_2)=
\Lambda^4 (-\frac{b z_1 z_2}{(-2 a b+b^2-1)^2 (4 a^2 b-6 a b^2+2 a+2 b^3-b)}
-\frac{b^2}{z_1 z_2 (2 a b+b^2+1)^2 (2 a b+b^2+2) (2 a b+2 b^2+1)})\\
+\Lambda^6 (-\frac{2 b^4}{ (2 a b+b^2+1)^3(2 a b+b^2+2) (2 a b+2 b^2+1) (2 a b+3 b^2+1)}(\frac{1}{z_1^2 z_2}+\frac{1}{z_1 z_2^2})\\
-\frac{2 b^3 (z_1 z_2^2+z_1^2 z_2)}{(2 a-b) (2 a b-3 b^2+1) (2 a b-2 b^2+1) (2 a b-b^2+1)^3}+{\cal O}(\Lambda^8).
\end{array}
\eeq
Under the limit $a \rightarrow \frac{a}{\hbar}$, $\Lambda \rightarrow \frac{\Lambda}{\hbar}$ $\hbar \rightarrow 0$, we have
\beq
-\frac{\Lambda^4 (z_1^2 z_2^2+1)}{16 a^4 b^2 z_1 z_2}
-\frac{\Lambda^6 (z_1+z_2) (z_1^3 z_2^3+1)}{32 a^6 b^2 z_1^2 z_2^2}+{\cal O}(\hbar).
\label{N02}
\eeq
This agrees with the B model results.

\noindent $\bullet$ The case $N=3$: For the simplest fusion path $a_{i}=a+ \frac{i}{2b}$, we have
\beq
\begin{array}l
\displaystyle
Y_0(z)=1,\quad
Y_1=\frac{z_1+z_2+z_3}{2 a b-b^2+2}-\frac{1}{\left(2 a b+b^2+1\right)}(\frac{1}{z_1}+\frac{1}{z_2}+\frac{1}{z_3}),\\[6mm]
\displaystyle
Y_2=c_1
+c_2(\frac{1}{z_2^2}+\frac{1}{z_3^2}+\frac{1}{z_1^2})
+c_3(z_1^2+z_2^2+z_3^2)
+c_4(\frac{z_1}{z_2}+\frac{z_1}{z_3}+\frac{z_3}{z_2}+\frac{z_2}{z_3}+\frac{z_2}{z_1}+\frac{z_3}{z_1}),\\
\displaystyle
+c_5(z_1z_2+z_3z_2+z_1z_3)
+c_6(\frac{1}{z_1z_3}+\frac{1}{z_1z_2}+\frac{1}{z_3z_2})\\
c_1=\frac{2b^2+3}{(-2ab+b^2-2)(2ab+b^2+1)},\quad
c_2=\frac{1}{2(2ab+b^2+1)(2ab+2b^2+1)},\\
c_3=\frac{1}{4(ab-b^2+1)(2ab-b^2+2)},\quad
c_4=-\frac{1}{(2ab-b^2+2)(2ab+b^2+1)},\\
c_5=\frac{2ab-2b^2+1}{2(ab-b^2+1)(2ab-b^2+1)(2ab-b^2+2)},\quad
c_6=\frac{2(ab+b^2+1)}{(2ab+b^2+1)(2ab+b^2+2)(2ab+2b^2+1)},\\[6mm]
\displaystyle
Y_3=
c_{1}\big(\frac{1}{z_2^3}+\frac{1}{z_3^3}+\frac{1}{z_1^3}\big)
+c_{2} \big(\frac{1}{z_2}+\frac{1}{z_3}+\frac{1}{z_1}\big)
+c_{3} (z_1+z_2+z_3)\\
\displaystyle
+c_{4} \big(z_1^3+z_2^3+z_3^3\big)
+c_{5} \big(\frac{1}{z_2^2 z_3}+\frac{1}{z_2 z_3^2}+\frac{1}{z_2^2 z_1}+\frac{1}{z_3^2 z_1}+\frac{1}{z_2 z_1^2}+\frac{1}{z_3z_1^2}\big)\\
\displaystyle
+c_{6} \big(z_2 z_1^2+z_3 z_1^2+z_2^2 z_1+z_3^2 z_1+z_2 z_3^2+z_2^2 z_3\big)
+c_{7} \big(\frac{z_1}{z_2^2}+\frac{z_1}{z_3^2}+\frac{z_3}{z_2^2}+\frac{z_2}{z_3^2}+\frac{z_2}{z_1^2}+\frac{z_3}{z_1^2}\big)\\
\displaystyle
+c_{8} \big(\frac{z_1^2}{z_2}+\frac{z_1^2}{z_3}+\frac{z_3^2}{z_2}+\frac{z_2^2}{z_3}+\frac{z_2^2}{z_1}+\frac{z_3^2}{z_1}\big)
+c_{9} \big(\frac{z_1}{z_2 z_3}+\frac{z_3}{z_2 z_1}+\frac{z_2}{z_3 z_1}\big)\\
\displaystyle
+c_{10} \big(\frac{z_3 z_2}{z_1}+\frac{z_1z_2}{z_3}+\frac{z_1 z_3}{z_2}\big)
+c_{11} \frac{1}{z_1 z_2 z_3}
+c_{12} z_1 z_2 z_3,
\end{array}
\eeq
where
\beq
\begin{array}l
c_{1}= -\frac{1}{6 (2 a b+b^2+1) (2 a b+2 b^2+1) (2 a b+3 b^2+1)},\\
c_{2}= -\frac{8 a b^3+10 a b+8 b^4+17 b^2+10}{2 (-2 a b+b^2-2) (2 a b+b^2+1) (2 a b+b^2+2) (2 a b+2 b^2+1)},\\
c_{3}= -\frac{8 a b^3+10 a b-8 b^4-5 b^2+5}{4 (a b-b^2+1) (2 a b-b^2+1) (2 a b-b^2+2) (2 a b+b^2+1)},\\
c_{4}= \frac{1}{12 (2 a b-3 b^2+2) (a b-b^2+1) (2 a b-b^2+2)},\\
c_{5}= -\frac{2 a b+3 b^2+2}{2 (2 a b+b^2+1) (2 a b+b^2+2) (2 a b+2 b^2+1) (2 a b+3 b^2+1)},\\
c_{6}= \frac{2 a b-3 b^2+1}{4 (2 a b-3 b^2+2) (a b-b^2+1) (2 a b-b^2+1) (2 a b-b^2+2)},\\
c_{7}= -\frac{1}{2 (-2 a b+b^2-2) (2 a b+b^2+1) (2 a b+2 b^2+1)},\\
c_{8}= -\frac{1}{4 (-2 a b+b^2-2)(-a b+b^2-1) (2 a b+b^2+1)},\\
c_{9}= -\frac{2 (a b+b^2+1)}{(-2 a b+b^2-2) (2 a b+b^2+1) (2 a b+b^2+2) (2 a b+2 b^2+1)},\\
c_{10}= -\frac{2 a b-2 b^2+1}{2 (a b-b^2+1) (2 a b-b^2+1) (2 a b-b^2+2) (2 a b+b^2+1)},\\
c_{11}= -\frac{2 (2 a^2 b^2+5 a b^3+5 a b+3 b^4+7 b^2+3)}{(2 a b+b^2+1) (2 a b+b^2+2) (2 a b+b^2+3) (2 a b+2 b^2+1) (2 a b+3 b^2+1)},\\
c_{12}= \frac{4 a^2 b-10 a b^2+2 a+6 b^3-b}{2 (2 a-b) (2 a b-3 b^2+2) (a b-b^2+1) (2 a b-b^2+1) (2 a b-b^2+2)}.
\end{array}
\eeq

In the free energy $F=\log Y$, the relevant terms at order $\Lambda^6$ are
\beq
\begin{array}l
\displaystyle
\frac{4 b^3}{(2 a-b) \left(2 a b-3 b^2+2\right) \left(a b-b^2+1\right) \left(2 a b-b^2+1\right) \left(2 a b-b^2+2\right)^3} z_1z_2z_3\\
\displaystyle
-\frac{8 b^4}{\left(2 a b+b^2+1\right)^3 \left(2 a b+b^2+2\right) \left(2 a b+b^2+3\right) \left(2 a b+2 b^2+1\right) \left(2 a b+3 b^2+1\right)}
\frac{1}{z_1z_2z_3}.
\end{array}
\eeq
Under the limit $a \rightarrow \frac{a}{\hbar}$, $\hbar \rightarrow 0$, this gives
\beq
\hbar^7\frac{1}{16 a^7 b^3}(z_1z_2z_3-\frac{1}{z_1z_2z_3})+{\cal O}(\hbar^8).
\label{N03}
\eeq
This is consistent with the B model results.
   
\subsection{$N_f=1$ case}

We put $\langle A |=\langle \Delta_{-},\Lambda,m |$ and $|B\rangle=|\Delta_{+},\Lambda \rangle$, then (\ref{eq:cftNpt}) takes the form
\beq
\begin{array}l
\displaystyle
\Big[\Big(L_0\Big)_0+\frac{\Lambda^2}{z_1}+\Big(b^2 z_1^2 \partial_{z_1}^2-z_1 \partial_{z_1}\Big)
+\sum_{j=2}^{N}\Big(\frac{z_1z_j}{z_1-z_j}\partial_{z_j}+\frac{z_1^2}{(z_1-z_j)^2}h_{1,2}\Big)\\
\displaystyle
+z_1 (-2m\Lambda)+z_1^2 (-\Lambda^2)\Big]\Psi=0,
\end{array}
\eeq
where the action of the first term is given as
\beq
\Big(L_0\Big)_0=\frac{\Lambda}{3}\partial_{\Lambda}+\frac{\Delta_{-}+2\Delta_{+}}{3}-\frac{1}{3}
\sum_{i=1}^{N}(z_i\partial_{z_i}+h_{1,2}),
\eeq
by using the relations
$L_0 |B\rangle=(\Delta_{+}+\frac{\Lambda}{2}\partial_{\Lambda})|B\rangle$ and 
$\langle A | L_0=(\Delta_{-}+{\Lambda}\partial_{\Lambda})\langle A |$.

The equation has a solution such as $Y(z)=1+Y_1 \Lambda+Y_2 \Lambda^2+\cdots$ with the same
pre-factor as $N_f=0$ case.
The first terms are as follows
\beq
\begin{array}l
Y_0=1,\quad Y_1=-\frac{2 m (z_1+z_2)}{-b^2+2 a b+1},\\[6mm]
Y_2=
-\frac{(-b^2+2 a b-4 m^2+1)(z_1^2+z_2^2)}{2 (-2 b^2+2 a b+1) (-b^2+2 a b+1)}
+\frac{(-8 m^2 b^2-b^2+8 a m^2 b+2 a b+1)z_2z_1}{(2 a-b) b (-2 b^2+2 a b+1) (-b^2+2 a b+1)}
-\frac{1}{(b^2+2 a b+1)}(\frac{1}{z_1}+\frac{1}{z_2}),\\[6mm]
Y_3=
\frac{m (-5 b^2+6 a b-4 m^2+3) (z_1^3+z_2^3)}{3 (-3 b^2+2 a b+1) (-2 b^2+2 a b+1) (-b^2+2 a b+1)}
+\frac{m (3 b^4-8 a b^3+4 a^2 b^2+12 m^2 b^2+b^2-8 a m^2 b-2 a b-2)(z_1z_2^2+z_2z_1^2)}{(2 a-b) b (-3 b^2+2 a b+1) (-2 b^2+2 a b+1) (-b^2+2 a b+1)}\\
+\frac{2 mz_1}{(-b^2+2 a b+1) (b^2+2 a b+1) (z_1+z_2)}
+\frac{4 (b^2+1) m}{(-b^2+2 a b+1) (b^2+2 a b+1)},\\[6mm]
Y_4=
\frac{(9 b^4-24 a b^3+12 a^2 b^2+56 m^2 b^2-12 b^2-48 a m^2 b+12 a b+16 m^4-24 m^2+3)(z_1^4+z_2^4)}{24 (-4 b^2+2 a b+1) (-3 b^2+2 a b+1) (-2 b^2+2 a b+1) (-b^2+2 a b+1)}\\
-\frac{(80 m^2 b^4+9 b^4-136 a m^2 b^3-24 a b^3+64 m^4 b^2+12 a^2 b^2+48 a^2 m^2 b^2-12 m^2 b^2-12 b^2-32 a m^4 b+12 a b-12 m^2+3)(z_1z_2^3+z_2z_1^3)}
{6 (2 a-b) b (-4 b^2+2 a b+1) (-3 b^2+2 a b+1) (-2 b^2+2 a b+1) (-b^2+2 a b+1)}\\
+\frac{C_1 z_2^2z_1^2}{4 (a-b) (2 a-b) b^2 (-4 b^2+2 a b+1) (-3 b^2+2 a b+1) (-2 b^2+2 a b+1) (-b^2+2 a b+1)}\\
+\frac{(-b^2+2 a b-4 m^2+1)z_2^2}{2 (-2 b^2+2 a b+1) (-b^2+2 a b+1) (b^2+2 a b+1)}(\frac{1}{z_1}+\frac{1}{z_2})\\
+\frac{(32 m^2 b^4+5 b^4-32 a m^2 b^3-12 a b^3+4 a^2 b^2+20 m^2 b^2-3 b^2-24 a m^2 b-2 a b-2)(z_1+z_2)}{2 (2 a-b) b (-2 b^2+2 a b+1) (-b^2+2 a b+1) (b^2+2 a b+1)}\\
+\frac{2 (b^2+a b+1)}{(b^2+2 a b+1) (b^2+2 a b+2) (2 b^2+2 a b+1)z_2 z_1}
+\frac{1}{2 (b^2+2 a b+1) (2 b^2+2 a b+1)(z_1^2+z_2^2)},
\end{array}
\eeq
{\footnotesize
$C_1=-8 b m^2 (2 a-3 b) ((2 a^2 b^2-5 a b^3-a b+2 b^4+b^2-1)
+((2 a b-3 b^2+1) ((2 a b-b^2+1) ((2 a^2 b^2-5 a
   b^3+2 b^4+1)+16 b^2 m^4 (2 a-3 b) (a-2 b)$}.

Then the free energy $F=\log Y=g(z_1)+g(z_2)+g(z_1,z_2)$ is given by
\beq
\begin{array}l
g(z_1)=-\frac{2 m z_1 L}{-b^2+2 a b+1}
+(-\frac{(-b^2+2 a b-2 m b+1) (-b^2+2 a b+2 m b+1) z_1^2}{2 (-2 b^2+2 a b+1) (-b^2+2 a b+1)^2}
-\frac{1}{(b^2+2 a b+1) z_1}) \Lambda^2\\
+(-\frac{4 b^2 m (b^2-2 a b-2 m b-1) (b^2-2 a b+2 m b-1) z_1^3}{3 (b^2-2 a b-1)^3 (2 b^2-2 a b-1) (3 b^2-2 a b-1)}
-\frac{2 b^2 m}{(b^2-2 a b-1) (b^2+2 a b+1)}) \Lambda^3\\
+(\frac{b^2 (-b^2+2 a b-2 m b+1) (-b^2+2 a b+2 m b+1) C_2

z_1^4}{4 (-4 b^2+2 a b+1) (-3 b^2+2 a b+1) (-2 b^2+2 a b+1)^2 (-b^2+2 a b+1)^4}\\
-\frac{2 b (-b^2+2 a b-2 m b+1) (-b^2+2 a b+2 m b+1) z_1}{(2 a-b) (-2 b^2+2 a b+1) (-b^2+2 a b+1)^2 (b^2+2 a b+1)}
-\frac{b^2}{2 (b^2+2 a b+1)^2 (2 b^2+2 a b+1) z_1^2}) \Lambda^4+{\cal O}(\Lambda^5),\\[6mm]
g(z_1,z_2)=
+\frac{(-b^2+2 a b-2 m b+1) (-b^2+2 a b+2 m b+1) z_1 z_2 \Lambda^2}{(2 a-b) b (-2 b^2+2 a b+1) (-b^2+2 a b+1)^2}
-\frac{4 b m (-b^2+2 a b-2 m b+1) (-b^2+2 a b+2 m b+1) z_1 z_2 (z_1+z_2) \Lambda^3}{(2 a-b) (-3 b^2+2 a b+1) (-2 b^2+2 a b+1) (-b^2+2 a b+1)^3}\\
+(-\frac{b^2}{(b^2+2 a b+1)^2 (b^2+2 a b+2) (2 b^2+2 a b+1) z_1 z_2}
-\frac{(-b^2+2 a b-2 m b+1) (-b^2+2 a b+2 m b+1) C_2
 z_1 z_2 (z_1^2+z_2^2) b}{(2 a-b) (-4 b^2+2 a b+1) (-3 b^2+2 a b+1) (-2 b^2+2 a b+1)^2 (-b^2+2 a b+1)^4}\\
-\frac{(-b^2+2 a b-2 m b+1) (-b^2+2 a b+2 m b+1) C_3 z_1^2 z_2^2}{4 (a-b) (2 a-b)^2 (-4 b^2+2 a b+1) (-3 b^2+2 a b+1) (-2 b^2+2 a b+1)^2 (-b^2+2 a b+1)^4 b}) \Lambda^4
+{\cal O}(\Lambda^5),
\end{array}
\eeq
{\footnotesize $C_2=
-3 b^6+14 a b^5-20 a^2 b^4+44 m^2 b^4+7 b^4+8 a^3 b^3-40 a m^2 b^3-20 a b^3+12 a^2 b^2-20 m^2 b^2-5 b^2+6 a b+1$,
$C_3=-3 b^{10}+26 a b^9-88 a^2 b^8+172 m^2 b^8-11 b^8+144 a^3 b^7-664 a m^2 b^7+54 a b^7-112 a^4 b^6
-84 a^2 b^6+784 a^2 m^2 b^6+20 m^2 b^6+40 b^6+32 a^5 b^5+40 a^3 b^5-288 a^3 m^2 b^5+24 a m^2 b^5
-150 a b^5+168 a^2 b^4-32 a^2 m^2 b^4-52 m^2 b^4-36 b^4-56 a^3 b^3+64 a m^2 b^3+82 a b^3-44 a^2 b^2
+4 m^2 b^2+11 b^2-12 a b-1$}.

Under the limit $a \rightarrow \frac{a}{\hbar}$, $m \rightarrow \frac{m}{\hbar}$, $\Lambda \rightarrow \frac{\Lambda}{\hbar}$ and $\hbar \rightarrow 0$, we have
\beq
\begin{array}l
\displaystyle
g(z_1,z_2)\rightarrow 
\frac{ z_1 z_2 \big(a^2-m^2\big)}{4 a^4 b^2}\Lambda^2
-\frac{ m z_1 z_2 (z_1+z_2) \big(a^2-m^2\big)}{4 a^6 b^2}\Lambda^3
-\Lambda^4 \big(\frac{1}{16 a^4 b^2 z_1 z_2}\\[6mm]\displaystyle
+\frac{z_1^2 z_2^2 \big(a^2-9 m^2\big) \big(a^2-m^2\big)}{32 a^8 b^2}
+\frac{z_1 \big(z_1^2+z_2^2\big) z_2 \big(a^2-5 m^2\big) \big(a^2-m^2\big)}{16 a^8 b^2}\big)+{\cal O}(\Lambda^5).
\label{N12}
\end{array}
\eeq
Again, this recovers the B model results correctly.



\Section*{Appendix C : Schur functions and topological vertex}
\renewcommand{\theequation}{C.\arabic{equation}}\setcounter{equation}{0}
\renewcommand{\thesubsection}{C.\arabic{subsection}}\setcounter{subsection}{0}

The Schur function satisfies the following properties 
\cite{Macdonald}:
\begin{eqnarray}
&& s_{\mu}(cx)=c^{|\mu|}s_{\mu}(x), 
\quad s_{\mu}(q^{\rho})=q^{\kappa_{\mu}/2}s_{\mu^t}(q^{\rho}),
\quad s_{\mu}(q^{\rho})=(-1)^{|\mu|}s_{\mu^t}(q^{-\rho}), 
\label{f3}\\
&& s_{\mu}(q^{\rho})s_{\nu}(q^{\rho+{\mu}})
=s_{\nu}(q^{\rho})s_{\mu}(q^{\rho+{\nu}}),
\label{f4} 
\end{eqnarray}
where $|\mu|$ and $\kappa_{\mu}$ are
\begin{eqnarray}
&& |\mu|:=\sum_i \mu_i, \label{d1}\\
&& \kappa_{\mu}:=|\mu|+\sum_i \mu_i(\mu_i-2 i)=2\sum_{(i,j)\in \mu}(j-i), 
\quad \kappa_{\mu^t}=-\kappa_{\mu}.
\label{d2}
\end{eqnarray}
The Cauchy formulas for the Schur functions are 
\begin{eqnarray}
&& \sum_{\mu} s_{\mu}(x)s_{\mu}(y)=\prod_{i,j}\frac{1}{1-x_i y_j}
=\exp\left[\sum_{n,i,j}\frac{1}{n}x_i^n y_j^n\right],
\label{s1}\\
&& \sum_{\mu} s_{\mu}(x)s_{\mu^t}(y)=\prod_{i,j}(1+x_i y_j)
=\exp\left[-\sum_{n,i,j}\frac{(-1)^n}{n}x_i^n y_j^n\right].
\label{s2} 
\end{eqnarray}

The topological vertex in the canonical framing is \cite{Aganagic:2003db}
\beq
C_{\mu_1\mu_2\mu_3} (q) =q^{(\kappa_{\mu_2}+\kappa_{\mu_3})/2}
s_{\mu_2^t}(q^{\rho})
\sum_{\eta}s_{\mu_1/\eta}(q^{\rho+\mu_2^t})s_{\mu_3^t/\eta}(q^{\rho+\mu_2}),
\label{v1}
\eeq
where $s_{\mu/\nu}$ is the skew Schur function defined by
\begin{eqnarray}
s_{\mu/\nu}=\sum_{\eta}c^{\mu}_{\nu\eta}s_{\eta}.
\end{eqnarray}
We denote the Littlewood-Richardson coefficient by $c^{\mu}_{\nu\eta}$.
The topological vertex enjoys the cyclic symmetry
\beq
C_{\mu_1\mu_2\mu_3} (q) =C_{\mu_3\mu_1\mu_2} (q) =C_{\mu_2\mu_3\mu_1} (q).
\label{v2}
\eeq
If some of $\mu_i$'s are the trivial representation $\emptyset$,
the topological vertex simplifies as follows:
\begin{eqnarray}
&& C_{\mu\emptyset\emptyset}=s_{\mu}(q^{\rho}),
\label{v3}\\
&& C_{\mu\nu\emptyset}=q^{\kappa_\nu/2}s_{\mu}(q^{\rho})s_{\nu^t}(q^{\rho+\mu})
=s_{\nu}(q^{\rho})s_{\mu}(q^{\rho+\nu^t}).
\label{v4}
\end{eqnarray}

The gluing rule for the topological vertex is 
\begin{eqnarray}
 \sum_{\mu}C_{\mu\eta_1\eta_2}
(-Q)^{|\mu|}
(-1)^{n|\mu|}q^{-n\kappa_{\mu}/2}
C_{\mu^t\mu_1\mu_2},
\end{eqnarray}
where the integer $n$ is defined by the exterior product of 
the vectors $v_{\mu_1}$ and $v_{\eta_1}$
\begin{eqnarray} 
n=v_{\mu_1}\wedge v_{\eta_1}
=\det\left(
\begin{array}{cc}
v_{\mu_1}^1 & v_{\mu_1}^2 \\
v_{\eta_1}^1 & v_{\eta_1}^2
\end{array}
\right).
\end{eqnarray}
The vectors $v_{\mu_i}:=(v_{\mu_1}^1 , v_{\mu_1}^2 )$ and 
$v_{\eta_i}:=(v_{\eta_1}^1 , v_{\eta_1}^2 )$ are
the directions of the corresponding legs in the toric diagram.
\begin{figure}[tbp]
 \begin{center}
   \includegraphics[width=150mm,clip]{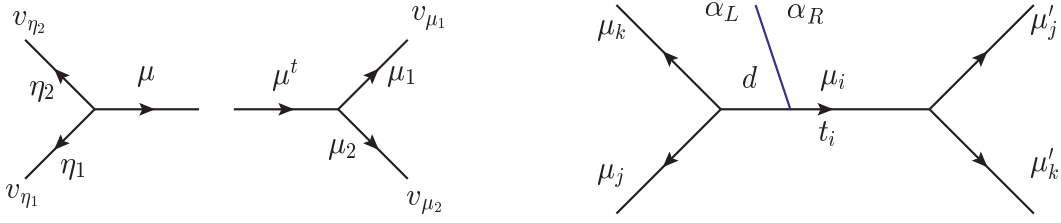}
  \caption{Gluing rule for topological vertex}
 \end{center}
\end{figure}
In particular for inner branes, the gluing rule is generalized as follows:
\beq
\sum_{\mu_i,\alpha^L,\alpha^R}
C_{\mu_j\mu_k(\mu_i\otimes \alpha^L)}(-1)^{s(i)}q^{f(i)}e^{-L(i)}
C_{(\mu_i^t\otimes \alpha^R)\mu_j^{\prime}\mu_k^{\prime}}~{\rm Tr}_{\alpha^L}V~{\rm Tr}_{\alpha^R}V^{-1},
\eeq
 with 
 \begin{eqnarray} 
&&L(i)=|\mu_i|t_i+|\alpha^L|r+|\alpha^R|(t_i-r),
\\
&&f(i)=p\kappa_{\mu_i\otimes \alpha^L}/2+(n+p)\kappa_{\mu_i^t\otimes \alpha^R}/2,\\
&&s(i)=|\mu_i|+p_i|\mu_i\otimes \alpha^L|+(n+p)|\mu_i^t\otimes \alpha^R|,
\end{eqnarray}
where $|\alpha\otimes\beta|=|\alpha|+|\beta|$
and $\kappa_{\alpha\otimes\beta}=\kappa_{\alpha}+\kappa_{\beta}$.


%

\end{document}